\documentclass[twoside,a4paper,12pt,openright]{book} 

\usepackage{amsmath,amssymb,amsthm}
\usepackage{eucal}
\usepackage{latexsym}
\usepackage{multicol}
\usepackage[latin1]{inputenc}
\usepackage{cite}

\usepackage{graphicx}
\usepackage{subfigure}
\usepackage{morefloats}
\usepackage{fancyhdr} 
\usepackage{appendix}

\newcommand{\ls}{\hspace{4mm}}

\newcommand{\clearemptydoublepage}{\newpage {\pagestyle{empty} \cleardoublepage}}

\renewcommand\caption[1]{\refstepcounter{figure} \linespread{1}\footnotesize  {\bf \figurename\ \thefigure} \ #1 \addcontentsline{lof}{figure}{\thefigure \ls #1}} 

\renewcommand{\P}{\textnormal{Pr}}
\newcommand{\bone}{\mathbf{1}}
\newcommand{\beq}{\begin{equation}}
\newcommand{\eeq}{\end{equation}}
\newcommand{\beqa}{\begin{eqnarray}}
\newcommand{\eeqa}{\end{eqnarray}}
\newcommand{\dfz}{\stackrel{def}{=}}
\newcommand{\bPhi}{\mbox{\boldmath{$\Phi$}}}

\newcommand{\bI}{\mbox{\boldmath{$I$}}}
\newcommand{\bT}{\mbox{\boldmath{$T$}}}
\newcommand{\bt}{\mbox{\boldmath{$t$}}}
\newcommand{\by}{\mbox{\boldmath{$y$}}}
\newcommand{\be}{\mbox{\boldmath{$e$}}}

\newcommand{\bW}{\mbox{\boldmath{$W$}}}

\newcommand{\bx}{\mbox{\boldmath{$x$}}}
\newcommand{\bz}{\mbox{\boldmath{$z$}}}

\newcommand{\bM}{\mbox{\boldmath{$M$}}}
\newcommand{\SNR}{\textnormal{SNR}}
\newcommand{\ARE}{\textnormal{ARE}}
\newcommand{\VAR}{\textnormal{VAR}}

\newcommand{\bzero}{\mbox{\boldmath{$0$}}}

\newcommand{\bs}{\mathbf{s}}
\newcommand{\bC}{\mathbf{C}}
\newcommand{\bphi}{\mathbf{\Phi}}

\newcommand{\N}{{\cal N}}

\newcommand{\E}{{\textnormal E}}

\begin{document}

\title{Running Consensus for Decentralized Detection}
\author{Paolo Braca}

\pagenumbering{gobble}
\begin{flushleft}
\includegraphics[width = 1.2\textwidth,angle=0]{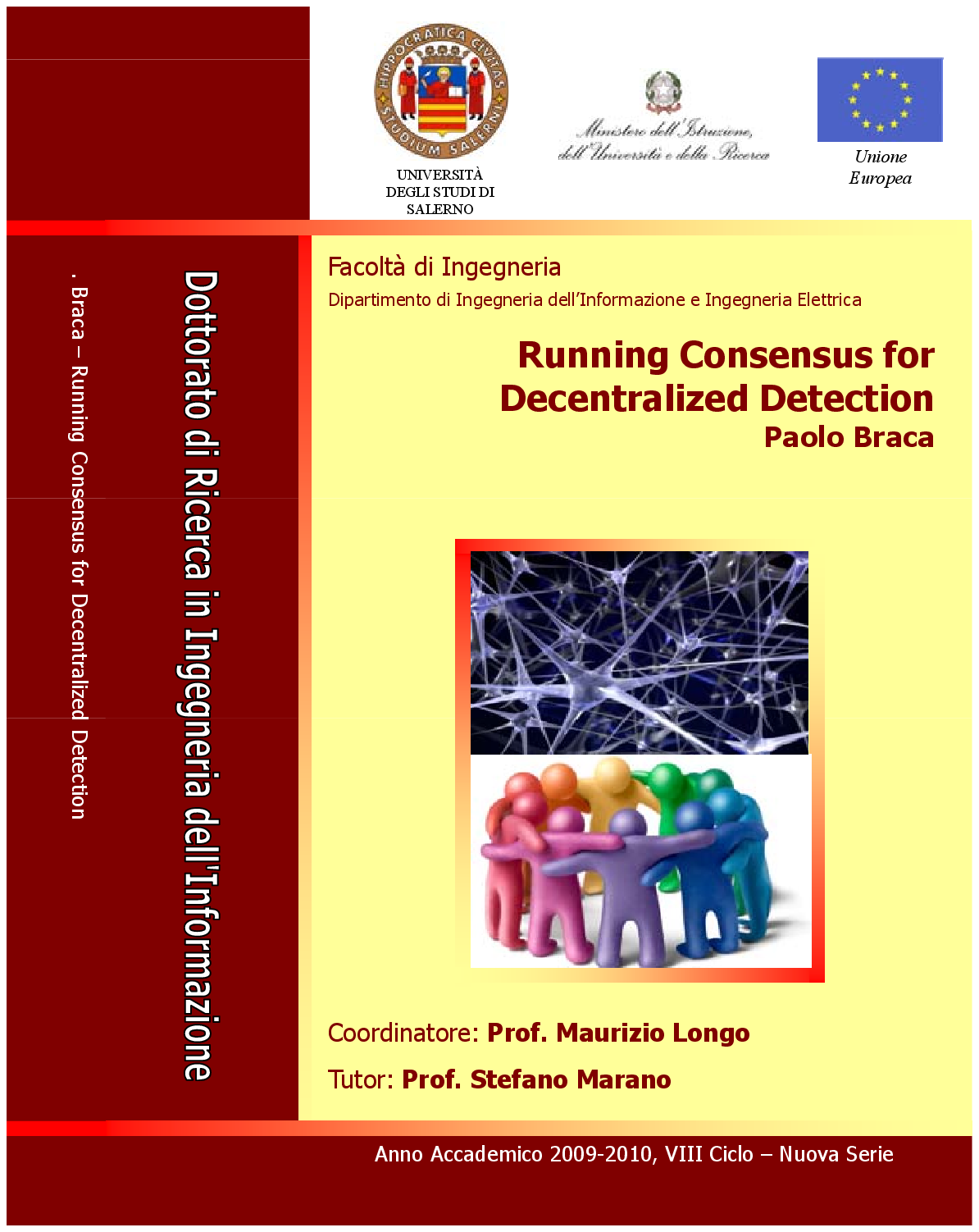}
\end{flushleft}
\newpage

\pagenumbering{gobble}
\begin{flushleft}
\includegraphics[width = 1.2\textwidth,angle=0]{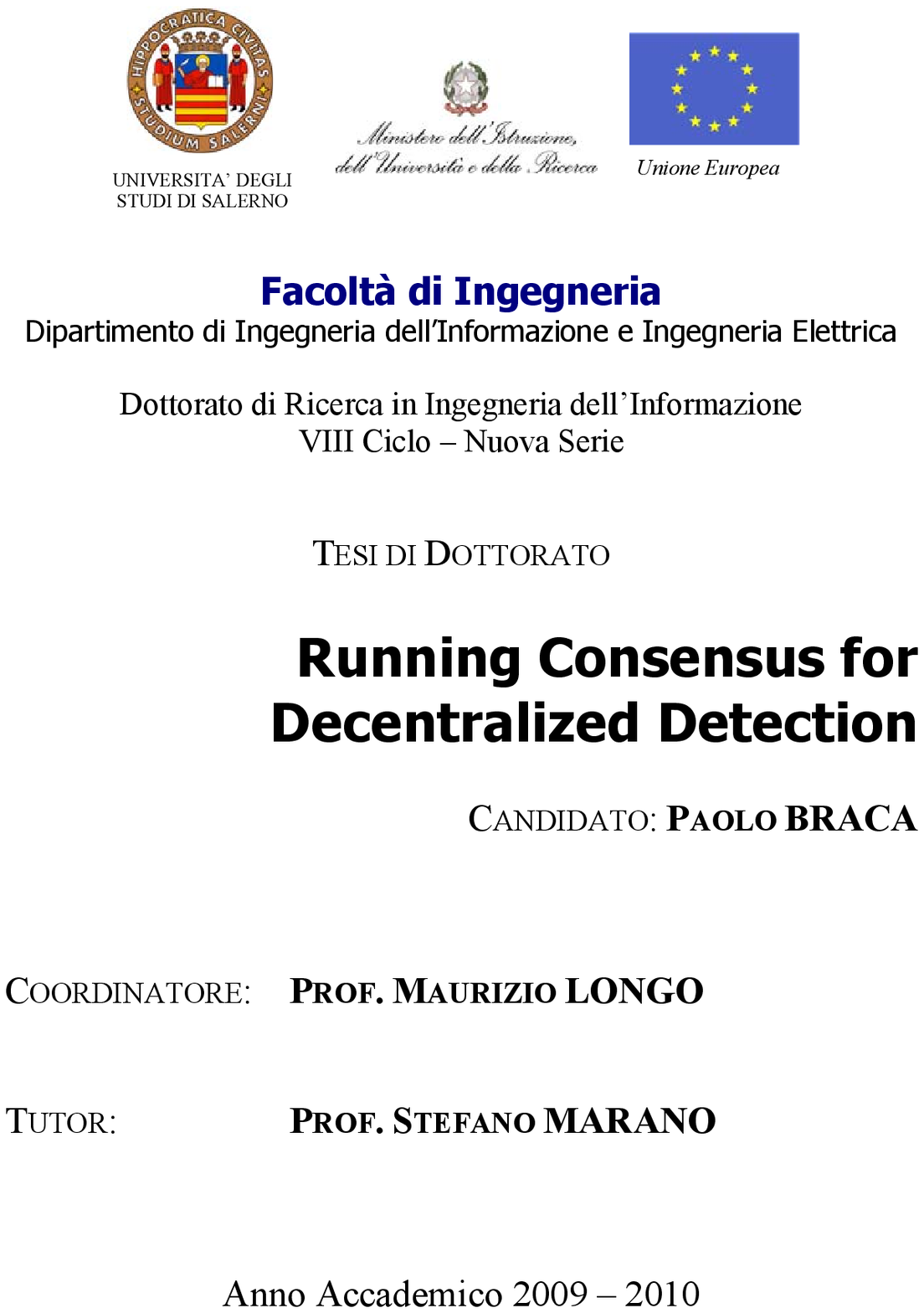}
\end{flushleft}
\newpage

\pagenumbering{roman}
\tableofcontents       

\clearemptydoublepage




\renewcommand{\sectionmark}[1]{\markright{\thesection.\ #1}}
\renewcommand{\chaptermark}[1]{\markboth{\thechapter.\ #1}{}}



\renewcommand{\headrulewidth}{0.5pt}
\renewcommand{\footrulewidth}{0pt}
\addtolength{\headheight}{0.5pt}


\pagenumbering{arabic} 

\theoremstyle{plain}
\newtheorem{thm}{Theorem}
\newtheorem{cor}{Corollary}
\newtheorem{lem}{Lemma}
\newtheorem{prop}{Proposition}
\theoremstyle{definition}
\newtheorem{defn}{Definition}
\theoremstyle{remark}
\newtheorem{rem}{Remark}


\chapter*{Preface}

\textit{This thesis represents a culmination of work and learning that has taken place over a period of almost three years (2007 - 2010) at the University of Salerno, and at the University of Connecticut (2009). It is mostly an unified mathematical dissertation of the \emph{running consensus} procedures~\cite{fusion2008,running-cons,braca-spawc09,asymptotic-rc,Bracaetal-Pageconsensus,braca_EUSIPCO11}, which has been applied to the problem fields of sequential estimation (see~\cite{fusion2008,running-cons} and Chapter~\ref{ch:estimation}), sequential/non-sequential detection (see~\cite{braca-spawc09,asymptotic-rc} and Chapter~\ref{ch:CH4}), and change detection (see~\cite{Bracaetal-Pageconsensus,braca_EUSIPCO11} and Chapter~\ref{ch:quickest detection}).}

\textit{In the recent years, the detection using the paradigm of the running consensus has been recognized as one of the three possible classes of distributed detection (e.g. see~\cite{Bajovic2011,cattivelli2011,Kar2011,Bajovic2012}) in which the phases of sensing and communication need not be mutually exclusive, i.e., sensing and communication occur simultaneously.} 

\textit{Considering that the running consensus paradigm is just an intuitive inference procedure, i.e. suboptimal w.r.t. an ideal centralized system scheme which is optimal, the most important result is that it asymptotically reaches the performance of this ideal scheme. There are two asymptotic frameworks. In the first one the running consensus is \emph{locally} efficient~\cite{lehmann-testing,kassam} as the centralized procedure, see~\cite{asymptotic-rc} and Chapter~\ref{ch:CH4}. The limit is in the number of observations (which is proportional to the time duration of the algorithm), and while this number diverges the two statistical hypotheses are closer and closer.  
In the second framework, that of large deviations~\cite{Dembo-Zeitouni}, while the procedure time duration diverges the two statistical hypotheses are fixed and it is studied the convergence rate of the error probability. Interestingly, this rate can be equal or below that of the ideal system depending on the connectivity of the network~\cite{Bajovic2011,Bajovic2012}}.

\textit{Recently, the running consensus has been also extended and generalized in order to cover more general problems, related for instance to the noisy networks~\cite{jakovetic2012}.} 

\textit{Beyond the detection problems some connections with the running consensus have been highlighted w.r.t the so-called ``consensus+innovations'' distributed inference procedures, e.g. see~\cite{Kar2013}. 
One of the last consensus+innovations distributed algorithms in developed in~\cite{FantacciPHD2013}, and faces the multitarget tracking problem, where the number of targets is unknown and time-varying, generalizing the work in~\cite{Olfati-Saber-CDC}, where just a single target has been considered. In~\cite{FantacciPHD2013} the multisensor Probability Hypothesis Density (PHD) filter is approximately computed in a distributed fashion using the consensus paradigm. It is important to note that the ``optimality'' in a consensus tracking problem cannot be reached as in~\cite{asymptotic-rc,Bajovic2011,Bajovic2012} just waiting a sufficiently large time because the state of the nature (targets' positions, velocities, etc.) evolves in time instead to be fixed.
In this case only increasing the number of sensors allows to obtain an asymptotic optimality property~\cite{BracaPHDFusion12,bracaAsymPHD}.}

\vspace{75pt}

\hspace{225pt} \textit{Paolo Braca}

\noindent\textit{\footnotesize May, 2013,}\\
\textit{\footnotesize La Spezia, Italy.}

\chapter*{Acknowledgments}

It is anything but easy to thank all the people who contributed to my personal and professional development
during my Ph.D. studies. More than anyone else, however, my Tutor Stefano Marano and Vincenzo Matta have been
the ones who made me think research was what I really wanted to do. I learned a lot, not only on a professional plane, and I owe to
them most of what I know about doing research. A special thanks goes to Peter Willett, he allowed me to be exposed to a stimulating
research environment during my stay at the University of Connecticut. 

I should have never started my Ph.D. studies without the support of my Department (DIIIE), and in particular its director Maurizio Longo.  

Besides the above, my colleagues Marco Guerriero, Gianluca Gennarelli, Domenico Attianese, Paolo Addesso have also been good friends and wonderful people to work with.

I am grateful to friends inside and outside the professional environment, people I simply shared good moments and laughter with:
Antonio De Costanza, Igor Negri, Giuseppe Zambrano, Antonio Fortunato, Domenico De Pascale, Christian Berger, Vishal Cholapadi Ravindra, Fabio Postiglione, Rocco Restaino, Roberto Conte, Fabio Mazzarella, Luigi Bruno, Pierpaolo D'Agostino, Pasquale Pistillo, Michele Grimaldi, CoRiTeL boys, Ozgur Erdinc, Ramona Georgescu, Bill Donat, Naomi Thonakkaraparayil, Sumit Narayan, Xin Tian, Javier and Carol Areta.

More than anyone else, however, my family has been fundamental. Even when we had to face problems way more important than anything I could write in this thesis, I never missed your unconditional support to my professional development. Eventually, I hope to have a chance to return what you gave me. 

Last but not least, a very special thanks to Loredana, without her love, encouragement
and support, I would not have finished this Thesis.

\vspace{75pt}
Thank you.

\hspace{225pt} Paolo

\noindent{\footnotesize March, 2010,}\\
{\footnotesize Salerno, Italy.}

\chapter{Introduction}

Recent advances in hardware technology have led to the emergence of small, low-power, and possibly mobile devices with limited on-board processing and wireless communication capabilities. Typically, these devices, called sensors, consist of a radio frequency circuit, a low-power digital signal processor, a sensing unit, and a battery. Due to their low cost and low complexity design requirement, individual  sensors can only perform simple local computation and communicate over a short range at low data rates. But when deployed in a large number across a spatial domain, these primitive sensors can form an \emph{intelligent network} to monitor the physical environment with high performance.

Sensor networks are suited for situation awareness applications such as environmental monitoring (air, water, and soil), biomedical engineering, home applications, radar, smart factory instrumentation, military surveillance, precision agriculture, space exploration, and intelligent transportation.

On the other side, the design of sensor network poses new challenges and requires multifaceted, interdisciplinary, and cross-layer approaches. 
In particular it becomes of paramount importance to develop energy and bandwidth efficient signal processing algorithms that can be implemented in a fully distributed manner. Distributed signal processing in a wireless sensor network differs from the traditional signal processing framework in several important aspects.
Sensor measurements are collected in a distributed fashion across a network, and this needs an \emph{appropriate data sharing} among the sensors to save the energy and bandwidth, that are critical resources for this kind of system. Indeed, in wireless sensor networks applications, the typical goal is to study inference problems with bandwidth and energy constraints. The sensor measurements are appropriately transformed and transmitted to a \emph{fusion center}, that can be static or mobile, see \emph{sensor network with mobile access} (SENMA) concept \cite{SENMA,tong:C-SENMA-LDPC,doasplet05,doasp06}. The research has been focused on finding good quantization at sensors and suitable fusion strategies. A very small list of papers would include \cite{GershoIT,Gish,Gubner,Ishwar,pradhan03,GershoISIT,TongSP07}. Achievable performance, in terms of estimation error as a function of the data rate  is of particular interest, and in the Gaussian case this has become known as the CEO problem \cite{CEO} for which the answers are both simple and elegant. The distributed target tracking problem with power-limited sensors is studied in \cite{ZhuGiannakis09}, while the static target estimation problem with data association and bandwidth constraint is studied in \cite{Marano-Matta-data-ass,fusion2009,braca2010}.
 
In a wireless sensor network, sensors may enter or leave the network dynamically, resulting in unpredictable changes in network size and topology. Sensors may disappear permanently either due to damage to the nodes or drained batteries. Temporarily due to topology, traffic, and channel conditions. This makes it necessary for distributed signal processing algorithms to be \emph{robust} to the changes in network topology. This motivates us to consider fully decentralized sensor network \emph{without fusion center}, where the remote units sense the 
environment and collect data, but, due to the lack of the fusion center, they are also programmed to run a \emph{consensus} protocol aimed at corroborating the local measurements with observations made at the neighboring nodes. The process of data exchange updates the locally computed statistics and (asymptotically) leads to the agreement about a common value, shared by all the nodes, that represents the final statistic. 

The ``consensus'' problem is studied from the '70s in the social sciences, in which the behavior of some people that tray to reach consensus is mathematically modeled~\cite{DeGroot}. Then this mathematical framework was transfered in the engineer field to design the ``behavior'' (cooperation and flocking) of artificial agents, for example sensors or mobile vehicles. A consensus algorithm is, in general, the set of cooperation rules in a network of agents. 

The consensus scheme for the computation of statistic functions has also attracted much interest in the last years, especially for its properties of robustness and scalability. We mention as useful entry points to the topical literature~\cite{Olfati-Saber,ren-beard-atkins,lynch,willett-parley,tsitsiklis-athans,boyd-gossip-IT,boyd-infocom,scaglione-varshney,scaglione-isit04,Jin-Murray,Olfati-Saber-CDC,barbarossa-magazine07,basar-quantized-consensus,consensus-propagation-IT}.

In this thesis we propose an innovative modification of classical consensus protocols, which we call \emph{running consensus}, see~\cite{fusion2008,running-cons,braca-spawc09,asymptotic-rc,Bracaetal-Pageconsensus}. While in the classical scheme, first the environmental sensing stage is performed, and only successively the acquired data are exchanged among nodes to reach the consensus, in the running consensus paradigm, this limitation is relaxed and the data acquisition proceeds simultaneously with data exchange. In this way, there is no need to fix in advance when the environmental monitoring stage is to be terminated, which may represent a drawback in application where the system operates in dangerous environments. Indeed, with running consensus, even though many nodes of the system are abruptly destroyed or impaired, the information they collected up to that time still contributes to the final statistic being shared across the system on-the-fly.

Both in the classical scheme and its modification, one would like that the final statistic achieved asymptotically be the same statistic that an ideal centralized entity would compute, provided that all the measurements collected by the entire sensor network are available to it.
As we will see, with the consensus protocols, the state of each node (namely, the locally computed statistic) approaches asymptotically the state of the ideal centralized system, as the number of consensus steps (data exchanges) diverges. However, the exponentially fast convergence that the classical consensus algorithms usually exhibits is lost and the convergence, exhibited by the running consensus, is much slower due to the presence of the new measurements made at each step, that must be incorporated on-the-fly into the statistics. 

The above consideration raises the following question.
Suppose that the ideal centralized statistic is asymptotically optimal, in the sense of achieving unbeatable performance as the number of overall measurements grows. Is such asymptotic optimality retained by a sensor network implementing the running consensus algorithm?
The main contribution of this thesis is to answer (affirmatively) the above question. In more generality, we show under what conditions 
the statistic computed by running consensus attains the same asymptotic performance as an ideal centralized system.

The remainder of this work is so organized. In chapter~\ref{ch:2} the mathematical properties of the running consensus are developed, with the related physical insights deferred to chapters. 
In chapter~\ref{ch:estimation} the sequential estimation problem, implemented via the running consensus
is considered~\cite{running-cons,fusion2008}. Analytical performances are characterized in terms of bounds, and it is shown that each sensor reaches asymptotically the consensus and the ideal estimation performance.
In chapter~\ref{ch:CH4} the detection problem is considered using the running consensus~\cite{asymptotic-rc,braca-spawc09}. Here the sensors are cooperating by the running consensus scheme to discriminate among two simple hypotheses. First, we consider the Neyman-Pearson setup in which the number of measurements used for the decision is fixed, and for this reason this is also called the FSS (fixed sample size) setup. Then, we focus on the the sequential case in which there is a virtually unbounded number of data available, and an appropriate number of these is used to make the final decision according to prescribed error probabilities. In this case, the asymptotic scenario is essentially that where the average number of samples needed for a decision tends to infinity. 
In chapter~\ref{ch:quickest detection} the quickest change detection problem is addressed~\cite{Bracaetal-Pageconsensus}. The problem is to detect an event (that is modeled by a change in the data distribution) as soon as possible, see also~\cite{basseville-book}. Approximate performance evaluation is considered and the running consensus performance is close to the ideal system. The comparison with a bank of parallel Page's test is also provided.

\chapter{Mathematical Properties of Running Consensus}
\label{ch:2}

In this chapter the mathematical properties of the running consensus are developed and the main properties are formally proved. In Section~\ref{sec:basic} we introduce the running consensus procedure. In Section~\ref{sec:ICS} we define the performance benchmark for the running consensus: an ideal centralized system that observes all the data shared in the network. In Section~\ref{sec:bounds} upper and lower bounds on the performance metrics are developed, then the asymptotic optimality for hypothesis testing problems is proved in Section~\ref{sec:binarytest}.

\section{Basic Procedure}
\label{sec:basic}
Let us consider a slotted system where in the $n^{th}$ time slot 
each sensor of the network collects a new measurement, 
and these measurements are modeled as iid (independent, identically distributed) random variables. 
The collected random values are stored for successive processing; the 
value stored by a node is hereafter also referred to as the \emph{state} of that sensor.
We denote by $\bx_n$ the column vector whose entries are the measurements collected by the sensors at time $n$. Its size is $M$, the number of nodes in the network. 

Let $\bs_n$ be the $M$-vector whose entries represent the state 
of the nodes at the $n^{th}$ time slot. 
We assume that this (random) state vector is updated as follows:
\beq
\left ( \begin{array}{c} 
s_{n,1} \\ s_{n,2} \\ \vdots \\ s_{n,M} \\
\end{array} \right ) 
=\bW_n \alpha_n
\left ( \begin{array}{c} 
s_{n-1,1} \\ s_{n-1,2} \\ \vdots \\ s_{n-1,M} \\
\end{array} \right ) 
+\bW_n\ \beta_n
\left ( \begin{array}{c} 
t(x_{n,1}) \\ t(x_{n,2}) \\ \vdots \\ t(x_{n,M}) \\
\end{array} \right ) , \label{eq:basic_CH2}
\eeq
the above can be cast in a compact vector form as
\beq
\bs_n=\bW_n \left(\alpha_n\,\bs_{n-1}+\beta_n\,\bt (\bx_n) \right), \label{eq:basic0}
\eeq
note that $t(x)$ is a real function and $\alpha_n,\ \beta_n$ are deterministic time-varying weights.
Specifically, it is assumed that $(i)$ data are acquired by sensors at time instants $n=1,2,\dots$, and $(ii)$  for each $n$, a randomly chosen subset of nodes is allowed to exchange data, according to the so-called gossip algorithm~\cite{boyd-gossip-IT}. The matrices $\bW_n$, $n=1,2,\dots,$ are assumed iid and doubly stochastic, so that
the connection protocol among nodes (formalized by the product $\bW_n \bs_{n-1}$ in the above equation)
amounts to a weighted average of the nodes' states~\cite{boyd-gossip-IT}. 
The $\bW_n$'s are also assumed statistically independent of the sensors' observations.

The basic assumption made throughout this work is that the statistical average $E[\bW_n \bW_n^T]$
(which is obviously doubly stochastic as well) has unitary eigenvalue
with algebraic multiplicity $1$, see also~\cite{boyd-gossip-IT}. 
Due to the identical distribution with respect to the time slot $n$, to simplify the notation, in the following we write $E[\bW \bW^T]$ for $E[\bW_n \bW_n^T]$.

\section{The Ideal Centralized System}
\label{sec:ICS}
An ideal system to which all the collected observations would be made available is able to compute the following quantity:
\beq
s^{(c)}_n = \chi_n \sum_{i=1}^{n} \bone^T \bt(\bx_{i}),
\label{eq:ics}
\eeq
where the suffix $c$ emphasizes that this is the \emph{ideal centralized system} state and $\bone$ is the vector of all ones.
The choice of $\chi_n$ and $t(x)$ depends on the application. Now if we are interested to estimate $\mu=E[x_{i,j}]$ then, choosing $\chi_n=1 /( n\,M)$ and $t(x)=x$, the ideal centralized estimator is the simple arithmetic mean:
	\beq
	\hat{\mu}_n \dfz s^{(c)}_n, \quad \chi_n=\frac{1}{n\,M},\quad t(x)=x; 
	\eeq
If we are interested to compute the random walk $\sum_{i=1}^{n}\sum_{j=1}^{M} t(x_{i,j})$ (which will play when we introduce the sequential tests), then we choose $\chi_n=1$. 

Now our interest is to analyze the differences between the ideal centralized system state and the running consensus state.
Let us define $\bphi_{n,i}$ and $\widetilde{\bphi}_{n,i}$ as the product of $\bW_n$ and $\widetilde{\bW}_n\dfz \bW_n - \bone\bone^T/M$, respectively, \emph{i.e.}:
\beq
\bphi_{n,i} \dfz \left \{ 
\begin{array}{lcl}
\bW_n & \textnormal{ if } & i=n \, ,\\
\bW_n \bW_{n-1} \dots \bW_{i} & \textnormal{ if } & i<n \, ;\\
\end{array}
\right.
\label{eq:phi}
\eeq

\beq
\widetilde{\bphi}_{n,i} \dfz \left \{ 
\begin{array}{lcl}
\widetilde{\bW}_n & \textnormal{ if } & i=n \, ,\\
\widetilde{\bW}_n \widetilde{\bW}_{n-1} \dots \widetilde{\bW}_{i} & \textnormal{ if } & i<n \, .\\
\end{array}
\right.
\label{eq:phi_tilde}
\eeq

The properties of $\bphi_{n,i}$ and $\widetilde{\bphi}_{n,i}$ are investigated in~\cite{boyd-gossip-IT}, here we stress that $\bphi_{n,i}$ 
is doubly stochastic, being the product of doubly stochastic matrices, this implies from the above definitions 
\beq
\widetilde{\bphi}_{n,i}= \bphi_{n,i} - \frac{\bone\bone^T}{n\,M}.
\label{eq:phi_tilde2}
\eeq

To quantify the difference between the sensor state $s_{n,j}$ and the centralized state, we introduce the \emph{running consensus error} \cite{running-cons,asymptotic-rc}.
\noindent 
\begin{defn}
The running consensus error is defined as
\beq
\be_n \dfz \beta_n\ \sum_{i=1}^n 
\widetilde{\bPhi}_{n,i}\  \bt(\bx_i).
\label{eq:yerr_2}
\eeq
\end{defn}
The following proposition remarks the relationship between the running consensus state (\ref{eq:basic_CH2}) and the ideal centralized system state (\ref{eq:ics}). 
\vspace*{5pt}
\noindent
\begin{prop}{$\left.\right.$} \label{prop:1}
\begin{enumerate}	
\item $\bs_n$ is a linear combination of $\bt(\bx_i)$ and $\bphi_{n,i}$:

for $\alpha_n=(n-1)/n$ and $\beta_n=1/n$ we have
\beq
\bs_n=\frac{1}{n}\sum_{i=1}^{n}\bphi_{n,i}\, \bt(\bx_i) ; \label{eq:ricorsion}
\eeq
for $\alpha_n=1$ and $\beta_n=M$ we have
\beq
\bs_n=M\,\sum_{i=1}^{n}\bphi_{n,i}\, \bt(\bx_i) ; \label{eq:ricorsion_2}
\eeq
\item the state vector $\bs_n$ is the sum of the ideal centralized system state $s_n^{(c)}$ and the error term $\be_n$
\beq
\bs_n= s_n^{(c)}\bone+\be_n;
\label{eq:error_cons}
\eeq
with
\beq
\chi_n=
\left\{
\begin{array}{ll}
\frac{1}{n\,M} & \alpha_n=\frac{n-1}{n},\beta_n=\frac{1}{n}; \\ \\
1      &\alpha_n=1,\beta_n=M
\end{array}
\right.
\eeq
%
\item The expected value of $\bs_n$ is:
	\beq
	E[\bs_n] = \left\{
\begin{array}{ll}
 \bone\, \mu & \alpha_n=\frac{n-1}{n},\beta_n=\frac{1}{n}; \\ \\ 
 \bone\, n\,M\,\mu		&\alpha_n=1,\beta_n=M
\end{array}
\right.
	\eeq
	and consequently the running consensus error is a zero-mean random variable: 
	\beq
	E[\be_n] = \bzero. 
	\eeq
\end{enumerate} 
~\hfill~$\bullet$
\end{prop}

\vspace*{5pt} \noindent
{\em Proof.}
Equations (\ref{eq:ricorsion})-(\ref{eq:ricorsion_2}) follow immediately from the (\ref{eq:basic_CH2}) with the particular choose of $\alpha_n$ and $\beta_n$. Suppose $\alpha_n=(n-1)/n$ and $\beta_n=1/n$ (the proof for $\alpha_n=1$ and $\beta_n=M$ is similar), from the equation (\ref{eq:ricorsion}) we have 
$\bs_n = \frac{\bone\bone^T}{n\,M} \sum_{i=1}^{n} \bt(\bx_{i}) + \frac{1}{n}\ \sum_{i=1}^n 
\left(\bPhi_{n,i}\ - \frac{\bone\bone^T}{M}\right)\bt(\bx_i)$ that gives us equation 
(\ref{eq:error_cons}) by the error definition (\ref{eq:yerr_2}) and by (\ref{eq:phi_tilde2}).
Finally we have  
$E[\bs_n] = E \left [ \frac{1}{n}\sum_{i=1}^{n}\bphi_{n,i} \bx_i\right ]=
\frac{1}{n}\sum_{i=1}^{n} E \left [ \bphi_{n,i} \right ] \bone \mu$.
Since $\bphi_{n,i}$ is doubly stochastic, the same property holds for its statistical expectation, 
thus implying that $E \left [ \bphi_{n,i} \right ] 
\bone$ $= \bone$, yielding $E [\bs_n]= E[\bx_n]=\bone \mu$. 
~\hfill$\bigtriangledown$

\section{Some Bounds}
\label{sec:bounds}
In this section the basic properties of the error $e_{n,j}$, defined in~(\ref{eq:yerr_2}), are investigated. This will be key to prove many results in the optimality of running consensus.
Let us define:
\begin{itemize}
  \item $\sigma^2\dfz\VAR[t(x)]$;
	\item $\xi^3\dfz
\E\left[\left\|\bt(\bx)- \mu\,\bone\right\|^3\right]$, where $\|\bz\|$ is the Euclidean norm of the vector $\bf z$; 
	\item $\lambda_U$ the second largest eigenvalue of $\E[{\bf W}^T {\bf W}]$;
	\item $\lambda_L$ the minimum eigenvalue of $\E[{\bf W}^T {\bf W}]$;
\end{itemize}
where $\VAR[t(x)]$ is the variance of $t(x)$. The fundamental hypothesis $\lambda_U<1$ means that the graph of the network is connected~\cite{boyd-gossip-IT}, \emph{i.e.} the data sharing involves each sensor.
\vspace*{5pt}
\noindent
\begin{prop}
\label{prop:2}
Suppose $\alpha_n=1,\,\beta_n=M$ and $\lambda_U<1$ then for all $n=1,2,\dots$ and $j=1,\dots,M$ we have 
\beqa
\E[e^2_{n,j}]&\leq& C_1(M,\lambda_U)\, \sigma^2,
\label{eq:Varerr}\\
\E\left[\left|e_{n,j}\right|^3\right]&\leq&
C_1(M,\lambda_U)\,\xi^3 + C_2(M,\lambda_U)\,\sigma^3,
\label{eq:thirderr}
\eeqa
where 
\beqa
C_1(M,\lambda_U)&=&M^3\,\frac{\lambda_U}{1-\lambda_U};\\
C_2(M,\lambda_U)&=&\frac{M^{\frac 9 2}}{1-\sqrt{\lambda_U}} \left(
\frac{\lambda_U}{1-\sqrt{\lambda_U}}
+
\frac{1}{1-\lambda_U}
\right).
\eeqa
\end{prop}
~\hfill~$\bullet$

\vspace*{5pt} \noindent
{\em Proof.}
See appendix \ref{sec:lemma}.
~\hfill$\bigtriangledown$

Suppose, without loss of generality, that $\mu=0$. Let us define the covariance matrix of the state vector
\beq
\bC_n\dfz\E\left[\bs_n\bs^T_n\right],
\eeq
denoting by $(\bC_n)_{ij}$ its entries.
Now we define
\[
\rho^c_{n,ij} \dfz \frac{(\bC_n)_{ij}}{\sqrt{(\bC_n)_{ii} \, (\bC_n)_{jj}}} \; , \quad \textnormal{ and} \quad
\rho^e_{n,ij} \dfz \frac{\sqrt{(\bC_n)_{ii} \, (\bC_n)_{jj}}} { \displaystyle{\frac{(\bC_n)_{ii} + (\bC_n)_{jj}}{2} }} \; .
\] 
The first quantity is the standard statistical correlation coefficient, and the second is the ratio between the geometric and the arithmetic 
mean of two diagonal entries of the matrix $\bC_n$. 

\noindent 
\begin{defn} 
The consensus coefficient between nodes $i$ and $j$ is defined as:
\beq
\rho_{n,ij} \dfz \rho^c_n \, \rho^e_n 
=\frac{2 \; (\bC_n)_{ij}} { (\bC_n)_{ii} + (\bC_n)_{jj}}.
\label{eq:rho}
\eeq
\end{defn} 
\begin{defn} 
The performance coefficient of sensor $i$ is defined as:
\beq
\gamma_{n,i} \dfz \frac{(\bC_n)_{ii}}{\sigma^2_n}.
\label{eq:gamma}
\eeq
where $\sigma^2_n \dfz \sigma^2/(n \, M)$.
\end{defn}
The rationale of the above definition is explained in Chapter~\ref{ch:estimation}. 
\noindent
\vspace*{5pt}
\begin{thm}
\label{prop:3}
If $\lambda_U<1$ then for all $i=1,\dots,M$ and $j\neq i$ the following bounds hold
\beqa
(M-1) \, \psi^L_n &\le \gamma_{n,i} -1 \le&  (M-1) \; \psi^U_n , \label{eq:claim1}\\
\frac{M \, \psi^L_n}{1+\left(M-1\right)\psi^U_n} \label{eq:claim2}
&\le
1- \rho_{n,ij}
\le&
\frac{M \, \psi^L_n}{1+\left(M-1\right)\psi^L_n},
\eeqa
where we define
\[
\psi^U_n \dfz \displaystyle{\frac{\lambda_U}{n}\; \frac{1-\lambda_U^{n}}{1-\lambda_U}}
\quad \textnormal{ and } \quad 
\psi^L_n \dfz\displaystyle{\frac{\lambda_L}{n}\; \frac{1-\lambda_L^{n}}{1-\lambda_L}} .
\]
~\hfill~$\bullet$
\vspace*{5pt}
\end{thm}
 \noindent
{\em Proof.}
See appendix \ref{sec:lemma}.
~\hfill$\bigtriangledown$

\section{Binary Hypothesis Test by Running Consensus}
\label{sec:binarytest}
Consider the problem to discriminate between the hypotheses
\beq
\begin{array}{lcl}
{\cal H}_0&:& x_{i,j}\sim f_{\theta_0}(x),\\
{\cal H}_1&:& x_{i,j}\sim f_{\theta}(x),
\end{array}
\label{eq:test0_CH2}
\eeq
where $i=1,2,\dots, n$, $j=1,2,\dots, M$, the symbol $\sim$ stands for ``is distributed according to'', and $f_{\theta_0}(x)$, $f_{\theta}(x)$ are the marginal probability density functions of the data, parametrized by $\theta$. Note that the moments of $t(x)$ will depend on $\theta$, then
\beqa
\mu(\theta)&=&\E_{\theta}[t(x)]; \\
\sigma^2(\theta)&=&\VAR_{\theta}[t(x)];\\
\xi^3(\theta)&=&\E_{\theta}\left[\left\|\bt(\bx)- \mu(\theta)\,\bone\right\|^3\right]; 
\eeqa
where $\E_\theta[\cdot]$ stands for expectation under $f_\theta(x)$, including as particular case $\theta=\theta_0$.
We define the efficacy $d=\sqrt{M}\mu^{'}(\theta_0)/\sigma(\theta_0)$, the role of the efficacy will be explained in Chapter \ref{ch:estimation}. The statistic $T_n$ of the observations 
\[
[\bx_{1},\bx_{2},\dots,\bx_{n}],
\]
can be $T_n=s_n^{(c)}$ and $T_n=s_{n,j}$, with $\alpha_n=1,\,\beta_n=M,$ and $\chi_n=1$.

We are interested to study the asymptotic behaviour ($\theta\rightarrow\theta_0$) of:
\begin{itemize}
	\item the fixed sample size test or Neyman-Pearson paradigm~\cite{kassam}, in which the number of measurements is fixed;
	\item the sequential test~\cite{lai-AMS}, in which the number of measurements is a random variable.
\end{itemize}

\subsection{Fixed Sample Size Test}
\label{subsec:FSS}
For the fixed sample size case we use the model~(\ref{eq:test0_CH2}), with~\cite{kassam}
\beq
\theta_n=\theta_0+\frac{\gamma}{\sqrt{n}},~~~\gamma>0,
\label{eq:FSSscaling_CH2}
\eeq
and the performances are studied in the limit of $n\rightarrow\infty$.
For a given $n$ the detector is identified by the pair $(T_n,\delta_n)$, in the sense that ${\cal H}_1$ is declared whenever $T_n \ge \delta_n$ and ${\cal H}_0$ is declared otherwise. 
The detection and false alarm probabilities are accordingly defined as $p_{dn}=\P[T_n\ge \delta_n|{\cal H}_1]$,
$p_{fn}=\P[T_n \ge \delta_n|{\cal H}_0]$.
\vspace*{5pt}
\noindent
\begin{thm}
 Suppose that:
\begin{itemize}
	\item[-] $\lambda_U<1$;
	\item[-] $\mu(\theta)$ is differentiable with $\mu^{\prime}(\theta_0)\not =0$, $\sigma(\theta)$ is a continuous function;
	\item[-] the centralized test statistic $s^{(c)}_n$ fulfills 	
\beq
\begin{array}{lcl}
&&\displaystyle{\frac{s^{(c)}_n-n\,M\mu(\theta_0)}{\sqrt{n\,M}\,\sigma(\theta_0)}}
\stackrel{f_{\theta_0}}{\longrightarrow} {\cal N}(0,1),
\\
&&\displaystyle{\frac{s^{(c)}_n-n\,M\mu(\theta_n)}{\sqrt{n\,M}\,\sigma(\theta_n)}}
\stackrel{f_{\theta_n}}{\longrightarrow} {\cal N}(0,1),
\label{eq:centrasynorm_CH2}
\end{array}
\eeq
where $\stackrel{f_{\theta_n}}{\longrightarrow}$ (resp. $\stackrel{f_{\theta_0}}{\longrightarrow}$) denotes convergence in distribution under $f_{\theta_n}(x)$ (resp. under $f_{\theta_0}(x)$) as $n$ diverges. 
\end{itemize}
Then, the decentralized detection statistic $s_{n,j}$, for any sensor $j=1,2,\dots,M$, is asymptotically equivalent to $s^{(c)}_n$ in the sense that,  for any fixed asymptotic probability of false alarm $p_f=\lim_{n \rightarrow \infty}p_{fn}$, they achieve the same asymptotic probability of detection
\beq
p_d=\lim_{n \rightarrow \infty}p_{dn}= Q(Q^{-1}(pf)-\gamma\,d)
\eeq
~\hfill$\bullet$
\vspace*{5pt} 
\end{thm}
\noindent{\em Proof.} 
The proof is in Appendix \ref{app:th1}.~\hfill$\bigtriangledown$

\subsection{Sequential Test}
\label{subsec:seq}
For the sequential case we use the model~(\ref{eq:test0_CH2}), with~\cite{lai-AMS}
\beq
\theta_r=\theta_0+\frac{1}{\sqrt{r}},
\label{eq:test_CH2}
\eeq
where $r$ is a positive number, and the asymptotic performances will be studied in the limit $r \rightarrow \infty$.
Consider the sequential test
\beq
\left\{
\begin{array}{ll}
T_n - n\,M\,\eta_r \ge b_r,& \textnormal{declare } {\cal H}_1,\\
T_n - n\,M\,\eta_r \le a_r,& \textnormal{declare } {\cal H}_0,\\
T_n - n\,M\,\eta_r\in \left(a_r,b_r\right),& \textnormal{take another sample}
\end{array}
\label{eq:decseq_CH2}
\right.
\eeq
with stopping time
\beq
N_r=\inf\left\{n:T_n - n\,M\,\eta_r \notin \left(a_r,b_r\right) \right\} .
\label{eq:stop_CH2}
\eeq
The false alarm and detection probabilities are
\beqa
p_{dr}&=&\P[\textnormal{declare } {\cal H}_1|{\cal H}_1],\\
p_{fr}&=&\P[\textnormal{declare } {\cal H}_1|{\cal H}_0],
\eeqa
and their asymptotic values are
\beqa
p_d &=& \lim_{r \rightarrow \infty} p_{dr}, \\
p_f &=& \lim_{r \rightarrow \infty} p_{fr}.
\label{eq:pfpd_CH2}
\eeqa
In (\ref{eq:decseq_CH2}) we define
\beqa
\eta_r  &\dfz& \frac{\mu(\theta_r)+\mu(\theta_0)}{2}, \\
a_r     &\dfz& \frac{\sqrt{r\,M}\,\sigma(\theta_0)}{d}\log\frac{1-p_d}{1-p_f}, \\
b_r     &\dfz& \frac{\sqrt{r\,M}\,\sigma(\theta_0) }{d}\log\frac{p_d}{p_f}.
\label{eq:ar_br_CH2}
\eeqa
\vspace*{5pt}
\noindent
\begin{thm}
 Suppose that:
\begin{itemize}
	\item[-] $\lambda_U<1$;
	\item[-] $\mu(\theta)$ is differentiable, $\mu^{\prime}(\theta_0)\not =0$ and $\sigma(\theta)$ is continuous;
	\item[-] the centralized test statistic $s^{(c)}_n$ fulfills 
\beq
\begin{array}{lcl}
&&\displaystyle{\frac{s^{(c)}_{[rt]} - [rt]\,M\,\eta_r}{\sqrt{r\,M}\,\sigma(\theta_0)}}
\stackrel{f_{\theta_0}}{\longrightarrow} {\cal W}_{-d/2}(t),
\\
&&\displaystyle{\frac{s^{(c)}_{[rt]} - [rt]\,M\,\eta_r}{\sqrt{r\,M}\,\sigma(\theta_0)}}
\stackrel{f_{\theta_r}}{\longrightarrow} {\cal W}_{+d/2}(t),
\label{eq:asyWiener_CH2}
\end{array}
\eeq
where $\stackrel{f_{\theta_{r}}}{\longrightarrow}$ (resp. $\stackrel{f_{\theta_{0}}}{\longrightarrow}$) denotes weak convergence to a random process~\cite{billingsley-book}, under $f_{\theta_r}(x)$ (resp. $f_{\theta_0}(x)$) as $r$ diverges, and ${\cal W}_{z}(t)$ is a Wiener proces with drift $z$.
  \item[-] for any $\epsilon>0$, there exist $r_\epsilon$ and a function $g_\epsilon(t)>0$ such that,
for all $r\geq r_\epsilon$ and $t\geq 1$, 
\beq
\begin{array}{lrl}
\textnormal{P}_{\theta_0}\displaystyle{\left[\frac{s^{(c)}_{[rt]}-[rt]\,M\,\mu(\theta_0)}{\sqrt{r\,M}\,\sigma(\theta_0)}>\epsilon t\right]}
&\leq& g_\epsilon(t), \\
\textnormal{P}_{\theta_r}\displaystyle{\left[\frac{s^{(c)}_{[rt]}-[rt]\,M\,\mu(\theta_r)}{\sqrt{r\,M}\,\sigma(\theta_r)}\leq -\epsilon t\right]}
&\leq& g_\epsilon(t),
\end{array}
\label{eq:addcond_CH2}
\eeq
with $\int_1^\infty g_\epsilon(t)\,dt<\infty$, and by $\textnormal{P}_\theta$ we mean the probability corresponding to distribution $f_\theta(x)$.
\end{itemize}
Then, the decentralized statistic $s_{n,j}\, \forall j=1,2,\dots,M$, is asymptotically equivalent to $s^{(c)}_n$ in the sense that, for any fixed $p_d$ and $p_f$, they achieve the same asymptotic expected stopping time

\beq
\begin{array}{rcl}
{\displaystyle\lim_{r\rightarrow\infty}r^{-1}\, \E_{\theta_0}[N_r]}&=&2\frac{\displaystyle {\cal D}_b(p_f,p_d)}{d^2},\\
{\displaystyle\lim_{r\rightarrow\infty}r^{-1}\, \E_{\theta_r}[N_r]}&=&2\frac{\displaystyle {\cal D}_b(p_d,p_f)}{d^2},
\end{array}
\label{eq:ASN_CH2}
\eeq
where ${\cal D}_b(p,q)$ stands for the Kullback-Leibler divergence between the binary probability mass functions $[p,1-p]$ and $[q,1-q]$~\cite{CT}.

~\hfill$\bullet$
\vspace*{5pt} \noindent
\end{thm}
{\em Proof.} The proof is in Appendix~\ref{app:S}.~\hfill$\bigtriangledown$
\vspace*{5pt}

\chapter{Sequential Estimation}
\label{ch:estimation}

In this chapter the sequential estimation problem is solved by the running consensus procedure and a new suitable definition of consensus is proposed related to the asymptotic optimality of our scheme, see~\cite{running-cons,fusion2008}. The chapter is so organized. The problem is precisely formalized in Section~\ref{sec:estimation1}, the performance metrics are defined in Section~\ref{sec:Performance Metrics} and the related analytical bounds are developed in Section~\ref{sec:Asymptotic Optimality}. Examples of applications are provided in Section~\ref{sec:examples}.

\section{Parameter Estimation}
\label{sec:estimation1}
Let $n=1,2,\dots$ and $M$, be respectively the discrete time index and the number of sensors. 
For each time slot $n$, the sensors of the network collect measurements that are iid realizations of a random variable with mean $\mu$ and with variance $\sigma^2$. These observations are iid over time and across the sensors. 
The goal is to estimate $\mu$. The arithmetic mean of the $n\,M$ measurements is the statistic of the ideal centralized system, and it represents the benchmark for the running consensus estimator\footnote{Note that the arithmetic mean is a consistent estimator of $\mu$.}.
The ideal centralized system, defined in~(\ref{eq:ics}) with $\chi_n=1/n$ and $t(x)=x$, can be recast as
\beq
s_n^{(c)}=\frac{n-1}{n} s_{n-1}^{(c)}+\bone^T\,\frac{\bx_n}{n}.
\eeq
 Consider the running consensus update rule with $\alpha_n=(n-1)/n,\beta_n=1/n$
\beq
\bs_n=\frac{n-1}{n}\bW_n \bs_{n-1}+\frac{\bx_n}{n} , \label{eq:basic_CH3}
\eeq
in which there is a light difference with~(\ref{eq:basic_CH2}): the new measurements taken at the moment time $n$ are not exchanged among the sensors. It is clear that all the asymptotic results and the physical insights are the same in both the cases. Note that the first term on the right-hand side (RHS) of~(\ref{eq:basic_CH3}) enforces consensus among the nodes in other words one would like that for $n$ sufficiently large 
\beq
\bW_n \bs_{n-1} \approx \bone\, s_{n-1}^{(c)},
\label{eq:desire}
\eeq
that means that the state of each node is close to each other and close to ideal centralized state. 
The second term on the RHS of~(\ref{eq:basic_CH3}) accounts for the new measurements. These two terms are \emph{properly} weighted by $(n-1)/n$ and $1/n$, respectively. 
The state $s_{n,j}$, for the generic sensor $j$, is the local estimator for the parameter $\mu$ at time $n$. 

\begin{figure}
\centerline{\includegraphics[width = 0.95\textwidth,angle=0]{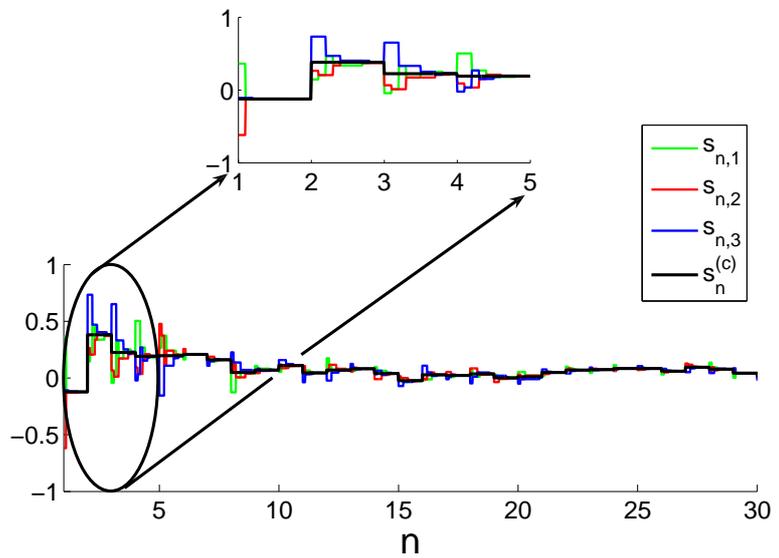}}
\caption{State evolution of the sensors in a network made of $M=3$ nodes, with $v=10$ pairwise averages per single time slot. $s^{(c)}_n$ is the arithmetic mean of all processed measurements in the network, up to time $n$. The zoom emphasizes the behavior in the first five time slots.}
\label{fig3_1}
\end{figure}

In Figure~\ref{fig3_1} the system evolution is depicted with $\mu=0$. Note that for $n$ sufficiently large the running consensus algorithm seems to force $s_{n,j} \approx s_{n}^{(c)}$ as expressed in eq.~(\ref{eq:desire}). The mathematical explanation of this phenomenon is given in the next sections.

In the example it is supposed
\beq
\bW_n=\bW_{i_1 j_1}\bW_{i_2 j_2}\dots \bW_{i_v j_v},
\label{eq:example}
\eeq
in which the random matrices $\{\bW_{ij} \}$ can be expressed as 
\beq
\bW_{ij}=I-\frac{(e_i-e_j)(e_i-e_j)^T}{2} , \label{eq:Wij}
\eeq
where $e_i$ denotes a vector of zeros with only the $i^{th}$ entry equal to~1 and, accordingly, the product $\bW_{ij} \bs_{n-1}$ amounts to replace the entries $i$ and $j$ of the vector with their arithmetic mean, which is just the pairwise averaging.
From equation~(\ref{eq:example}) in the $n$th slot the pairs of nodes performing the $v$ pairwise averages are $(i_1,j_1)$, $\dots$,$(i_v,j_v)$.
The fact that $\bW_n$ is actually the product of several pairwise matrices (\emph{i.e.}, $v>1$) is a minor aspect, while the important fact with eq.~(\ref{eq:basic_CH3}) is the simultaneous presence of the sensing stage and of the averaging step.
The state of each sensor at time $n$ encompasses the current measurement made in that time slot. On the other hand, note that
the $v$ pairwise averages performed within the $n$ time slot involves the state of the node at the previous time slot.
In these regards, different formalizations are possible, leading to minor modifications of the formulas while leaving substantially unchanged the physical insights.

\section{Performance Metrics}
\label{sec:Performance Metrics}
By observing Figure~\ref{fig3_1}, it seems that when the time is large, the state of each sensor is approximately close to each other and close to the ideal centralized state. We can quantify analytically this effect by the performance figures $\gamma_{n,i}$ and $\rho_{n,ij}$ defined in Chapter~\ref{ch:2}. In the following two subsections these metrics are discussed.

\subsection{Consensus in the Network}
Consider the consensus coefficient $\rho_{n,ij}= \rho^c_{n,ij} \, \rho^e_{n,ij} $, defined in (\ref{eq:rho}). 
As well known, $\rho^c_{n,ij}$ quantifies the degree of statistical dependence between random variables (state of the nodes, in our case), 
and attains its maximum value of~$1$ only if the variables are linearly dependent almost everywhere (a.e.). On the other hand, $\rho^e_{n,ij}$
is an index of equality between two positive numbers (here variances), and takes the value~$1$ only when these numbers are equal. 
Consequently, the product $\rho^c_{n,ij} \, \rho^e_{n,ij}$ belongs to $(-1, 1)$ and the value~$1$ is attained only when the two random variables
coincide a.e.: having the same mean and being linearly related ($\rho^c_{n,ij}=1$) they can only differ for a scale factor, which must by unitary because the variables share the same variance ($\rho^e_{n,ij}=1$).
This legitimates the adoption of $\rho_{n,ij}$ as a quantitative measure of the consensus degree:
when $\rho_{n,ij}=1$ the state of the two nodes is a.e.\ identical. 
While a unitary consensus coefficient for all the pair of nodes
would mean that all nodes share the same state (a.e.), it is also necessary to check that
such state is that desired. 

\subsection{Comparison with the Ideal Centralized System}
If all the observations collected by the entire network up to time slot $n$
were simultaneously available to a single device, this latter could compute the following arithmetic mean, which represents the state of an ideal centralized system: $s^{(c)}_n = 
\frac{\sum_{i=1}^n\bone^T\, \bx_i}{n\,M}$, see Section~\ref{sec:ICS} for the mathematical details. 
Computing $s^{(c)}_n$ turns out to be the main goal of
many inference problems in WSNs where the optimal decision/estimation statistic is obtained by averaging the network observations, or functions thereof, see, \emph{e.g.},~\cite{barbarossa-magazine07}.
Accordingly, in our fully decentralized architecture, the ideal goal would be that 
any node attained the same performances of the centralized scheme.
To this aim, let us refer to the mean square error. 
As to the statistical mean, $E[s^{(c)}_n]$ is just $\mu$ (that we set to zero for simplicity), and being $E[\bs_n] = \bone \mu$ we recognize that the state of all the nodes takes the same mean value of the optimal centralized scheme, at any time $n$. Consequently, all the nodes should hopefully share the same \emph{variance} of the centralized scheme $\sigma^2_n = \sigma^2/(n \, M)$. When $\gamma_{n,i}=\frac{(\bC_n)_{ii}}{\sigma^2_n} \rightarrow~1$ 
the \emph{optimal} centralized performance is reached for the node $i$.

\section{Asymptotic Optimality}
\label{sec:Asymptotic Optimality}
Here we introduce upper and lower bounds for the metrics $\rho_{n,ij}$ and $\gamma_{n,i}$, using
the analogous of Theorem~\ref{prop:3}, when the update rule is (\ref{eq:basic_CH3}).  
\begin{thm}
\label{prop:3_2}
If $\lambda_U<1$ then for all $i=1,\dots,M$ and $j\neq i$ the following bounds hold
\beqa
(M-1) \, \psi^L_n &\le \gamma_{n,i} -1 \le&  (M-1) \; \psi^U_n , \label{eq:claim1_2}\\
\frac{M \, \psi^L_n}{1+\left(M-1\right)\psi^U_n} \label{eq:claim2_2}
&\le
1- \rho_{n,ij}
\le&
\frac{M \, \psi^L_n}{1+\left(M-1\right)\psi^L_n},
\eeqa
where
\[
\psi^U_n = \displaystyle{\frac{1}{n}\; \frac{1-\lambda_U^{n}}{1-\lambda_U}}
\quad \textnormal{ and } \quad 
\psi^L_n =\displaystyle{\frac{1}{n}\; \frac{1-\lambda_L^{n}}{1-\lambda_L}} .
\]
~\hfill~$\bullet$
\end{thm}
\vspace*{5pt} \noindent
{\em Proof.} The proof is similar to that of Theorem~\ref{prop:3}.~\hfill$\bigtriangledown$
\vspace*{5pt}

Some comments on the theorem are following.
\begin{itemize}
  	\item Since $\psi^{U,L}_n \rightarrow 0$ when $n \rightarrow \infty$, we have that
	$\rho_{n,ij} \rightarrow 1$, and $\gamma_{n,i} \rightarrow 1$:
	asymptotically, the consensus is reached, and the performance of the optimal centralized system is attained. 
	Furthermore, $\psi^{U,L}_n$ both go to zero essentially as $n^{-1}$, implying the same speed of convergence
	to $1$ for $\rho_{n,ij}$ and $\gamma_{n,i}$. 
\item 
	For large $n$, we have $\lambda_{U,L}^{n} \ll 1$. 
	Then, confusing $M-1$ with $M$, which is the case of interest, we obtain the approximate bounds
\beq
{\cal B}^L_n\dfz\displaystyle{\frac M n \, \frac{1}{1-\lambda_L}} \le  
\epsilon_n \le \displaystyle{\frac M n \, \frac{1}{1-\lambda_U}}\dfz {\cal B}^U_n ,
\label{eq:claim3}
\eeq
where $\epsilon_n$ is a compact notation for both the error figures $\gamma_{n,i}-1$ and $1-\rho_{n,ij}$.
Note that the bounds on $\epsilon_n$ depend upon the normalized time $M/n$, and
large values of $M$ slow down the convergence, as one might expect.
Note also that $M/n$ seems like the rate at which the new observations become negligible to the
average made up to time $n$, and the system performance seems hence to be dominated by such rate.

\item The upper bound in the previous equation
gives a conservative (worst case) estimate of the rate of convergence, namely
\beq
r \dfz	\lim_{n \rightarrow \infty} 
\; n \, \epsilon_n  = \frac{M}{1-\lambda_U} \; .
\label{eq:rate}
\eeq
Using the update rule~(\ref{eq:basic_CH2}), with $\alpha_n=(n-1)/n,\beta_n=1/n$, the approximate bounds are
\beq
{\cal B}^{U,L}_n = \displaystyle{\frac M n \, \frac{\lambda_{U,L}}{1-\lambda_{U,L}}},
\eeq
with rate
\beq
r=M\,\frac{\lambda_U}{1-\lambda_U}.
\eeq

\item Useful insights about the differences with classical consensus algorithms can be gained by allowing multiple 
averaging steps per time slot, namely, by assuming that the weighting matrix $\bW_n$ in eq.~(\ref{eq:basic_CH3}) is actually the product of $v>1$
iid doubly stochastic matrices, say $\bW_n=\bM_{n,1} \dots \bM_{n,v}$. Now we can obtain bounds like those in the theorem, in terms of the smallest and the second largest eigenvalues of $E[\bM \bM^T]$, say
$\xi_L \ge 0$ and $\xi_U<1$: the final result
amounts to substitute in the claims of the theorem $\lambda_{U,L}$ with $\xi_{U,L}^v$, respectively.
In the classical consensus scenario, see \emph{e.g.}~\cite{boyd-gossip-IT}, $v$ diverges so that $\xi_{U,L}^v$ become negligible
with exponential rate. 

\item
In the same spirit of the previous item, one may also refer to the case that 
each sensor just spends time $n$ to gather data and then exchange the locally averaged
data using classic algorithms~\cite{boyd-gossip-IT} which reach consensus 
in a time negligible compared to the data gathering time $n$.
Of course, this scheme does not take into account possible sensors' failures occurring before the data exchange process, which is the main motivation of our analysis.

\item
A distinct feature of the running consensus scheme is the speed of convergence of the performance figures.
In fact, this is substantially different from the exponential law 
that governs the classic consensus algorithms~\cite{boyd-gossip-IT}. 
Furthermore, in our setup, the specific network topology/connectivity (which rules the system eigenvalues) is less crucial with respect to the classical case. Indeed, in this latter, a network design yielding a larger value of $\xi_U$ 
is expected to {\em exponentially} outperform a system with a smaller eigenvalue. Oppositely, in our case, the \emph{universal} scaling law is $n^{-1}$, and only the value of the rate coefficient 
can be tuned by the eigenvalues of the system.

\end{itemize}

\section{Examples: Pairwise Averaging}
\label{sec:examples}

\begin{figure}
\centerline{\includegraphics[width = 0.95\textwidth,angle=0]{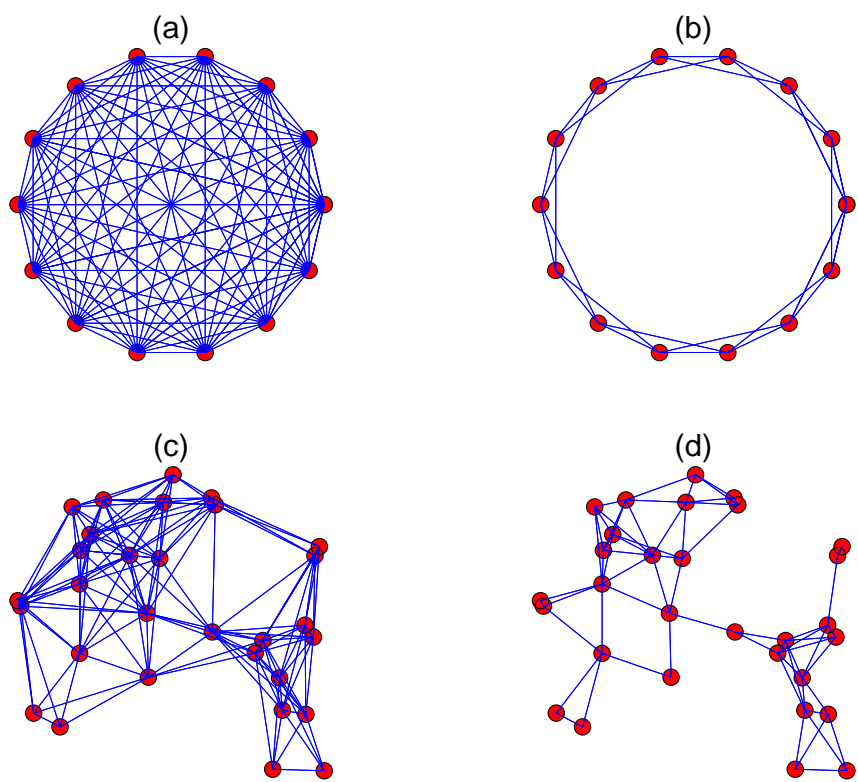}}
\caption{Architectures of the networks used in the examples. Panel $(a)$ represents a completely connected
ring of $M=15$ sensors. The same ring is considered in panel $(b)$ where each node is connected only to four neighbors.
In panels $(c)$ and $(d)$, which refer to $M=30$, the node position is that typical of randomly deployed sensors, 
with the network in $(c)$ having a larger number of admissible pairs than that depicted in $(d)$.}
\label{fig:topology}
\end{figure}

In this section we limit the analysis to the simple pairwise protocol described in section~\ref{sec:estimation1}.
Specifically, we make reference to the network topologies schematically
depicted in Figure~\ref{fig:topology}, and we assume that only the \emph{admissible} pairs of nodes (those
connected by straight lines) can be selected for the pairwise averaging. 
Any such pair is selected with one and the same probability, so that
$\bW_n=\bW_{ij}$ and any realization of such random matrix (any choice of an admissible pair $(i,j)$) has the same chance
of occurrence.

In this case, the eigenvalues appearing in Theorem~\ref{prop:3_2}
admit a simple interpretation. Indeed, from eq.~(\ref{eq:Wij}) we immediately see that
$\bW_{ij}$ is doubly stochastic, symmetric and idempotent.
The last two properties imply that $E[\bW\bW^T]=E[\bW]$, with the consequence that the eigenvalues $\lambda_L$ 
and $\lambda_U$ can be equivalently referred to this latter matrix. 
In addition, it can be easily shown that our basic requirement, 
namely $\lambda_U < 1$, is fulfilled provided that the graph associated to $E[\bW]$ 
is strongly connected~\cite{Johnson-Horn},
and this is certainly true in the architectures of Figure~\ref{fig:topology}. 

\begin{figure}
\centerline{\includegraphics[width = 0.95\textwidth,angle=0]{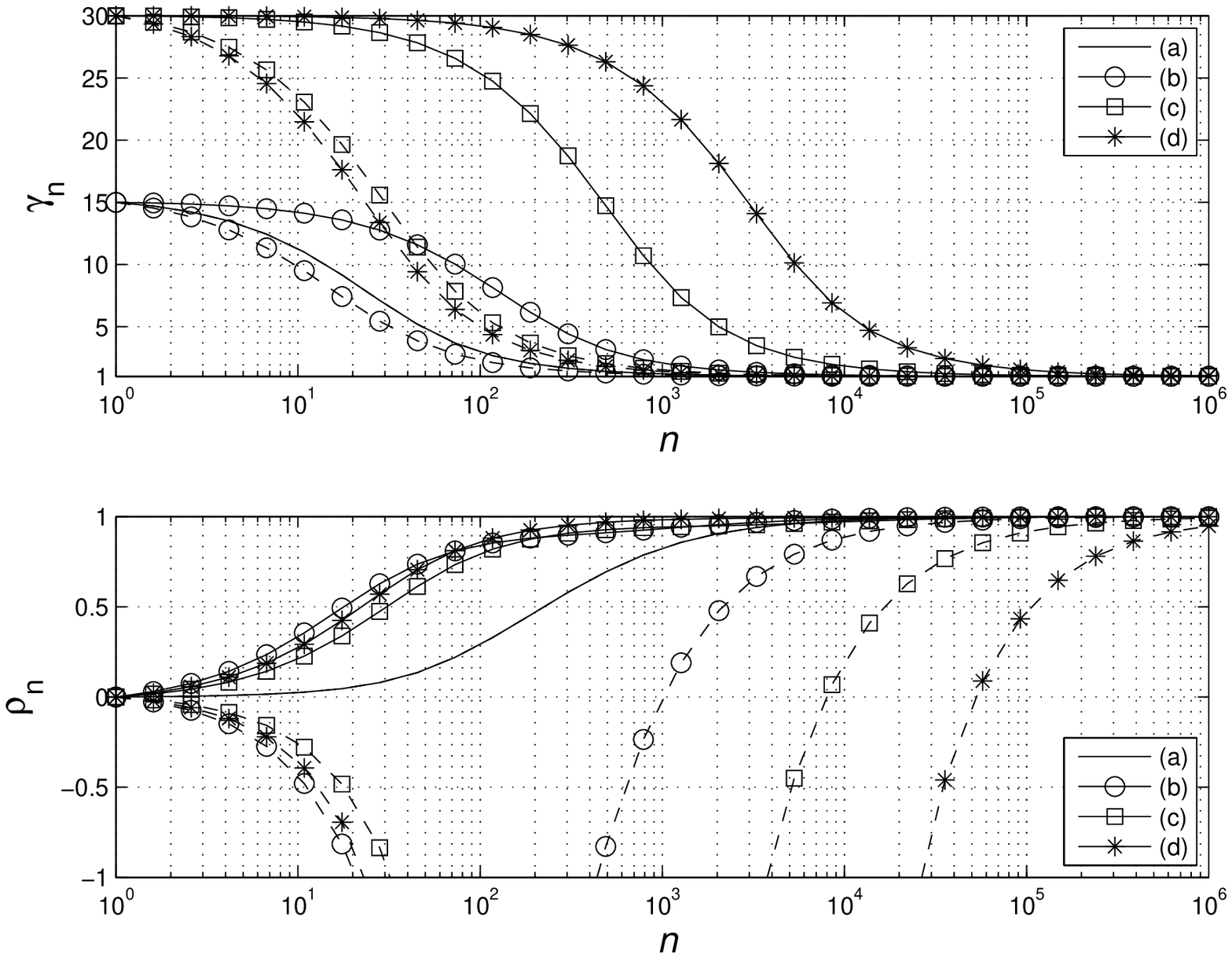}}
\caption{Bounds for the normalized variance $\gamma_{n,i}$ and for the consensus coefficient $\rho_{n,ij}$,
as provided by Theorem~\ref{prop:3}. The upper bounds are drawn as solid curves, while the
lower bounds as dashed lines. The labeling $(a)$-$(d)$ refers to the networks in Figure~\ref{fig:scheme} (in $(a)$ the two bounds coincide). Recall that $(a)$ and $(b)$ refer to $n=15$, while $(c)$ and $(d)$ refer to $n=30$, whence the differences of $\gamma_{n,i}$ at the initial point $n=1$.}
\label{fig:bound1}
\end{figure}

In Figure~\ref{fig:topology} $(a)$, $M=15$ sensors are arranged to form a ring, and all 
the pairs of node are admissible. 
In Figure~\ref{fig:topology} $(b)$ the same ring topology of Figure~\ref{fig:bound1} $(a)$ 
is considered, with the difference that
each sensor can only communicate with four neighbors, two in one direction and two in the opposite one,
and this result in a lower number of admissible pairs.
Similarly, Figures.~\ref{fig:topology} $(c)$-$(d)$ refer 
to a network made of $M=30$ sensors with a topology typical of randomly deployed sensors, and
in $(c)$ more pairs of nodes are admissible than in $(d)$.

The considered architectures determine $E[\bW]$ and,
specifically, its eigenvalues $\lambda_U$ and $\lambda_L$.
For the scenario in Figure~\ref{fig:topology}$(a)$, 
we have $\lambda=\lambda_U=\lambda_L=(M-2)/(M-1)$ ($\approx 0.9286$ in our case).
In fact, we can easily find $E[\bW]= \lambda \bI + \frac{1-\lambda}{M} \bone \bone^T$, and the eigenvalues of 
such matrix are $\lambda, \lambda, \dots, \lambda, 1$. 
The bounds in both the eqs.~(\ref{eq:claim1}) and~(\ref{eq:claim2}) coincide, 
implying that $\gamma_{n,i}$ and $\rho_{n,ij}$ can be computed exactly. 
These functions are drawn in Figure~\ref{fig:bound1} as solid lines without markers in top and bottom panels, respectively. 
We see that $\gamma_{n,i}$ starts from $M=15$ and decreases monotonically 
toward~$1$, while $\rho_{n,ij}$ grows monotonically from~$0$ to~$1$.

Consider now the network of Figure~\ref{fig:topology} $(b)$: here we have $\lambda_U \approx 0.9868$
and $\lambda_L \approx 0.8921$. The bounds on $\gamma_{n,i}$, see eq.~(\ref{eq:claim1}), 
are shown in the top plot of Figure~\ref{fig:bound1}, while the bounds in eq.~(\ref{eq:claim2})
for the consensus coefficient $\rho_{n,ij}$ are drawn in the bottom plot.
The vertical axis of this latter is limited to the meaningful range $(-1,1)$; in fact
the lower bound for $\rho_{n,ij}$ may occasionally fall below~$-1$, thus loosing significance.

The eigenvalues $(\lambda_U,\lambda_L)$ of the networks depicted in Figure~\ref{fig:topology} $(c)$ and $(d)$ 
are $(0.9964, 0.9426)$ and $(0.9994, 0.9262)$, respectively. The bounds in 
eqs.~(\ref{eq:claim1}) and ~(\ref{eq:claim2}) are also drawn in Figure~\ref{fig:bound1}.
Note that the performance index $\gamma_{n,i}$ now starts from $M=30$.

\begin{figure}
\centerline{\includegraphics[width = 0.95\textwidth,angle=0]{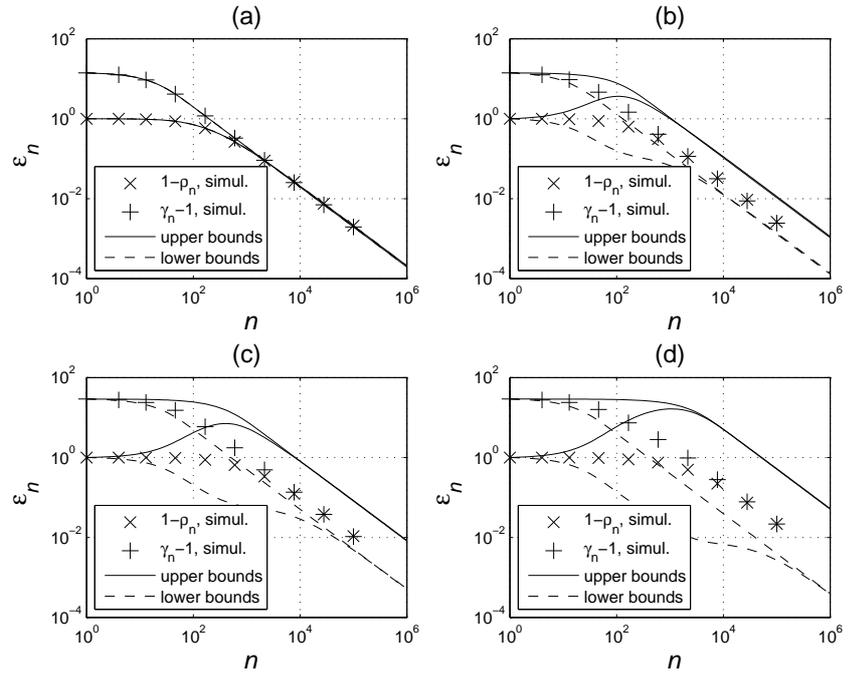}}
\caption{Bounds for the performance indices $\epsilon_n=\gamma_{n,i}-1$ and $\epsilon_n=1-\rho_{n,ij}$ as function of $n$,
for the same four scenarios $(a)$-$(d)$ of Figure~\ref{fig:scheme}. For large $n$,
the bounds of the two performance figures coincide, as prescribed by eq.~(\ref{eq:claim3}).
The upper bounds are drawn as solid curves, while the
lower bounds as dashed lines. The points marked with ``$+$'' and ``$\times$'' result from computer simulations for estimating respectively $\gamma_n=\sum_i\gamma_{n,i} / M$ and $\rho_n=\sum_{ij}\rho_{n,ij} /J$, where $J$ is the number of admissible pairs.}
\label{fig:bound2}
\end{figure}

The asymptotic behavior of the network performances is better highlighted in Figure~\ref{fig:bound2}, where the panels $(a)$-$(d)$ refer to the their analogue in Figure~\ref{fig:topology}, and the bounds for $\gamma_{n,i}-1$ and $1-\rho_{n,ij}$ are drawn on the same plot. Note that, for large $n$, the bounds simplify as in eq.~(\ref{eq:claim3}), 
giving the portion of the curves marked by dots.
Clearly, in Figure~\ref{fig:bound2} $(a)$ the upper and the lower bound coincide so that only two curves are drawn. 
In Figure~\ref{fig:bound2} we also check the derived
bounds by means of computer simulations based on a standard Monte Carlo counting procedure. 
For each time slot $n=1,2,\dots$, the shown simulation points involve
$10^3$ program runs for computing the entries  $(\bC_n)_{ij}$ of the covariance matrix. Then, 
the estimated values of $\rho_{n,ij}$ and $\gamma_{n,i}$ are obtained as arithmetic averages of the pertinent entries;
for instance, the values of $\gamma_{n,i}$ result from averaging the $n$ diagonal entries $(\bC_n)_{ii}$, see eq.~(\ref{eq:gamma}).

\chapter{Sequential Detection}
\label{ch:CH4}

The detection problem is considered using the running consensus~\cite{asymptotic-rc,braca-spawc09}. Here the sensors are cooperating by the running consensus scheme to discriminate among two hypotheses. The problem formalization is presented in Section~\ref{sec:form_CH4}, then the decentralized statistic is developed in Section~\ref{sec:form_running_CH4}. The asymptotic optimality is proved in Section~\ref{sec:asy}, where we study the performance for the Neyman-Pearson setup, in which the number of measurements used for the decision is fixed, and for the sequential setup in which there is a virtually unbounded number of data available, and an appropriate number of these is used to make the final decision according to prescribed error probabilities. Examples of application and
computer experiments are given in Section~\ref{sec:appli_CH4}, while Section~\ref{sec:bo} addressed some specific issues of practical relevance. 

\section{Centralized Hypothesis Testing}
\label{sec:form_CH4}

We assume that a network of wireless sensors monitors a phenomenon of physical interest, modeled as a binary state of the nature. Each sensor collects $n$ samples; all data are iid. By denoting with $x_{i,j}$ the $i^{th}$ sample collected by the $j^{th}$ sensor, the goal of the network is to discriminate between the two hypotheses
\beq
\begin{array}{lcl}
{\cal H}_0&:& x_{i,j}\sim f_{\theta_0}(x),\\
{\cal H}_1&:& x_{i,j}\sim f_{\theta}(x),
\end{array}
\label{eq:test0}
\eeq
where $i=1,2,\dots, n$, $j=1,2,\dots, M$ and $f_{\theta_0}(x)$, $f_{\theta}(x)$ are the marginal probability density functions of the data, parametrized by $\theta$.

A centralized statistic $T_n^{(c)}$ must be able to discriminate among ${\cal H}_0$ and ${\cal H}_1$. Two main different kinds of detection strategies are available: the Neyman-Pearson setup~\cite{lehmann-testing,kaydetection} and the sequential setup~\cite{poorbook,wald}. 

In the Neyman-Pearson setup the time index $n$ is fixed and non-random, then the number of measurements is $n\,M$. The centralized detector is identified by the pair $(T^{(c)}_n,\delta_n)$, in the sense that ${\cal H}_1$ is declared whenever $T^{(c)}_n \ge \delta_n$ and ${\cal H}_0$ is declared otherwise. Typically the choice of $\delta_n$ takes account of the desired false alarm probability.

\begin{figure}
\centerline{\includegraphics[width = 1.2\textwidth,angle=0]{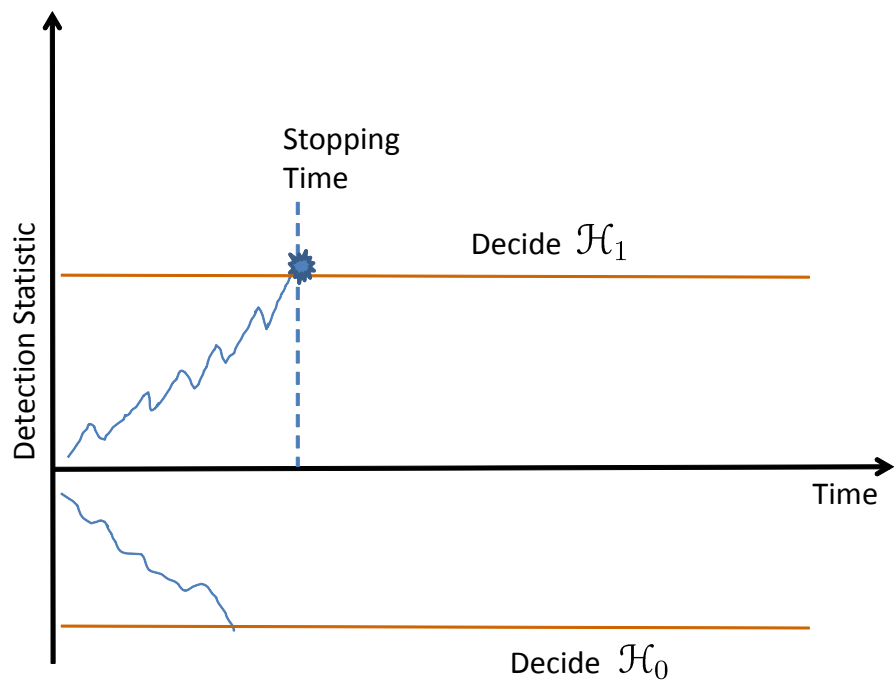}}
\caption{The sequential test, defined in~(\ref{eq:decseq_intro}) is depicted here. Under ${\cal H}_0$ the detection statistic tends to decrease down to eventually declare ${\cal H}_0$ when the lower threshold is crossed. Under ${\cal H}_1$ the detection statistic tends to increase up to eventually declare ${\cal H}_1$ if the upper threshold is crossed. The stopping time is defined as the first instant index in which a threshold is crossed.}
\label{fig:seq_test}
\end{figure}

In the sequential paradigm the number of measurements cannot be fixed in advance, and the decision rule is defined as~\cite{poorbook,wald}
\beq
\left\{
\begin{array}{ll}
T^{(c)}_n \ge b,& \textnormal{declare } {\cal H}_1,\\
T^{(c)}_n \le a,& \textnormal{declare } {\cal H}_0,\\
T^{(c)}_n \in \left(a,b\right),& \textnormal{take another sample},
\end{array}
\label{eq:decseq_intro}
\right.
\eeq
where $a$ and $b$ determine the error probabilities of the test~\cite{poorbook}.
The time index, in which a decision is taken, is called stopping time. In Figure~\ref{fig:seq_test} an example of sequential test, defined in~(\ref{eq:decseq_intro}), is schematically depicted. Under ${\cal H}_0$ the detection statistic tends to decrease to eventually declare ${\cal H}_0$ when the lower threshold is crossed. Under ${\cal H}_1$ the detection statistic tends to increase up to eventually declare ${\cal H}_1$ if the upper threshold is crossed. The stopping time is the first instant index in which a threshold is crossed.

\section{Hypothesis Testing in Sensor Networks with Running Consensus}
\label{sec:form_running_CH4}

The focus is on detectors that are asymptotically equivalent when the hypotheses come close to each other, as the number of samples increases. We consider, as is standard in these contexts~\cite{kassam}, detection statistics of the form:
\beq
T^{(c)}_n=\sum_{i=1}^n \sum_{j=1}^M t(x_{i,j});
\label{eq:centrg}
\eeq
note that $T^{(c)}_n$ is the ideal centralized state, defined in (\ref{eq:ics}) choosing $\chi_n=1$. 
Since data are iid, the first two moments of the ideal centralized statistic are the sum of the moments of $t(x_{i,s})$:
$\E_{\theta}[T^{(c)}_n]=n\,M\,\mu(\theta)$, and $\VAR_{\theta}[T^{(c)}_n]=n\,M\,\sigma^2(\theta)$, 
where $\E_\theta[\cdot]$ stands for expectation under $f_\theta(x)$, including as particular case $\theta=\theta_0$.
In order to compute statistics of the form~(\ref{eq:centrg}), all data should be available at a common site. The focus is on a fully decentralized flat architecture, without a fusion center and we would like to obtain some surrogate form of $T^{(c)}_n$ made available to {\em any} sensor of the network.

We assume that sensors measure and exchange data according to the running consensus procedure, as detailed in Chapter~\ref{ch:2}. 
Let us denote with $T_{n,j}$ the state (detection statistic) computed at the $n^{th}$ epoch by the $j^{th}$ node. The updating rule for the states of the nodes according to the eq.~(\ref{eq:basic_CH2}) with $\alpha_n=1$ and $\beta_n=M$
\beq
\bT_n=\bW_n\left[ \bT_{n-1} +M\ {\bt}({\bx}_n) \right], \label{eq:vectorform}
\eeq 
where the definition of vectors $\bT_n$ and ${\bt}({\bx}_n)$ is equivalent of that in Chapter~\ref{ch:2}.

For our purposes, it is convenient to make explicit the relationships between the statistic available at the $j$ sensors and the centralized statistic $T^{(c)}_n$: we can always write, $\forall \, n$,
\beq
\bT_{n}=T^{(c)}_n\,\bone + \be_{n},
\label{eq:Tsn}
\eeq
or, by zooming on a single sensor:
\beq
T_{n,j}=T^{(c)}_n + e_{n,j},
\label{eq:Tsn2}
\eeq
$e_{n,j}$ being the running consensus error at instant $n$ for the $j^{th}$ sensor, see (\ref{eq:yerr_2}).

The main result we are going to present is that the decentralized statistic $T_{n,j}$, {\em for any node} $j$, is asymptotically equivalent to $T^{(c)}_n$, as claimed by the theorems in the next section.
Such result follows from the basic properties of the error $e_{n,j}$, see Chapter~\ref{ch:2}. Roughly speaking, the consensus error becomes less relevant with respect to the centralized statistic $T^{(c)}_n$ when the time index $n$ grows. In fact $T^{(c)}_n$ has an expectation proportional to $n$ while the average square of the consensus error is bounded by a constant, independent from $n$, see equation~(\ref{eq:Varerr}). Now supposing that $n$ is sufficiently large the consensus error becomes negligible with respect to the centralized statistic. 

\section{Asymptotic Performances} 
\label{sec:asy}

A meaningful setup for the asymptotic design and characterization of the detector is obtained by studying test~(\ref{eq:test0}), as the parameter $\theta$ of the alternative hypothesis approaches $\theta_0$, and we investigate such issue by considering two different kinds of detection strategies: the Neyman-Pearson setup (FSS test) and the sequential paradigm (see \emph{e.g.},~\cite{poorbook}). 

In the former case, we refer to the standard framework of Pitman ARE~\cite{PitmanARE}, and asymptotic optimality is proved by simply exploiting the boundedness of the second moment of $e_{n,j}$. 
Addressing the sequential case is considerably more involved. Therefore, we first introduce the proper asymptotic framework proposed by Lai~\cite{lai-AMS}, and then offer a rigorous proof of asymptotic optimality, as detailed in Appendices~\ref{sec:lemma} and~\ref{app:S}.

\subsection{Fixed Sample Size Test} \label{sec:FSS}

Let us consider the hypothesis test of a simple alternative $\theta=\theta_0$ against the one-sided alternative $\theta>\theta_0$, with $\theta$ being a real, unknown parameter.
Formally, we use model~(\ref{eq:test0}), with~\cite{kassam}
\beq
\theta_n=\theta_0+\frac{\gamma}{\sqrt{n}},~~~\gamma>0,
\label{eq:FSSscaling}
\eeq
and the performance figures are studied in the limit of $n\rightarrow\infty$.
For a given $n$ the ideal centralized detector is identified by the pair $(T^{(c)}_n,\delta_n)$, in the sense that ${\cal H}_1$ is declared whenever $T^{(c)}_n \ge \delta_n$ and
${\cal H}_0$ is declared otherwise. 
The detection and false alarm probabilities are accordingly defined as $p_{dn}=\P[T^{(c)}_n\ge \delta_n|{\cal H}_1]$,
$p_{fn}=\P[T^{(c)}_n \ge \delta_n|{\cal H}_0]$, 
and the detector is said of (asymptotic) size $p_f$ if $\lim_{n\rightarrow \infty} p_{fn}=p_f$~\cite{kassam}.
\beq
\begin{array}{lcl}
&&\displaystyle{\frac{T^{(c)}_n-n\,M\mu(\theta_0)}{\sqrt{n\,M}\,\sigma(\theta_0)}}
\stackrel{f_{\theta_0}}{\longrightarrow} {\cal N}(0,1),
\\
&&\displaystyle{\frac{T^{(c)}_n-n\,M\mu(\theta_n)}{\sqrt{n\,M}\,\sigma(\theta_n)}}
\stackrel{f_{\theta_n}}{\longrightarrow} {\cal N}(0,1),
\label{eq:centrasynorm}
\end{array}
\eeq
where $\stackrel{f_{\theta_n}}{\longrightarrow}$ (resp. $\stackrel{f_{\theta_0}}{\longrightarrow}$) denotes convergence in distribution under $f_{\theta_n}(x)$ (resp. under $f_{\theta_0}(x)$) as $n$ diverges. 

Convergence in distribution requires that the cumulative distribution function of a sequence of random variables converges toward the cumulative distribution function (the standard normal, in our case) of the limiting random variable. 
Now, asymptotic normality under $\theta_0$ is expected by simple application of the Central Limit Theorem (CLT), due to the additive nature of the centralized statistic $T^{(c)}_n$. Under the alternative, the convergence is less immediate since the normalization terms and the underlying distributions vary with $n$. However, exploiting  
extensions of CLT to triangular arrays or Le Cam's contiguity theory, such convergence is usually met under very mild technical conditions, see \emph{e.g.},~\cite{lehmann-testing,billingsley-book}.

Under the assumption (\ref{eq:centrasynorm}), it can be shown that, for any detector of size $p_f$, 
$p_d = \lim_{n\rightarrow\infty}p_{dn}=Q\left(Q^{-1}(p_f)-\gamma d\right)$,
where $Q(\cdot)$ is the area under the right tail of a standard Gaussian function, $Q^{-1}(\cdot)$ is its inverse function, and 
$d=\sqrt{M}\;  \mu^{\prime}(\theta_0)/\sigma(\theta_0)$ is called efficacy\footnote{In the literature sometimes $d^2$ is referred to as the efficacy, see also footnote 3 in~\cite{kassam}. Note that, in general, the normalizing functions $n\, M\, \mu(\theta)$ and $\sqrt{n\,M} \, \sigma(\theta)$ yielding asymptotic normality need not to be the moments of the detection statistic. The definition of efficacy involves, in any case, just those normalizing functions.}~\cite{kassam,lehmann-testing}. We note explicitly that the asymptotic formula depends on the specific relationship between $\theta_n$ and $\theta_0$ through the factor $\gamma$.

Comparison of two detectors in the asymptotic regime is usually accomplished by their Asymptotic Relative Efficiency (ARE). The ARE of detector~2 with respect to detector~1 is defined as the asymptotic ratio of the sample size of detector~1, say $n_1$, divided by that of detector~2, $n_2$, for the same asymptotic probabilities $p_f$ and $p_d$. We have
$\ARE=(n_1/n_2)=(d_2/d_1)^2$, where $d_1$ and $d_2$ are the efficacies of detector $1$ and $2$, respectively~\cite{PitmanARE,Noether}. As a consequence, we shall say that two detectors are asymptotically equivalent if they share the same efficacy $d$, and that a detection statistic is asymptotically optimal if it reaches the best attainable efficacy, which, under suitable regularity conditions, is $d_{max}=\sqrt{M\;I(\theta_0)}$, where $I(\theta_0)$ is the Fisher information at $\theta_0$~\cite{lehmann-testing,vantrees-book1}

Since now, we have essentially summarized some known results referred to the ideal centralized statistic $T^{(c)}_n$. 
Now, we want to characterize the asymptotic detection performances of the decentralized statistic $T_{n,j}$ when the running consensus algorithm comes into scene. 
To this end, let us consider the detector $(T_{n,j},\delta_n)$, that is, we design the running consensus test using the same threshold $\delta_n$ which is used for the ideal system.

\vspace*{5pt}
\noindent
\textbf{Theorem 2.} 
{\em
Under mild technical conditions, see Subsection~\ref{subsec:FSS}, if the network graph is connected then the decentralized detection statistic $T_{n,j}$ is asymptotically equivalent to the centralized detection statistic $T^{(c)}_n$.
~\hfill$\bullet$}
\vspace*{5pt}

\noindent
Before ending this section, we would like to make a brief comment on the main hypotheses of the theorem.
The technical regularity conditions, detailed in Subsection~\ref{subsec:FSS}, are usually adopted in the context of asymptotic detection~\cite{kassam}, and focused on the asymptotic normality of the detection statistic. The condition on the graph connectivity is a basic requirement~\cite{boyd-gossip-IT} of having that the information can flow toward/from each sensor. 

\subsection{Locally Optimum Sequential Detection}
\label{sec:seq}
First, some known results about sequential testing between continuous time processes are recalled.
Let ${\cal W}(t)$ be a standard Wiener process, where $t$ ranges over the reals,
and suppose that one wants to test the hypothesis that the Wiener process has a negative drift $-d/2$ against
that of a positive drift $+d/2$. In formulas: 
\[
\begin{array}{lcl}
{\cal H}_0&:& {\cal W}(t) = {\cal W}_{-d/2}(t),\\
{\cal H}_1&:& {\cal W}(t) = {\cal W}_{+d/2}(t),
\end{array}
\]
Adopting a sequential approach, it is known that the optimum test, in the sense of minimizing the expected sample size for a prescribed pair of error probabilities $p_f$ and $p_d$, is 
\[
\left\{
\begin{array}{ll}
{\cal W}(t) \ge \beta,& \textnormal{declare } {\cal H}_1,\\
{\cal W}(t) \le \alpha,& \textnormal{declare } {\cal H}_0,\\
{\cal W}(t)\in \left(\alpha,\beta\right),& \textnormal{take another sample},
\end{array}
\right.
\]
where the thresholds exactly take the form~\cite{basseville-book}
\beq
\alpha=\frac{1}{d}\log\frac{1-p_d}{1-p_f},
\qquad
\beta=\frac{1}{d}\log\frac{p_d}{p_f}, 
\label{eq:thr}
\eeq
The above test implicitly defines a (continuous) stopping time, \emph{viz.}:
\[
\tau=\inf\left\{t:{\cal W}(t)\notin (\alpha,\beta) \right\}.
\]
and the average times for making a decision are
\beq
\E_{-d/2}[\tau]=2\frac{{\cal D}_b(p_f,p_d)}{d^2},
\quad
\E_{d/2}[\tau]=2\frac{{\cal D}_b(p_d,p_f)}{d^2},
\label{eq:WienerASN}
\eeq
where we used $\E_{\pm d/2}$ to denote expectation under the two hypotheses, and where ${\cal D}_b(p,q)$ stands for the Kullback-Leibler divergence between the binary probability mass functions $[p,1-p]$ and $[q,1-q]$~\cite{CT}. 

Let us now come back to our discrete-time problem, involving the statistic $T^{(c)}_n$, see~(\ref{eq:centrg}). 
In paralleling the reasoning used in the FSS case, we study the asymptotic properties of a sequential decision rule as the hypotheses come close to each other. However, in sequential tests the number of samples is a random quantity, and a  natural modification of~(\ref{eq:FSSscaling}), see~\cite{lai-AMS}, is 
\beq
\theta_r=\theta_0+\frac{1}{\sqrt{r}},
\label{eq:test}
\eeq
where $r$ is a positive number, and the asymptotic performances will be studied in the limit $r \rightarrow \infty$.

The sequential test is then implemented as follows. We fix a value of the parameter $r$ in~(\ref{eq:test}), and the decision rule is designed for such $r$. To this aim, it is expedient to shift the detection statistic such that its expectations under the two hypotheses are opposite in sign. 
This is easily accomplished by defining 
\[
\frac{\E_{\theta_r}[T^{(c)}_n]+\E_{\theta_0}[T^{(c)}_n]}{2}=n\,M\,\frac{\mu(\theta_r)+\mu(\theta_0)}{2}\dfz 
n\,M\,\eta_r,
\] 
so that a sequential decision rule can be formulated as
\beq
\left\{
\begin{array}{ll}
T^{(c)}_n - n\,M\,\eta_r \ge b_r,& \textnormal{declare } {\cal H}_1,\\
T^{(c)}_n - n\,M\,\eta_r \le a_r,& \textnormal{declare } {\cal H}_0,\\
T^{(c)}_n - n\,M\,\eta_r\in \left(a_r,b_r\right),& \textnormal{take another sample}.
\end{array}
\label{eq:decseq}
\right.
\eeq
Such a test clearly defines implicitly a random variable $N_r$ representing the number of samples needed to terminate the testing procedure:
\beq
N_r=\inf\left\{n:T^{(c)}_n - n\,M\,\eta_r \notin \left(a_r,b_r\right) \right\}.
\label{eq:stop}
\eeq
The false alarm and detection probabilities are
\beqa
p_{dr}&=&\P[\textnormal{declare } {\cal H}_1|{\cal H}_1],\\
p_{fr}&=&\P[\textnormal{declare } {\cal H}_1|{\cal H}_0].
\eeqa
A suitable mathematical tool for dealing with the asymptotic performance of sequential tests, in the limit of $r \rightarrow \infty$,
is provided in~\cite{lai-AMS}, which essentially formulates a LOD (Locally Optimum Detection) theory in the sequential framework.
The rationale behind the development used in~\cite{lai-AMS} is as follows. For a prescribed pair of error probabilities, as the two hypotheses come close to each other ({\em i.e.}, as $r$ diverges), the (average) number of samples needed by the detection statistic for exceeding one of the thresholds is expected to increase. Therefore, the time evolution of the statistic can be regarded as a random walk which moves inside two barriers, with single steps that become smaller and smaller with respect to the distance between the barriers, as the hypotheses approach each other.
Otherwise stated, as $r$ grows, the random walk approaches a continuous time process.

In~\cite{lai-AMS} the above simple intuition is made precise by considering the following detection statistic
\beq
T^{(c)}_{[rt]} - [rt]\,M\,\eta_r= T^{(c)}_n - n\,M\,\eta_r,
\label{eq:shiftt}
\eeq
where $[x]$ stands for the integer part of $x$, and $t$ defines a continuous time axis. 
This means that we are looking at a piecewise constant random process which changes values at the integer time instants $n/r$ (for integer $n$), where the elementary step $1/r$ goes to zero as $r$ diverges.
As to the limiting distribution of the above statistic for $r \rightarrow \infty$, we can invoke the functional central limit theorem, namely, the convergence to Wiener processes. In paralleling eq.~(\ref{eq:centrasynorm}) suppose that 
the following convergences hold ($d$ is the efficacy):
\beq
\begin{array}{lcl}
&&\displaystyle{\frac{T^{(c)}_{[rt]} - [rt]\,M\,\eta_r}{\sqrt{r\,M}\,\sigma(\theta_0)}}
\stackrel{f_{\theta_0}}{\longrightarrow} {\cal W}_{-d/2}(t),
\\
&&\displaystyle{\frac{T^{(c)}_{[rt]} - [rt]\,M\,\eta_r}{\sqrt{r\,M}\,\sigma(\theta_0)}}
\stackrel{f_{\theta_r}}{\longrightarrow} {\cal W}_{+d/2}(t),
\label{eq:asyWiener}
\end{array}
\eeq
where $\stackrel{f_{\theta_{r}}}{\longrightarrow}$ (resp. $\stackrel{f_{\theta_{0}}}{\longrightarrow}$) denotes weak convergence to a random process~\cite{billingsley-book}, under $f_{\theta_r}(x)$ (resp. $f_{\theta_0}(x)$) as $r$ diverges.
As previously discussed for the FSS test, the convergence under $\theta_0$ is expected by application of the (functional version of the) CLT, also known as Donsker's theorem~\cite{billingsley-book}. For the same reasons explained for the FSS test, proving the convergence under $\theta_r$, which is usually met in many practical cases, is more tricky.
 
In light of~(\ref{eq:asyWiener}), the results on testing two continuous Wiener processes ${\cal W}_{\pm d/2}(t)$ summarized at the beginning of this section, can be exploited in our discrete-time setup. 
Indeed, a threshold comparison of the form~$\frac{T^{(c)}_{[rt]} - [rt]\,M\,\eta_r}{\sqrt{r\,M}\,\sigma(\theta_0)}\notin(\alpha,\beta)$, yields explicit expression of the thresholds to be used in~(\ref{eq:decseq}): 
\beq
a_r=\sqrt{r\,M}\,\sigma(\theta_0)\,\alpha,\qquad
b_r=\sqrt{r\,M}\,\sigma(\theta_0)\,\beta.
\label{eq:ar_br} 
\eeq

It is possible to show that the detection and false alarm probabilities used for setting the thresholds $\alpha$ and $\beta$, are asymptotically attained by the designed detector~\cite{lai-AMS}. Namely,
\beq
\lim_{r\rightarrow\infty}p_{dr}=p_d,
\qquad
\lim_{r\rightarrow\infty}p_{fr}=p_f.
\label{eq:pfpd}
\eeq
Furthermore, in view of the relationship $n=[rt]$, one would expect that $N_r\sim [r\tau]$.
Indeed, it turns out that, in the light of eq.~(\ref{eq:WienerASN}), the expected sample size scales as~\cite{lai-AMS} 
\beq
\begin{array}{rcl}
\displaystyle{\lim_{r\rightarrow\infty}\frac{\E_{\theta_0}[N_r]}{r}}&=&
\displaystyle{2\frac{{\cal D}_b(p_f,p_d)}{d^2}},\nonumber\\ \, \\
\displaystyle{\lim_{r\rightarrow\infty}\frac{\E_{\theta_r}[N_r]}{r}}&=&
\displaystyle{2\frac{{\cal D}_b(p_d,p_f)}{d^2}}.
\end{array}
\label{eq:ASN}
\eeq
By defining the asymptotic relative efficiency as the ratio of the expected samples sizes, and accordingly taking the limit for large $r$, it is immediate to see that these two detectors are simply compared in terms of their efficacies, just as happens in the FSS case.
It is now natural to use, for the sequential framework, the same definitions of asymptotic equivalence and asymptotic optimality adopted in the FSS case. 

In addition, the unbeatable efficacy in the sequential case is the same of $d_{max}$ defined in the FSS context.
This result can be obtained by using the expressions of the average stopping time for the 
optimal Sequential Probability Ratio Test (SPRT~\cite{wald,wald-wolfowitz-AMS}), in the limit of large $r$, see~\cite{lai-AMS,basseville-book}. 

The above reasoning has been focused on the ideal centralized statistic $T^{(c)}_n$. As done before, we switch now on the decentralized statistic available at the $j^{th}$ node $T_{n,j}$ and, as for the FSS test, we consider a sequential detector using the same thresholds $a_r$ and $b_r$ which are used for the ideal system. The main results about the sequential scenario is that the sequential test
\beq
\left\{
\begin{array}{ll}
T_{n,j} - n\,M\,\eta_r \ge b_r,& \textnormal{declare } {\cal H}_1,\\
T_{n,j} - n\,M\,\eta_r \le a_r,& \textnormal{declare } {\cal H}_0,\\
T_{n,j} - n\,M\,\eta_r\in \left(a_r,b_r\right),& \textnormal{take another sample},
\end{array}
\label{eq:decseq2}
\right.
\eeq
behaves asymptotically as one in which the statistic $T_{n,j}$ is replaced by the ideal centralized statistic $T^{(c)}_n$.
This statement is made precise in the following claim.

\vspace*{5pt}
\noindent
\textbf{Theorem 3.} 
{\em Under mild technical conditions, see Subsection~\ref{subsec:seq}, if the network graph is connected then the decentralized detection statistic $T_{n,j}$ in eq.~(\ref{eq:Tsn}) is asymptotically equivalent to the centralized detection statistic $T^{(c)}_n$.
~\hfill$\bullet$}\vspace*{5pt}

\noindent
Before ending this section we would like to make a remark. 
Since now, we have considered a family of alternative hypotheses with parameter $\theta_r=\theta_0+1/\sqrt{r}$, and accordingly designed and characterized the asymptotic performances of tests~(\ref{eq:decseq}) and (\ref{eq:decseq2}), with thresholds~(\ref{eq:ar_br}), as $r$ diverges.
It is also of interest to consider $\theta_r=\theta_0 + \chi/\sqrt{r}$, with $\chi$ being unknown.
The lack of knowledge about $\chi$ prevents us from setting the thresholds on the actual value of $\theta_r$. 

One solution is to design tests~(\ref{eq:decseq}) and~(\ref{eq:decseq2}) with thresholds~(\ref{eq:ar_br}), {\em i.e.}, assuming a nominal value of $\chi=1$. 
Clearly, the form of the test being unchanged, the error probability and average sample number under the null hypothesis will not depend upon $\chi$.
Moreover, it is shown in~\cite{lai-AMS} that, provided the technical conditions compactly called \emph{uniform invariance principles}, this test exhibits an asymptotic detection probability and an expected sample number under the alternative hypothesis  still expressible in terms of the performances of a Wiener process, but with drift parameter $(\chi-1/2)\,d$.
As a result, the comparison between two different test statistics can be still made by essentially looking at their efficacies.

\section{Examples and Numerical Experiments}
 \label{sec:appli_CH4}

The previous analysis is now corroborated by computer experiments with the twofold goal of $(i)$ providing a sanity check for the developed theory, and $(ii)$ investigating the behavior of the running consensus detection in practical, \emph{i.e.}, non-asymptotic, scenarios. 

As to the running consensus scheme, among the many possible choices we consider a simple pairwise exchange protocol, wherein a pair of sensors $(j,k)$ is randomly and uniformly selected among all the possible pairs taken from the set $\{1,2,\dots,M \}$, see previous chapters for details. The implicit assumption of a fully connected network is also made.
If the pair of sensors is $(j,k)$, the consensus matrix $\bW_{jk}$ would take the following form:
\beq
\bW_{jk}=\bI-\frac{(\be_k-\be_j)(\be_k-\be_j)^T}{2} , \label{eq:Wkj}
\eeq
where $\be_k$ denotes a vector of zeros, but for the $k^{th}$ entry which equals to~1, and where $\bI$ is the identity matrix. 
Note that, when the above $\bW_{jk}$ is multiplied by a vector, its effect is to replace the vector entries $j$ and $k$ by their arithmetic mean. Actually we assume that, in a single consensus step, many pairs of sensors can average their data.
This amounts to a connection matrix given by the product of $v \ge 1$ pairwise matrices of the form~(\ref{eq:Wkj}). Clearly, such a consensus matrix  operates by averaging the states of $v$ pairs of randomly selected nodes.

We first address, both in the FSS case and in sequential framework, a Gaussian example which is particularly interesting since it naturally leads to optimal detectors. Then, a non-Gaussian test is considered, focusing on the sequential case. 

\begin{figure}
\centerline{\includegraphics[width = 0.95\textwidth,angle=0]{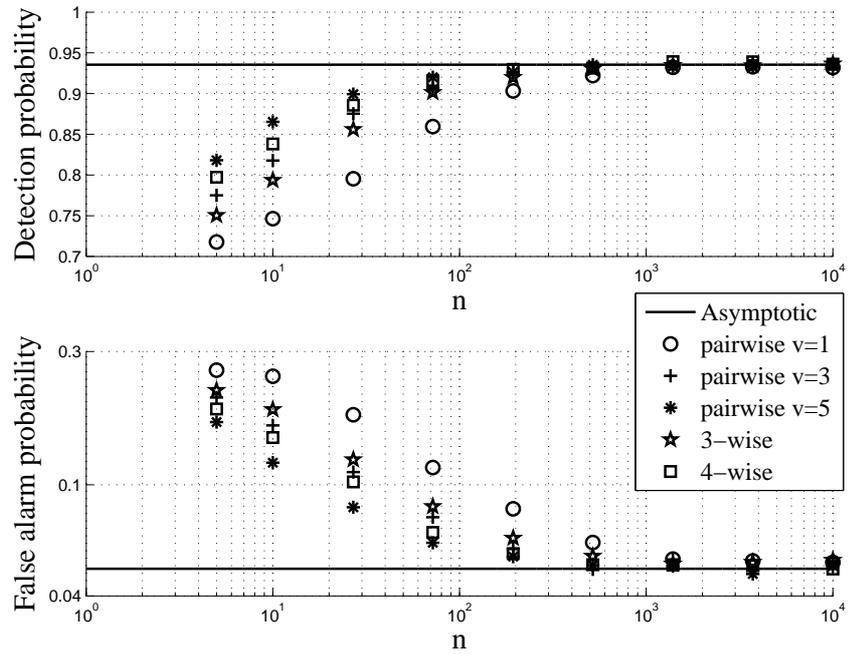}}
\caption{FSS test for the Gaussian example in a network made of $M=10$ sensors. We show the detection probability $p_{dn}$ of the running consensus versus $n$, for $v=1,10$, pairwise exchanges per epoch. According to the asymptotic framework here we set $\theta_n=1/\sqrt{n}$. Also shown is the performance of the ideal centralized system that, 
in this example, is constant with $n$ and represents the asymptotic performance. The number of Monte Carlo trials is $10^4$.}
\label{fig:FSS3}
\end{figure}

\subsection{Neyman-Pearson FSS Tests}

Let us consider the following Gaussian shift-in-mean hypothesis test: for $i=1,2,\dots,n$, and $j=1,2,\dots,M$,
\beq
\begin{array}{lcl}
{\cal H}_0&:& x_{i,j}\sim {\cal N}(0,\sigma^2) ,\\
{\cal H}_1&:& x_{i,j}\sim {\cal N}(\theta,\sigma^2) ,
\end{array}
\label{eq:exFSS1}
\eeq
where recall that the $x_{i,j}$ are \emph{iid}, and ${\cal N}(\theta,\sigma^2)$ is our shortcut for a Gaussian distribution with mean $\theta$ and variance $\sigma^2$.
For this scheme, clearly $\theta_0=0$, and
\[
\mu(\theta)=\theta,\quad \sigma^2(\theta)=\sigma^2,\quad d=\sqrt{\frac{M}{\sigma^2}}=\sqrt{M\,I(0)},
\]
where $I(0)=1/\sigma^2$ is the Fisher information.
In this particular case, the {\em optimal} detection statistic (\emph{i.e.}, the log-likelihood) is just in the additive form of~(\ref{eq:centrg}),
with $t(x_{i,j}) \propto x_{i,j}$. 
As a consequence, the ROC (Receiver Operating Characteristic) of the centralized statistic $T^{(c)}_n$ 
can be computed in closed form as
\[
Q \left ( Q^{-1} \left ( p_f\right )- \sqrt{\frac{\theta^2}{\sigma^2}\, n M}\right ).
\]
Note that, imposing $\theta=\theta_n=\gamma/\sqrt{n}$, as prescribed by the asymptotic theory, straightforwardly yields $p_d=Q\left(Q^{-1}(p_f)-\gamma d\right)$. 

We are now ready to comment on the simulation results pertaining to the considered Gaussian shift-in-mean detection problem, with and without the consensus stage. 
From the theory, we know that, in the asymptotic setting specified in the previous sections, with the hypotheses coming closer and closer as $n\rightarrow \infty$, the decentralized detector must approach the same performance as that of the ideal centralized one. 
Figure~\ref{fig:FSS3} accordingly shows the detection and false alarm probabilities of a test based on the decentralized statistic $T_{n,j}$ (for a generic $j$) of the running consensus scheme, as function of $n$, by assuming $\theta_n=\gamma/\sqrt{n}$ with $\gamma=1$. Therefore, Figure~\ref{fig:FSS3} provides a sanity-check for the asymptotic results and is useful to show the rate of convergence toward the ideal system. 

Figure~\ref{fig:FSS3} also emphasizes the role that different kinds of consensus schemes may have. With reference to the pairwise algorithm it shows that, by increasing the number $v$ of pairwise averaging per single time slot, the convergence become faster, as one may expect. 
In addition, gossip protocols with different number of sensors participating to the single-slot average are considered. Intriguingly, it is not easy to anticipate the relative merits of the different communication schemes. For instance, in Figure~\ref{fig:FSS3}, a pairwise with $v=3$, involving $6$ sensors per single time slot, is outperformed by the 4-wise scheme. 

A possible explanation of this behavior can be that, in order to achieve 4-wise averaging, we need at least 4 pairwise exchanges. 
For example, with four sensors, we can first take average of $x_1$ and $x_2$, and then of $x_3$ and $x_4$. Then, sensors $1$ and $3$, and sensors $2$ and $4$ can do pairwise averaging, which then gives each sensor the 4-wise average $(x_1+x_2+x_3+x_4)/4$. 

\begin{figure}
\centerline{\includegraphics[width = 0.95\textwidth,angle=0]{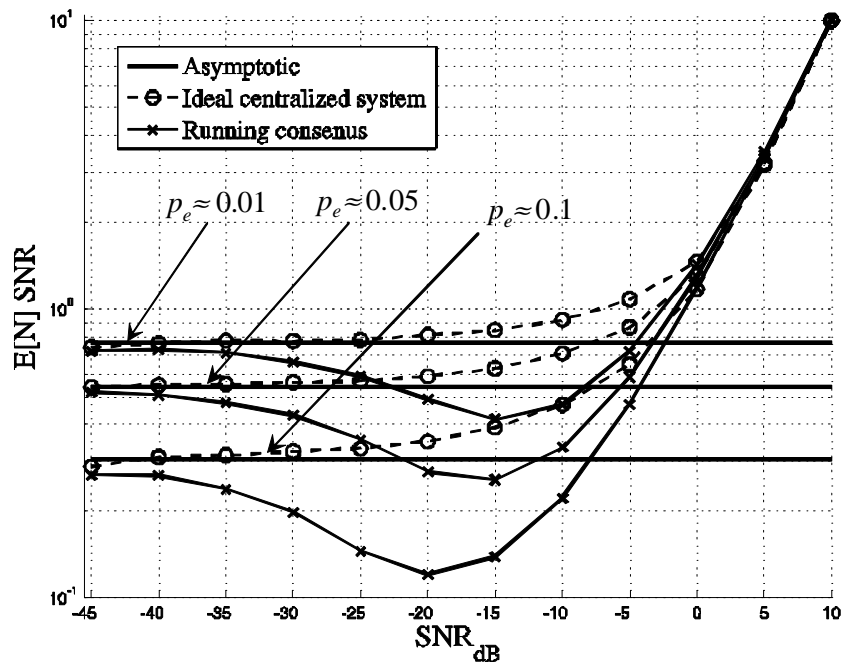}}
\caption{
Sequential tests for the Gaussian example, with $M=10$ sensors and $v=5$ pairwise exchanges per each time instant. The average sample number $\E[N]$ multiplied by the SNR is displayed as function of this latter, for three different values of the (nominal, asymptotic) error probability $p_e\approx [0.01,0.05,0.1]$. 
The three different curves refer to the ideal centralized strategy, the decentralized one with running consensus, and the asymptotic value.
The number of Monte Carlo trials is $10^4$.}
\label{fig:NmedGauss}
\end{figure}

\begin{figure}
\centerline{\includegraphics[width = 0.95\textwidth,angle=0]{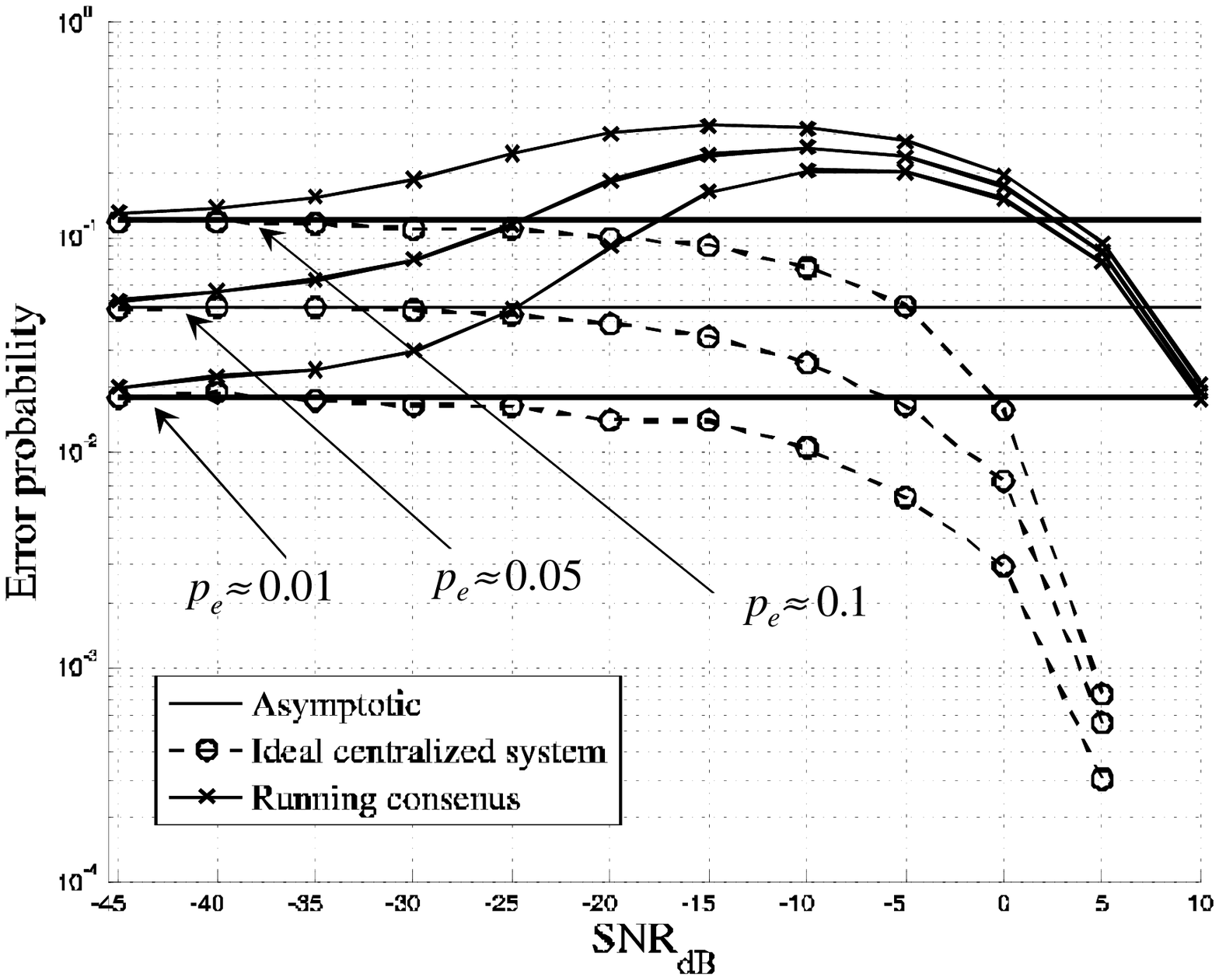}}
\caption{This figure complements Figure~\ref{fig:NmedGauss} by showing the actual error probability, as function of the SNR, for the same case study of the previous figure.}
\label{fig:PerrGauss}
\end{figure}

\subsection{Sequential Tests}
\subsubsection{Gaussian example} 
For the same shift-in-mean Gaussian problem studied in the previous section, we have implemented our sequential detection strategy with running consensus. 
Assuming $\theta_r=1/\sqrt{r}$, and using the detection statistic in eq.~(\ref{eq:centrg}), with decision rule and thresholds given by eq.~(\ref{eq:decseq}) and~(\ref{eq:ar_br}), the centralized system will compare the statistic 
\[
\sum_{i=1}^n\sum_{j=1}^M x_{i,j} - \frac{n\,M\,}{2\sqrt{r}},
\]
with thresholds
\[
a_r=\sqrt{r}\,\sigma^2\log\frac{1-p_d}{1-p_f},
\qquad
b_r=\sqrt{r}\,\sigma^2\log\frac{p_d}{p_f}.
\]
As in the FSS case, for this simple Gaussian problem, it is straightforward to show that the above (ideal, centralized) sequential test is nothing but the optimal SPRT.
In the computer simulations, and according to our theoretical findings, we use the above thresholds also for the running consensus scheme. Furthermore, for simplicity, we work under the assumption that $p_f=1-p_d$, that we call $p_e$, yielding symmetric thresholds and equality of the average sample numbers, that we accordingly denote by $\E[N]$.

As discussed in Section~\ref{sec:seq}, the asymptotic performance of the SPRT (and, in view of Theorem 3, also of the running consensus) is just ruled by $d=\sqrt{M/\sigma^2}$, such that,
following eq.~(\ref{eq:ASN}), 
\[
\E[N]\sim 
2\,r\,\frac{{\cal D}_b(1-p_e,p_e)}{d^2}=\frac 2 M \,\frac{{\cal D}_b(1-p_e,p_e)}{\SNR},
\]
where we used SNR$=\theta_r^2/\sigma^2$.

In Figure~\ref{fig:NmedGauss} the (scaled) expected stopping time is displayed, as function of the signal-to-noise ratio, for three different values of the nominal (asymptotic) error probabilities. It can be seen that, as the SNR goes to zero, the product $\E[N]\, \, \SNR$ of the sequential test with running consensus approaches that of the ideal centralized system, and both converge toward the asymptotic constant value $2\,{\cal D}_b(1-p_e,p_e)/M$.
Figure~\ref{fig:NmedGauss} also reveals that the expected sample number of the decentralized scheme is smaller than that of the optimal SPRT. This actually makes sense, in view of Figure~\ref{fig:PerrGauss}, where the other relevant performance index, that is, the error probability, is displayed.
Figure~\ref{fig:PerrGauss} shows that the error probability enforced by the decentralized scheme is always greater than that of the ideal system, thus explaining the decrease of $\E[N]$ for the running scheme. The simulation results also show that both the running consensus scheme and the ideal centralized entity exhibit error probabilities that approach their nominal asymptotic values, as predicted by~(\ref{eq:pfpd}).
\begin{figure}
\centerline{\includegraphics[width = 0.95\textwidth,angle=0]{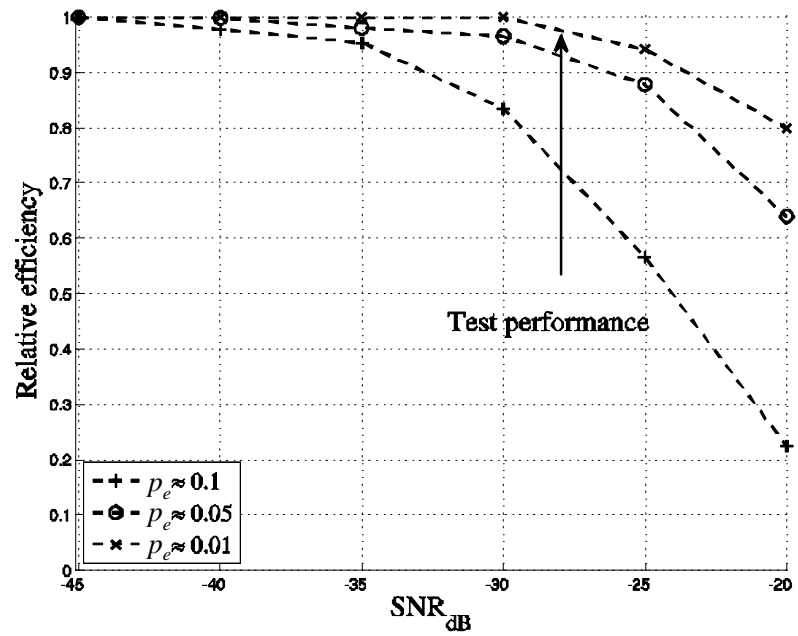}}
\caption{A summary of the evidences in Figures~\ref{fig:NmedGauss} and~\ref{fig:PerrGauss}: The (non asymptotic) relative efficiency of the decentralized versus an optimal SPRT using the same (actual) error probabilities reached by the decentralized strategy as in Figure~\ref{fig:PerrGauss}, see main text for details.}
\label{fig:AREGauss}
\end{figure}
The above considerations suggest making a further comparison between the two strategies in the following way: For each value of the error probability actually achieved by the decentralized scheme (see Figure~\ref{fig:PerrGauss}), we evaluate the correspondent average sample sizes obtained with the ideal centralized system ({\em i.e.}, an SPRT). The ratio between these numbers, namely, the relative efficiency,
is displayed in Figure~\ref{fig:AREGauss}, as function of the SNR, for the same three different values of error probability as in the previous figures. 
As it should be, being the comparison made for the same error probabilities, the SPRT outperforms our strategy, while being asymptotically equivalent (relative efficiency equal to one), when the SNR goes to zero.
Furthermore, it is seen that, the lower the error probability, the faster is the convergence toward the asymptotic value. This may be explained because smaller error probabilities mean larger thresholds and, for fixed SNR, the number of samples required to end the test grows. In these conditions, the relative impact of the consensus error is less predominant.

\begin{figure}
\centerline{\includegraphics[width = 0.95\textwidth,angle=0]{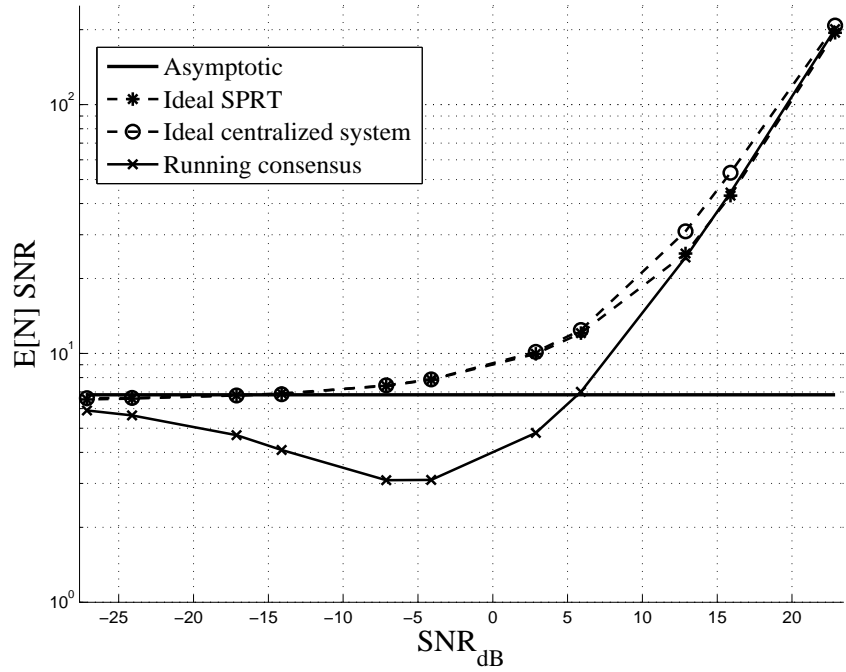}}
\caption{Sequential tests for the mixture-of-Gaussians example, with $M=10$ sensors and $v=5$ pairwise exchanges per each time instant. 
The parameters of the distribution in eq.~(\ref{eq:exFSS2}) are $p=0.3$, $\sigma_1=1$ and $\sigma_2=5$.
The arithmetic average $\E[N]$ of the sample numbers in eq.~(\ref{eq:arithm}) multiplied by the SNR is displayed as function of this latter, for $p_e\approx 0.1$. 
The four different curves pertain to the ideal centralized strategy with a score detector, the running consensus with score detector, the optimal SPRT and the asymptotic value.
The number of Monte Carlo trials is $10^4$.}
\label{fig:NmedMixt}
\end{figure}

\subsubsection{A Non-Gaussian Example} 
We switch now to a non-Gaussian shift-in-mean detection problem:
\beq
\begin{array}{lcl}
{\cal H}_0&:& x_{i,j}\sim p\,\N \left (0, \sigma_1^2 \right ) + (1-p)\,\N \left (0, \sigma_2^2 \right ) ,\\
{\cal H}_1&:& x_{i,j}\sim p\,\N \left (\theta, \sigma_1^2 \right ) + (1-p)\,\N \left (\theta, \sigma_2^2 \right ) ,
\end{array}
\label{eq:exFSS2}
\eeq
that is, a $p$-weighted mixture of two Gaussian random variables having different variances. 

We shall use, as nonlinearity characterizing the statistic in~(\ref{eq:centrg}), the well-known score function~\cite{lehmann-testing}:
\[
t(x)=
\left.\frac{\partial}{\partial\theta}\log f_\theta(x)\right|_{\theta=0}=x\,
\frac{\displaystyle{\frac{p}{\sigma_1^2}\,e^{-\frac{x^2}{2\sigma_1^2}}+\frac{1-p}{\sigma_2^2}\,e^{-\frac{x^2}{2\sigma_2^2}}}}
{\displaystyle{p\,e^{-\frac{x^2}{2\sigma_1^2}}+(1-p)\,e^{-\frac{x^2}{2\sigma_2^2}}}}.
\]

For the score detection statistic applied to our non-Gaussian setup, it is no longer easy to write explicit expressions for $\mu(\theta)$ and $\sigma(\theta)$, as well as for the Fisher information. As a consequence, we evaluate these quantities by numerical integration, and accordingly use the computed values for setting the thresholds.
We would like to stress that the test can be shown to attain the maximum efficacy $\sqrt{M\,I(0)}$, see, {\em e.g.},~\cite{basseville-book}.

\begin{figure}
\centerline{\includegraphics[width = 0.95\textwidth,angle=0]{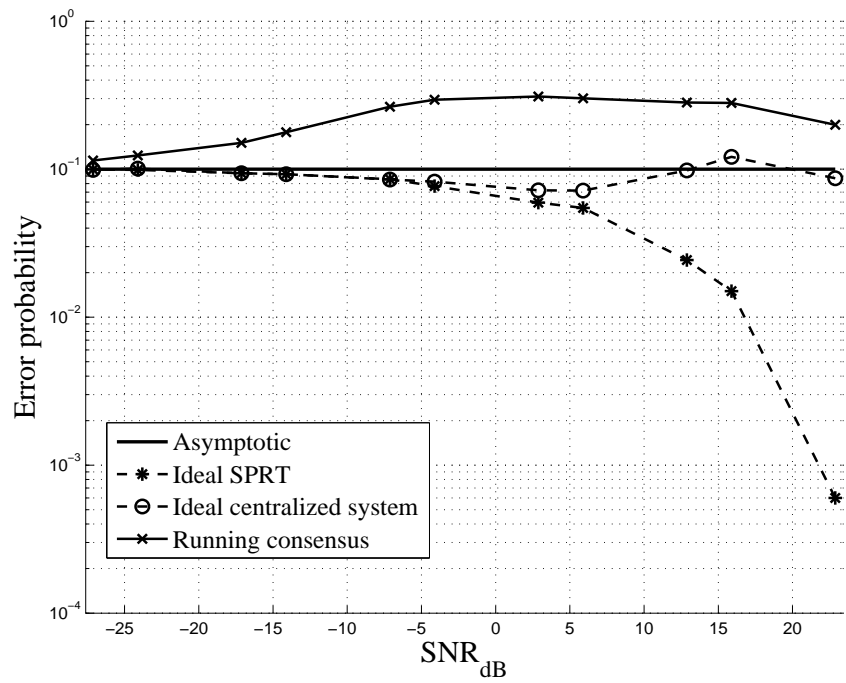}}
\caption{This figure complements Figure~\ref{fig:NmedMixt} by showing the actual error probability, as function of the SNR, for the same case study of the previous figure.}
\label{fig:PerrMixt}
\end{figure}

As done before, the simulations are designed to compare the centralized detector and the decentralized structure that implements the running consensus, under the scaling law~$\theta_r=1/\sqrt{r}$. 
In addition, an optimal (centralized) Wald's SPRT is also considered. 
Our Monte Carlo results are summarized in Figures~\ref{fig:NmedMixt} and~\ref{fig:PerrMixt}. 
Since in this case it is no longer possible to enforce perfect symmetry in terms of error probabilities and expected sample sizes, we display the arithmetic averages
\beq
p_e=\frac{p_f+(1-p_d)}{2},\quad
\E[N]=\frac{\E_{\theta_0}[N]+\E_{\theta_r}[N]}{2},
\label{eq:arithm}
\eeq
and further define the signal-to-noise ratio as 
\[
\textnormal{SNR}=\frac{\theta_r^2}{p\,\sigma_1^2+(1-p)\,\sigma_2^2}.
\]

As can be seen, for the score detector, the theoretical predictions are verified by the numerical evidence, in that the two statistics are seen to be asymptotically equivalent as the SNR decreases, and further they reach the best attainable performance $\sqrt{M I(0)}$, emphasizing how the consensus algorithm is effective for sequential detection.
Switching to the comparison with the optimal SPRT, we see that the differences between this latter and the centralized score detector are moderate, while the consensus error has an impact in non-asymptotic regimes. 

Similar considerations apply to the error probability $p_e$, as shown in Figure~\ref{fig:PerrMixt}. 
As found in the previous example, the centralized and the decentralized schemes do not achieve the same actual error probabilities. Therefore, we have again evaluated the (non-asymptotic) relative efficiencies of the proposed detectors and similar considerations as in in Figure~\ref{fig:AREGauss} apply, not reported here for sake of brevity.

\section{Some Practical Issues} \label{sec:bo}

Let us consider the sequential paradigm with running consensus, and note that in 
formulation~(\ref{eq:decseq2}) there is no explicit global stopping time for the detection task. In fact, when an individual sensor exceeds a threshold level we say that its individual decision is taken but, nonetheless, the sensor actually continues sensing the environment and participating in the data exchange. This is convenient from an analytic point of view because it preserves the problem symmetry, and allows easier analysis. 
Indeed, as already noted, the performance of the running consensus scheme can be computed with reference to an arbitrary sensor.

From a more practical perspective, however, a global termination rule is certainly needed.
One option is to stop the task of the node not at its threshold crossing but a little later, when it is very likely that all other nodes finished the task as well. Alternatively, there may be an additional flag in the sensors' data exchange that allows counting the number of sensors with individual decision taken. When such number reaches the total number of sensors, the system is turned off. Or, else, one could use broadcast messages to communicate the threshold crossings. 

\begin{figure}
\centerline{\includegraphics[width = 0.95\textwidth,angle=0]{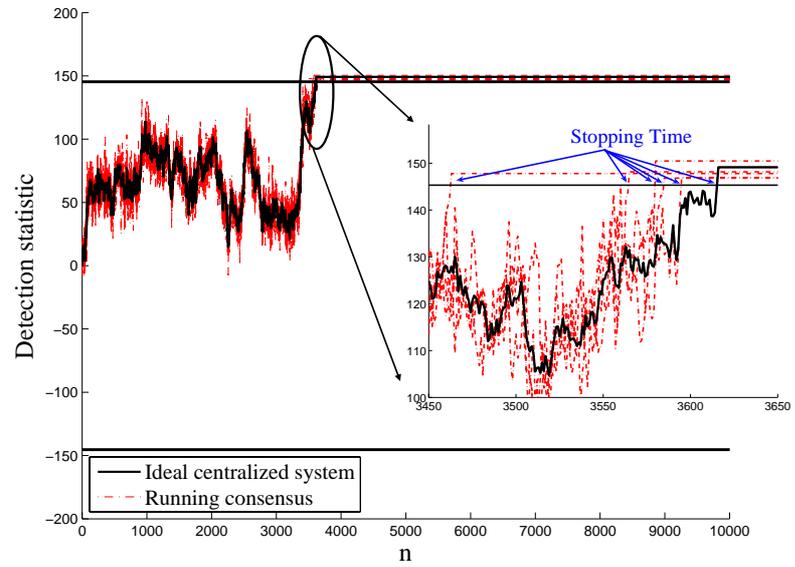}}
\caption{A realization, under ${\cal H}_1$, of the sequential test in~(\ref{eq:decseq2}), for the Gaussian shift in mean problem with $\SNR=-30$ dB. Here we consider a simple network made of $M=5$ sensors and the pairwise algorithm with $v=1$. The detection statistics of each sensor (dashed curves) are close to each other, and all closely follow the behavior of the ideal centralized statistic (solid curve). When the threshold is crossed the process is stopped to show how different stopping times cluster around a single value, as shown in the zoomed plot.}
\label{fig:stopping}
\end{figure}

The choice of one specific rule is expected to have a little impact on the system behavior,
in the light of the key point shown in the previous sections, that all nodes behaves asymptotically in the same manner.
In Figure~\ref{fig:stopping} we report an instance of a quantitative analysis carried out on this aspect. We see that the different stopping times at different sensors are close to each other, implying that the additional processing and sensing for a sensor whose statistic has already crossed the threshold is essentially negligible. 

A second issue that merits attention concerns the communication burden implied by the running consensus. Here we make specific reference to the sequential framework, but the arguments can be adapted to the FSS case as well.
The energy spent for the communication task certainly depends upon a number of factors, including the specific communication protocols, the physical resource available, the presence of feedback, the implementation of error control coding, the presence of synchronism signals, and so forth.
To simplify the matter, let us assume roughly that the energy spent in any time slot is proportional to the number of connections. For instance, if $v=1$, in the pairwise case the data exchange involves two sensors and hence costs $2$ energy units, while in the $3$-wise protocol the average of the sensors' state costs $6$ units, since each sensor must communicate with the remaining 2, and in the $4$-wise the cost is $12$ units. If $v>1$ these costs must be multiplied for $v$, the number of exchanges per time slot.\footnote{The classical consensus procedure is obtained by assuming that the number $v$ of exchanges goes to infinity, and the communication cost must be computed accordingly, see e.g.,~\cite{boyd-gossip-IT,scaglione-ITW08}.}

As to the ideal centralized system, this corresponds to the case where all the sensors share their state with all other sensors in any time slot: the energy cost amounts to $M(M-1) \approx M^2$ units. Alternatively, in an architecture equipped with a fusion center that implements the sequential test, the $M$ sensors at any time slot send their data to the fusion center and the communication cost reduces to $M$ energy units. In any case, we see that the ideal centralized system (ideal only with respect to the detection performance) implies a cost proportional or quadratic with $M$. Conversely the running consensus strategy is much more parsimonious in terms of energy.

Summarizing, with the running consensus, it is possible to achieve asymptotic optimal performances with a communication cost that does not grow with the network size $M$.
On the other hand, increasing the energy spent for the gossip protocol (that is, increasing the number of sensors sharing data in a single slot), might serve at speeding up the convergence toward the asymptotic performance.

As a final remark, we would like to stress that an implicit assumption made in this work is that the sensors exchange unquantized data, i.e., real numbers. A possible concern is that such working assumption can be only approximately met by practical systems. Furthermore, when the communication resources are particularly scarce, it would be certainly of interest to consider the case where sensors can only exchange severely quantized data and, to one extreme, only binary digits, i.e., local decisions corresponding to one-bit quantization of the local likelihoods. The problem can be conveniently put in the framework of the so-called ``quantized consensus'', see e.g.,\cite{willett-parley,basar-quantized-consensus}. Addressing the asymptotic performance of systems operating under a running quantized consensus paradigm is unlikely to be a simple generalization of the results proposed in this work, and probably requires quite different tools of analysis.

\chapter{Quickest Change Detection}
\label{ch:quickest detection}

The quickest change detection, implemented via running consensus, is addressed in this Chapter, see~\cite{Bracaetal-Pageconsensus}. The problem is to detect an event (that is modeled by a change in the data distribution) as soon as possible. Approximate performance evaluation is developed, moreover we show that the running consensus performance is close to the ideal centralized system.

The Chapter is organized as follows. In Section~\ref{sec:form} the problem is formalized, and classical results on Page's test are summarized. Section~\ref{sec:page_running} deals with the proposed decentralized counterparts thereof. In Section~\ref{sec:bank} a bank of parallel Page's test is provided. In Section~\ref{sec:RE} we give approximate formulas for the performance, in terms of relative efficiencies, of each system. In Section~\ref{sec:appli}, the algorithms are tested on a typical change detection problem, and the theoretical formulas are compared with the results of Monte Carlo simulations.

\section{Problem Formalization: Centralized Page's Test} 
\label{sec:form}

\begin{figure}
\centerline{\includegraphics[width = 1.2\textwidth,angle=0]{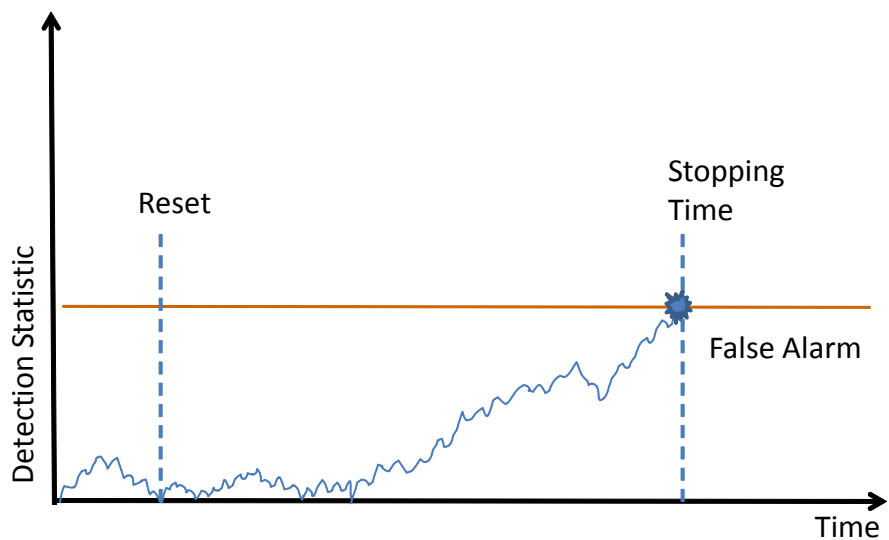}}
\caption{The Page's test, under ${\cal H}_0$, is schematically depicted here. The detection statistic has a negative drift, then it might be expected a large number of resets. The change of distribution is not present so a false alarm occurs when the statistic hits the threshold (red line).}
\label{fig:page_H0}
\end{figure}

\begin{figure}
\centerline{\includegraphics[width = 1.2\textwidth,angle=0]{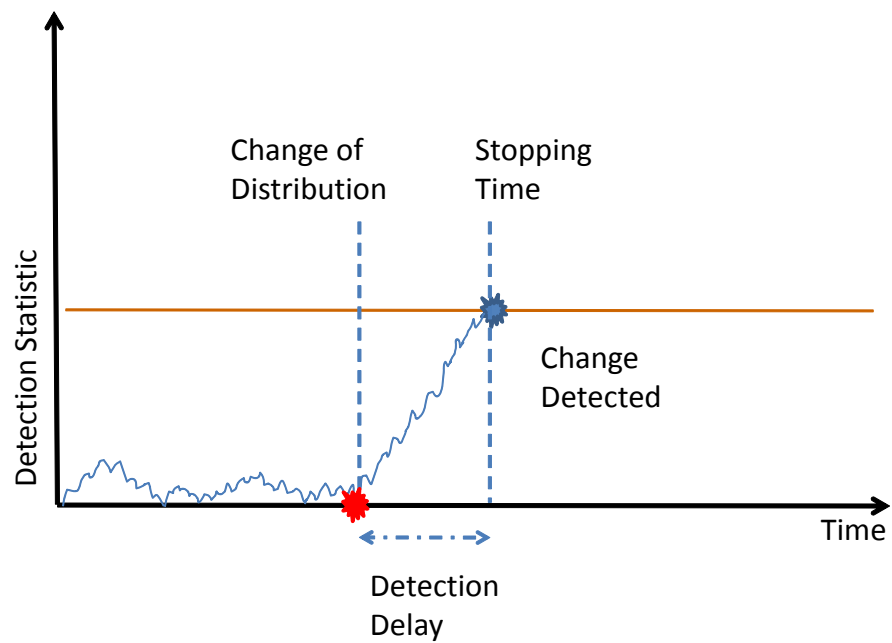}}
\caption{The Page's test is depicted here: initially the statistics often reset to zero while, after the change in distribution, it tends to increase up to eventually declare the change, when the threshold (red line) is crossed. The detection delay is the time required to declare the change, if this occurs.}
\label{fig:page_H1}
\end{figure}

We assume that a wireless sensor network is engaged in solving a change detection problem, 
which can be modeled as follows. Let $n\geq 1$ be a discrete time index, and let $j\in\{1,2,\dots,M\}$ denote the sensor of the network. The $n^{th}$ observation $x_{n,j}$ collected by the $j^{th}$ remote node obey a given distribution $f_0(x)$ until an unknown time $n_0$. From $n_0$ (included) on, the distribution at all sensors changes to $f_1(x)$.  The goal of the network is to detect the change as soon as possible, with a constraint on the average time between false alarms.
Formally, $\forall j$, we have
\beq \hspace*{-10pt}
\begin{array}{lcl}
f_0(x): x_{1,j},x_{2,j},\dots,x_{n_0-1,j}&&\\
&\searrow&\\
f_1(x): &&x_{n_0,j},x_{n_0+1,j},\dots
\end{array}
\label{eq:test0_CH5}
\eeq
All the involved random variables are assumed to be statistically independent.
According to this model, at each time slot~$n$, the network globally collects a vector of observations
\beq
\bx_n=[x_{n,1},x_{n,2},\dots,x_{n,M}].
\eeq 
This vector should be actually available to an ideal centralized entity, this latter would operate according to the well-known Page's test~\cite{Page}, which works as follows.
As detection statistic, it exploits the so-called CUSUM log-likelihood of the data
\beq
\sum_{i=1}^n l(\bx_i)=\sum_{i=1}^n \sum_{j=1}^M \log\frac{f_1(x_{i,j})}{f_0(x_{i,j})},
\label{eq:centrlog}
\eeq
where $l(\bx)$ denotes the log-likelihood pertaining to vector $\bx$.
The standard recursion for implementing Page's test is
\beq
L^{(c)}_n=\max\{ 0,L^{(c)}_{n-1}+l(\bx_n) \},
\label{eq:recursion}
\eeq
and the associated (random) stopping time is
\beq
N=\arg\min_n\{L^{(c)}_n\geq \gamma\}.
\label{eq:stop_CH5}
\eeq
This defines a decision rule: we say that a change happened when the first crossing of the threshold $\gamma$ takes place.
Note also that the recursion rule implies that the log-likelihood resets each time it falls below zero, which is thus the point from which it restarts. In Figures~\ref{fig:page_H0}-\ref{fig:page_H1} it is depicted schematically the behavior of the Page's test.

The usual optimality criterion in change detection is to minimize the delay needed for detecting the change, while diluting as much as possible the false alarms. 
Accordingly, the main performance metric will be the average delay for detection, corresponding to a prescribed average false alarm rate. As to the latter, it is usually defined as the reciprocal of the average sample size under the null hypothesis, that is, $1/\E_0[N]$, with $\E_{0,1}[\cdot]$ denoting expectation computed under distribution $f_{0,1}(x)$.

As for the delay to detection $\E_1[N]$, we as usual work with the upper bound that one obtains by assuming that the CUSUM is zero at time $n_0$~\cite{basseville-book}.
The exact computation, in fact, would require to account for the exact value of the CUSUM statistic at $n_0$, and is usually intractable. 

Approximate formulas for these relevant quantities are available in the literature, see e.g.,~\cite{Page,basseville-book,wald}. 
These mainly rely upon neglecting end effects, that is, the excess over the threshold of the test statistic at the stopping time. 
The relationship between the detection threshold $\gamma$ and the false alarm rate $R_c$ can be written as 
\beq
R_c(\gamma) \approx \frac {M \, \Delta_{01}}{e^\gamma -\gamma - 1}  \stackrel{\textnormal{ \footnotesize{large} }\gamma}{\approx} M \, \Delta_{01} \, e^{-\gamma}
\label{eq:centrthresh}
\eeq
where $\Delta_{01}$ is the Kullback-Leibler divergence between $f_0(x)$ and $f_1(x)$, see~\cite{CT}. 
Inverting the previous relationship defines the function $\gamma_c(R)$, which, for large thresholds, simplifies in
\beq
\gamma_c(R)\approx \log\left(M\,\Delta_{01}/R\right) .
\label{eq:gammac}
\eeq
The approximate relationship between the average delay $D_c$ and the threshold can be found as 
\beq
D_c(\gamma) \approx \frac {\gamma + e^{-\gamma} - 1}{M \, \Delta_{10}}  \stackrel{\textnormal{ \footnotesize{large} }\gamma}{\approx} \frac {\gamma}{M \, \Delta_{10}}
\label{eq:centrM}
\eeq
where $\Delta_{10}$ is the divergence between $f_1(x)$ and $f_0(x)$.
Note that in the previous equations we append a ``$c$'' for this \emph{centralized} detector.

It is seen that the delay essentially depends linearly from the threshold, while the false alarm has an exponential dependence thereof. Furthermore, in the large-$\gamma$ regime the system operating characteristic $D_c(R)$, i.e., the relationship between the detection delay and the false alarm rate, can be expressed in simple closed form:
\beq
D_c(R)\approx \frac{\gamma_c(R)}{M \,\Delta_{10}}
\approx \frac{\log \left ( M \,\Delta_{01} /R\right )}{M \,\Delta_{10}} .
\label{eq:operc}
\eeq

We note explicitly that the factor $M$ which multiplies the divergences takes into account the fact that, at each time slot, $M$ independent observations are collected, such that the overall divergence of the single-time observations is $M\,\Delta$.
For later use, let us report also the formulas pertaining to a single sensor, i.e., those of the idealized entity, specialized for $M=1$:
\beqa
R_s(\gamma) &\approx& \frac { \Delta_{01}}{e^\gamma -\gamma - 1}  \stackrel{\textnormal{ \footnotesize{large} }\gamma}{\approx} \Delta_{01} \, e^{-\gamma} ,
\label{eq:centrthresh1} \\
D_s(\gamma) &\approx& \frac {\gamma + e^{-\gamma} - 1}{ \Delta_{10}}  \stackrel{\textnormal{ \footnotesize{large} }\gamma}{\approx} \frac {\gamma}{\Delta_{10}} .
\label{eq:centr1}
\eeqa
For large $\gamma$, these yield
\beq
D_s(R)\approx \frac{\log \left ( \Delta_{01} /R\right )}{\Delta_{10}} .
\label{eq:opers}
\eeq

\section{Decentralized Page's Test by Running Consensus}
\label{sec:page_running}
Let us now switch to consider a genuinely decentralized scenario, and suppose, in particular, that a wireless sensor network engaged in a change detection has a fully flat architecture without fusion center. A running consensus protocol is implemented to reach agreement about the ideal centralized detection statistic, as the number of consensus steps grows.
The running consensus procedure is explained in the Chapter~\ref{ch:2} and will not be repeated here; in the following we only recall the basic elements needed to make this chapter self-consistent.

With reference to the running consensus strategy, we assume that sensors are able to ``simultaneously'' acquire, exchange and process data for eventually declaring the change. More in detail, during each time slot $n$, observations are first collected by the sensors, and then a randomly chosen subset thereof is selected for sharing their data according to a standard gossip algorithm~\cite{boyd-gossip-IT}. The exchanged data are not simply the measurements, but rather the current
detection statistic available to a sensor, also referred to as the state, say $L_{n,j}$, that summarizes its state of knowledge.

What one wants is, of course, that $L_{n,j}$ for all $j$ would converge to the the centralized detection statistic $S_n$, as the time index $n$ diverges. To this aim, the running consensus protocol prescribes using the following update, where $l(x)=\log {f_1(x)}/{f_0(x)}$ represents the log-likelihood pertaining to a single sample $x$: 
\beq 
\left ( \begin{array}{c} 
L_{n,1} \\ L_{n,2} \\ \vdots \\ L_{n,S} \\
\end{array} \right ) 
=\bW_n 
\left ( \begin{array}{c} 
L_{n-1,1} \\ L_{n-1,2} \\ \vdots \\ L_{n-1,S} \\
\end{array} \right ) 
+M\,\bW_n 
\left ( \begin{array}{c} 
l (x_{n,1}) \\ 
l (x_{n,2}) \\ \vdots 
\\ l (x_{n,M}) \\
\end{array} \right ) 
\label{eq:basic}
\eeq
and where the $M$ by $M$ matrices $\bW_n$, $n=1,2,\dots,$ are iid (independent identically distributed) and doubly stochastic. The equation~(\ref{eq:basic}) derives from~(\ref{eq:basic_CH2}) with $\alpha_n=1,\beta_n=M$ and $t(x)=l(x)$.
In Theorem~\ref{prop:3}, by a vary simple manipulation, it is possible to show that the squared error of the running consensus is bounded by a term\footnote{Note that in Proposition~\ref{prop:2} a similar bound is present in which the term $M^2$ is replaced with $M^3$.} 
\beq
\propto M^2 \, \E_{0,1}\left[l^2(x)\right]\,\frac{\lambda_U}{1-\lambda_U},
\label{eq:conserr}
\eeq
where the bound is independent of the time instant $n$. In the above expression $\lambda_U$ is the second largest eigenvalue of the matrix $\E[\bW_n \bW_n^T]$ and takes account of the connection properties of the network~\cite{running-cons}. In~(\ref{eq:conserr}) the error is bounded by a term that varies with the network topology and with $M$.


Let us compactly denote the update rule for the $j^{th}$ sensor by $L_{n,j}={\cal U}(L_{n-1,j})$.
By using update rule~(\ref{eq:basic}), together with recursion rule~(\ref{eq:recursion}), the overall recursion at the $j^{th}$ sensor can be written as follows 
\beq
L_{n,j}=\max\{0,{\cal U}(L_{n-1,j})\}.
\label{eq:recursion2}
\eeq
Note that, while the update rule ${\cal U}$ is linear, the addition of Page's reset rule introduce a nonlinear effect, which is not present in the classical gossip algorithms.

\begin{figure}
\centerline{\includegraphics[width = 0.95\textwidth,angle=0]{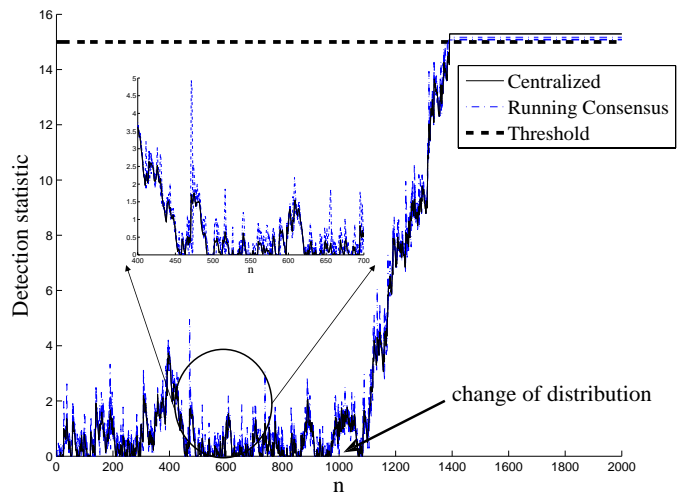}}
\caption{Behavior of the running consensus statistics for quickest detection, for a simple network made of three sensors (three dashed lines, almost superimposed each other). Initially, the statistics often reset to zero while, after the change in distribution, they tend to increase up to eventually declare the change, when the threshold is crossed. The times at which the threshold is crossed almost coincide for the three running consensus statistics, indeed, as seen in the zoomed plot, the three running consensus statistics always behave quite similarly. The continuous curve shows the statistic of an ideal centralized system. We see that the running consensus statistics closely track the statistic of the ideal system: their detection delays are close to that of the ideal system.}
\label{fig:scheme}
\end{figure}

A key observation should be made at this point. The nature itself of the running consensus scheme implies that all the nodes asymptotically share one and the same state (statistic). 
Thus, what one expect is that all the nodes report a detected change at approximately the same time instant. Running consensus introduces strong correlations among nodes by continuously propagating information across the system, see the metric behavior $\rho_{n,ij}$ for $n$ large in Theorem \ref{prop:3}. This correlation is responsible for detecting the change at almost equal times. Therefore, provided that the system evolves for a sufficiently long time, the estimated time at which the distribution change took place can be recovered by any of the $M$ nodes of the network, and there is no need to implement some ``broadcast'' form of halting in the system. Moreover, this also implies that the performance of the running consensus scheme can be computed with reference to \emph{any} sensor.

Since the running consensus protocol is designed to reach agreement about the ideal centralized statistic $L^{(c)}_n$, (see previous chapters), it is expected that the statistic $L_{n,j}$ (any $j$) tends to track $L^{(c)}_n$. Figure~\ref{fig:scheme} shows the behavior of these two statistics, and we indeed see that they behave similarly. One can hope that also the performances of the running consensus detector stay close to the theoretical limit represented by the performances of the centralized ideal system.
To elaborate on the issue, it is convenient to write the detection statistic available at the $j^{th}$ sensor as
\beq
L_{n,j}=L^{(c)}_n + e_{n,j},
\label{eq:Tsn_CH5}
\eeq
emphasizing the distance of the current state $L_{n,j}$ from its asymptotic value $L^{(c)}_n$, measured by the error term $e_{n,j}$. Note that this error is different from that defined in Chapter~\ref{ch:2} because of the non linearity in the update rule~(\ref{eq:recursion2}).
The corresponding stopping time at the $j^{th}$ sensor, which implicitly defines the decision rule, becomes
\beq
N_j=\arg\min_n\{ L_{n,j}\geq \gamma \}.
\label{eq:stop2}
\eeq

Approximate performance evaluation of the decentralized Page's test with running consensus is now in order, via the behavior of the error term $e_{n,j}$. To this aim, we shall exploit results from previous chapters, and try to apply them to the nonlinear update~(\ref{eq:recursion2}). We would like to stress that the arguments below are only heuristic, and their validity is checked by computer simulations in a later section.

That said, let us first work under the assumption that the absolute error is strictly bounded, that is, $|e_{n,j}|\leq\epsilon$, $\forall n$ and $\forall j$.
The system initially collects data distributed according to $f_0(x)$ and, until a threshold crossing occurs (either because a real change happened, or because a false alarm is going to occur), the $j^{th}$ sensor may have made a certain number of resets to zero, depending upon both $L^{(c)}_n$ and $e_{n,j}$. It is reasonable to assume that, for large values of $\gamma$ (i.e., $\gamma\gg\epsilon$), the threshold crossing will be typically determined by the behavior of the centralized statistic $L^{(c)}_n$, rather than by the error term.
Otherwise stated, when $L^{(c)}_n$ starts driving the random walk toward the upper threshold (again, either because a real change happened, or because a false alarm is going to occur), the error term that differentiates the states of the various sensors will become less and less influential compared to $L^{(c)}_n$, and the sensors will tend to agree toward declaring a change.

More formally, by noting that $L^{(c)}_n-\epsilon\leq L^{(c)}_n+e_{n,j}\leq L_n+\epsilon$, we can write (assuming $\gamma>\epsilon$):
\beq
\arg\min_n\{ L^{(c)}_n+\epsilon\geq \gamma \}
\leq
N_j
\leq
\arg\min_n\{ L^{(c)}_n-\epsilon\geq \gamma\}.
\label{eq:Nj}
\eeq
We can thus define
\[
\underline N=\arg\min_n\{ L^{(c)}_n\geq \gamma-\epsilon\},
\quad
\overline N=\arg\min_n\{ L^{(c)}_n\geq \gamma+\epsilon\},
\]
which are recognized to be the stopping times associated to classical centralized Page's tests having thresholds $\gamma-\epsilon$ and $\gamma+\epsilon$, respectively.
Thanks to eq.~(\ref{eq:Nj}), we can thus write 
\beq
\E_{0,1}[\underline{N}]\leq 
\E_{0,1}[N_j]\leq
\E_{0,1}[\overline{N}],\qquad \forall j=1,2,\dots M.
\label{eq:bounds}
\eeq
Thanks to the above, we can make an approximate performance evaluation of the running consensus test. 
As to the false alarm rate $R_j$ at sensor $j$, we use the lower bound $\E_0[\underline{N}]$ that yields
\[
R_j\leq R_c(\gamma-\epsilon)
\] 
with the function $R_c(\cdot)$ defined in~(\ref{eq:centrthresh}).
Now, let us fix $R$ and consider for the running consensus the threshold value $\gamma=\gamma_c(R)+\epsilon$, namely
the threshold of the centralized system plus $\epsilon$. In this way we get $R_j\leq R_c(\gamma-\epsilon)=R$, implying that such threshold setting provides a conservative choice ensuring that the prescribed false alarm $R$ is not exceeded. 

Switching to the analysis of the average delay at sensor~$j$, say $D_j$, using~(\ref{eq:bounds}), we have
\[
D_c(\gamma-\epsilon)\leq D_j \leq D_c(\gamma+\epsilon),
\] 
and using again $\gamma=\gamma_c(R)+\epsilon$, yields
\[
D_c(\gamma_c(R))\leq D_j \leq D_c(\gamma_c(R)+2\,\epsilon).
\] 
All these delays diverge when the threshold grows. 
However, since $D_c(\gamma)$ depends essentially linearly upon $\gamma$, we have that, in the regime of small $R$ (that is, large $\gamma_c(R)$), the ratio between $D_j$ and the average delay of a centralized test with false alarm rate $R$ approaches unity, for any $j$.
Accordingly, the overall performance of the running consensus test are well-described by the same operating characteristic of the centralized system:
\beq
D_r(R) \approx D_c(R) \approx \frac{\log \left ( M \,\Delta_{01} /R\right )}{M \,\Delta_{10}},
\label{eq:decentrM}
\eeq
where the subscript ``$r$'' stands for ``running'', versus
``$c$'' for ``centralized''.


The more important fact to be emphasized, however, is that the $\epsilon$ appearing in~(\ref{eq:Nj}) is unknown and, for this reason, a useful relationship between the threshold and the false alarm rate is actually not available. 
To avoid the impasse, one might simply neglect the $\epsilon$ and set the threshold of the running consensus detector equal to that of the centralized scheme, provided  by relationship~(\ref{eq:centrthresh}), or by approximation~(\ref{eq:gammac}); note that this does not ensure that the design is conservative.
A better option would be that of adjusting the threshold by adding a term related to~(\ref{eq:conserr}), in order to provide a more reliable protection against exceeding the false alarm constraint. This, however, requires a case-by-case analysis. 

At any rate, while the precise relationship to set the threshold is problematic, what remains true is that
any choice of the threshold leads to an operating point belonging to the optimal operating characteristic~(\ref{eq:decentrM}).

\section{Bank of Parallel Page's Detectors}
\label{sec:bank}
In this section we study the case that each sensor runs its own a Page's test, without exchanging data on-the-fly with other sensors.
Such a bank of parallel Page's processors evolves until the first of the $M$ sensors detects a change, and when this happens a broadcast message from the ``firing'' sensor is sent through the network in order to terminate the inference task. This allows energy saving, once that the change of the state of the nature has been observed.
On the other hand, it is just this latter broadcast which enforces some form of cooperation by exploiting the diversity among sensors' states, and might provide performance improvements with respect to a single Page's test.

The false alarm rate of the bank can be easily characterized. Let $L_j$ be the duration of the $j^{th}$ Page's test, i.e., the time at which the $j^{th}$ sensor would detect a change. The overall stopping time is defined as
\beq
N^{(bank)}=\min_{j\in\{1,\dots,M\}}{L_j}. \label{eq:Nbank}
\eeq
Assuming $f_0(\cdot)$, we resort to the following heuristic arguments. According to~\cite{basseville-book}, for $\gamma$ large enough, we can suppose that time interval between two successive false alarms is essentially ruled by the number of resets (a reset happens when the value of the CUSUM statistic is zero) times the length of a single path of the statistics evolving between two successive resets; the length of the path from the last reset to the threshold crossing may be neglected in this approximation (see also Figure~\ref{fig:page_H0}-\ref{fig:page_H1}).
Computing the time interval between false alarms as the random number of resets multiplied by the average length of a single inter-reset path, we are faced with characterizing such a random number. For a single Page's test the number of resets can be shown to follow a geometric distribution~\cite{basseville-book}; let $(1-p)p^k$ be the probability of $k$ resets, $p$ being the characteristic parameter of the geometric. Then, in the case of a bank of $M$ Page's tests the false alarm rate is ruled by the minimum of $M$ such random variables, which is a geometric variable itself, whose parameter changes to $p^M$. Accordingly, by replacing $p$ with $p^M$ in eq.~(5.2.22)	 of~\cite{basseville-book}, we finally have that the relationship between the threshold $\gamma$ of the bank and the false alarm rate $R_b$ is exactly the same already obtained with the ideal centralized system and with the running consensus scheme.
Roughly speaking, the false alarm of bank made of $M$ filters is $M$ times larger than that of a single filter. 
Therefore, we assume
\beq
R_b(\gamma) \approx R_c(\gamma) \approx \frac {M \, \Delta_{01}}{e^\gamma -\gamma - 1}  \stackrel{\textnormal{ \footnotesize{large} }\gamma}{\approx} M \, \Delta_{01} \, e^{-\gamma}
\label{eq:bankthresh}
\eeq

Switching our attention to the average delay, note that $D_b=\E_1[N^{(bank)}] $, 
and the approximate evaluation of this statistical expectation is now in order. 
Let $F_1(x)$ be the cumulative distribution function (CDF, hereafter) of $L_j$ when the observation model is $f_1(\cdot)$. Such a CDF can be approximated by that of the stopping time arising from a random walk with positive drift and a single barrier $\gamma>0$ (see also the discussion in~\cite{SP03-2}). This approximation involves the so-called
Wald or inverse Gaussian distribution, defined as 
\[
F_{W}(x;z)=\left[1-Q\left(\frac{x-1}{\sqrt{x}}\,\sqrt{z}\right)\right]
+e^{2\,z}\,Q\left(\frac{x+1}{\sqrt{x}}\,\sqrt{z}\right),
\]
$Q(\cdot)$ being the standard Gaussian exceedance probability function.
Indeed,~\cite{johnsonkotzbalakrishnan}
\[
F_1(\xi\,\E_1[L_j]) \approx F_{W}(\xi; \gamma \delta), \qquad \textnormal{with } \delta=\frac{\Delta_{10}} {\VAR_1\left[l(x)\right]} ,
\] 
where for $\E_1[L_j]$ we use~(\ref{eq:centr1}).
Therefore, from~(\ref{eq:Nbank}) and from the fact that the expectation of a nonnegative random variable can be computed as the integral of complementary CDF, we have
\beq
D_b(\gamma) \approx  \frac{\gamma + e^{-\gamma} - 1}{\Delta_{10}} \; \int_0^\infty \left [ 1- F_W \left (\xi;\gamma\delta \right )\right ]^M \, d\xi
\label{eq:integral}
\eeq
that can be easily used for numerical evaluation.

An explicit expression for operating characteristic of the bank can be found by approximating $\gamma + e^{-\gamma} - 1 \approx \gamma$ in (\ref{eq:integral}), and exploiting the last expression in~(\ref{eq:bankthresh}), yielding 
\beq
D_b(R) \approx \frac{\log \left ( M \, \Delta_{01}/R\right )}{ g(M,R) \; \Delta_{10}} 
\label{eq:operb}
\eeq
where the function $1/g(M,R)$ is defined as
\[
\int_0^\infty \left [ 1- F_W \left(\xi;\,\frac{\Delta_{10}}{\VAR_1\left[l(x)\right]} \; \log \left ( \frac{M\,\Delta_{01}}{R}\right) \right )\right]^M \, d \xi
\]

In this paper we mainly use the approximations of~(\ref{eq:integral}) and~(\ref{eq:operb}). Nonetheless, a brief digression on how a simpler formula can be also obtained for moderately large $M$, is now provided. 
Given $M$ iid random variables $L_j$, it holds true that
\[
\min_{j\in\{1,\dots,M\}}L_j \stackrel{d}{=}F_1^{-1}\left(\min_{j\in\{1,\dots,M\}}U_j\right),
\] 
where $U_j$ are iid $(0,1)$-uniform variates, and
$X\stackrel{d}{=}Y$ means that the two random variables $X$ and $Y$ have the same distribution. This implies
\[
\E_1\left[N^{(bank)}\right]=\E\left[F_1^{-1}\left(\min_{j\in\{1,\dots,M\}}U_j\right)\right].
\]
For large $M$, a legitimate approximation consists in exchanging the expectation operator with the function $F_1^{-1}(x)$ (see, e.g., Castillo~\cite{Castillobook}, p.\ 76), yielding
\beq
\E_1\left[N^{(bank)}\right]\approx
F_1^{-1}\left(\frac{1}{M+1}\right),
\label{eq:castillo}
\eeq
where we used the fact that the expected value of the minimum of $M$ uniform random variables is $1/(M+1)$.
Since, using Wald's distribution, $F_1(x)\approx F_W(x/\E_1[L_j];\gamma\delta)$, the above yields $F_1^{-1}(y)\approx \E_1[L_j]\,F_W^{-1}(y;\gamma\delta)$.
By substituting in~(\ref{eq:castillo}), the average delay can be expressed as 
\beq
D_b(\gamma)\approx\frac{\gamma}{\Delta_{10}}\,F_W^{-1}\left(\frac{1}{M+1};\gamma\delta\right),
\label{eq:N1bank}
\eeq
which, in addition to~(\ref{eq:integral}), is a further approximation of the delay, 
that could be useful in the regime of large $M$.

\section{Relative Efficiencies}
\label{sec:RE}
We use as a proxy to compare different detection schemes the so-called relative efficiency of strategy~$1$ with respect to strategy~$2$, for a given false alarm rate $R$. This is defined as
\[
\eta_{1,2}(R)=\frac{D_2(R)}{D_1(R)},
\]
$D_1(R)$ and $D_2(R)$ being the average delays of detectors~$1$ and~$2$, respectively, when their false alarm rate is fixed to the same value $R$. 
With this definition, $\eta_{1,2}(R)>1$ implies that strategy $1$ outperforms strategy $2$, in the sense that it exhibits a smaller delay, for the same false alarm rate, and vice versa for $\eta_{1,2}(R)<1$.
As to the notations, recall that we use the subscript ``$r$'' to denote the decentralized Page's test with running consensus, 
``$b$'' refers to the bank of parallel Page's processors, while ``$s$'' and ``$c$'' refer to the single Page's test and to the ideal centralized entity, respectively. The approximations derived in the previous sections allows us writing the following relationships.

The efficiency of the optimal centralized Page's detector with respect to its running consensus counterpart, thanks to~(\ref{eq:decentrM}), is simply:
\beq
\eta_{c,r}(R)=\frac{D_r(R)}{D_c(R)}\approx 1,
\label{eq:etacr}
\eeq
meaning that the two strategies are essentially equivalent. This result merits emphasis: within the limits of the approximation, the decentralized scheme is as efficient as the ideal centralized detector, in terms of detection quickness.

Note that, in our context of a flat network architecture, the best would be that each sensor had the same optimal detection performance of the ideal centralized system, and there is no doubt that any sensible implementation of such a system would require transferring a huge amount of data among the sensors. 
Therefore, the basic message of~(\ref{eq:etacr}) is that the same goal of mimic the ideal centralized system can be \emph{approximately} achieved, by means of the running consensus protocol based upon gossip algorithms that lead to much more parsimonious energy consumption for communications.

Recalling that the operating characteristic of the running consensus coincides approximatively with that of the centralized system, the efficiency of the single-sensor test with respect to the running consensus strategy can be computed by combining~(\ref{eq:centrthresh}) with~(\ref{eq:centrM}), and~(\ref{eq:centrthresh1}) with~(\ref{eq:centr1}).
When $\gamma$ is large enough, using the second approximation of~(\ref{eq:operc}) and~(\ref{eq:opers}), one gets
\beq
\eta_{s,r}(R)=\frac{D_r(R)}{D_s(R)}\approx\frac{1}{M}\,
\left( 1+\frac{\log M}{\log(\Delta_{01}/R) } \right).
\label{eq:etasr}
\eeq
For relatively small false alarm rate $R$, we see that using $M$ units essentially reduces the delay of a factor $M$, with respect to the use of a single detector.
We note explicitly that the gain in terms of detection performances is paid in the coin of the communication expense: the single sensor, by definition, requires no communication at all;
the running consensus, whose energy parsimony with respect to the centralized system has been emphasized above, requires a certain amount of communication among the sensors. Computing the exact communication cost requires specifying, among other things, the exact gossip procedure adopted.

The comparison between the two decentralized architectures, i.e., the bank of Page's tests and the running consensus, follows by combining~(\ref{eq:centrthresh}) with~(\ref{eq:centrM}), and~(\ref{eq:bankthresh}) with~(\ref{eq:integral}). Assuming $\gamma$ large, the relative efficiency can be derived from the second of~(\ref{eq:operc}) and~(\ref{eq:operb}) in the form:
\beq
\eta_{b,r}(R)=\frac{D_r(R)}{D_b(R)}\approx\frac{g(M,R)}{M}.
\label{eq:etabr}
\eeq

Note that the energy spent for communication by the bank is essentially negligible, amounting to a single broadcast at the end of the detection task. In this respect, it is almost equivalent to a single sensor.
Conversely, running consensus does involve sensor's communication, and thus one may expect that the bank is outperformed by the running consensus in terms of quickness.

It is also interesting to compare the behavior of the bank against the use of a single Page's test. 
From approximations (\ref{eq:opers}) and (\ref{eq:operb}) we have:
\beq
\eta_{b,s}(R)=\frac{D_s(R)}{D_b(R)}\approx\frac {g(M,R)} { \displaystyle{1+\frac{\log M}{\log(\Delta_{01}/R)}}} ,
\label{eq:etabs}
\eeq
while a more accurate expression would involve eqs.~(\ref{eq:centrthresh1}), (\ref{eq:centr1}),  (\ref{eq:bankthresh}), and~(\ref{eq:integral}).

\section{Examples and Numerical Experiments} 
\label{sec:appli}

The previous analysis is now corroborated by computer experiments, and to this aim we select a case study commonly used as benchmark in the context of model change detection.
Consider hence the detection problem formalized in~(\ref{eq:test0_CH5}) and suppose that two zero-mean Gaussian distributions, featuring different variances, are involved: 
\[
f_0(x)=\frac{1}{\sqrt{2\pi}}\,e^{-\displaystyle{\frac{\,\,\,x^2}{2}}},\qquad
f_1(x)=\frac{1}{\sqrt{2\pi\sigma^2}}\,e^{-\displaystyle{\frac{x^2}{2\sigma^2}}}.
\]

As to the specific running consensus protocol employed in the decentralized Page's test, we refer to a standard repeated pairwise averaging in which,
during each time slot $n$, the sensors have the chance of making $5$ successive pairwise exchanges; details can be found in, e.g.,~\cite{running-cons}. 

We report the results from $10^4$ Monte Carlo simulations for $M=10$ sensors, and $\sigma\approx 1.032$. This rather small value of $\sigma$ implies that the two hypotheses are quite close, namely, that the detection of the change in the statistical distribution is really a difficult task; we want to test our systems just in this challenging scenario.

\begin{figure}
\centerline{\includegraphics[width = 0.95\textwidth,angle=0]{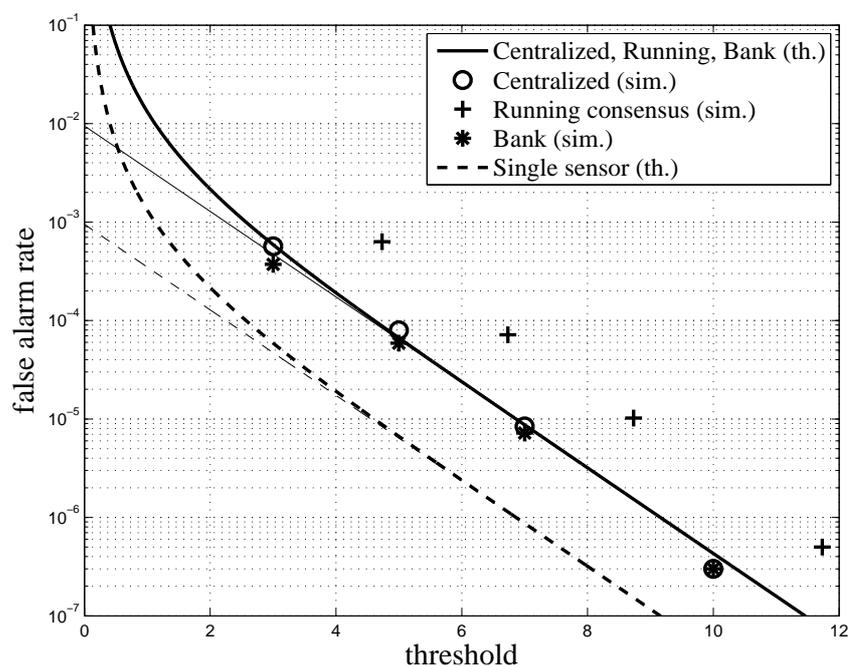}}
\caption{False alarm rate $R$ versus the threshold $\gamma$, for the four systems considered in the paper, and with  
specific reference to the Gaussian change in variance example, with the parameters detailed in the main text. The more accurate approximations are given as bold lines, while the simplifications obtainable with large $\gamma$ are depicted with the same linestyle, but with thinner lines.}
\label{fig:sim2}
\end{figure}

Figure~\ref{fig:sim2} reports the comparison between the simulated false alarm rate, and that obtained by exploiting the approximate analytical relationships between the threshold $\gamma$ and the false alarm $R$. The theoretical relationships for the ideal centralized, the running consensus and the bank coincide, as seen by~(\ref{eq:centrthresh}), (\ref{eq:bankthresh}), and by the discussion on the threshold setting for the running consensus detector,  while the curve of the single sensor is scaled by a factor $1/M$. In Figure~\ref{fig:sim2} the lines in bold represent the more precise approximations contained in the quoted equations, while the simplifications for large $\gamma$ are reported as thinner curves.

The simulation points of the studied systems (no simulation results are given for the case of a single Page's test) show a satisfying accuracy, except for the case of the running consensus.
Indeed, we have already pointed out that using the threshold of the centralized scheme is not very accurate, and at least a corrective term should be added. This could be certainly estimated by simulations, but a case-by-case analysis is required to address change detection problems with different distributions.

\begin{figure}
\centerline{\includegraphics[width = 0.95\textwidth,angle=0]{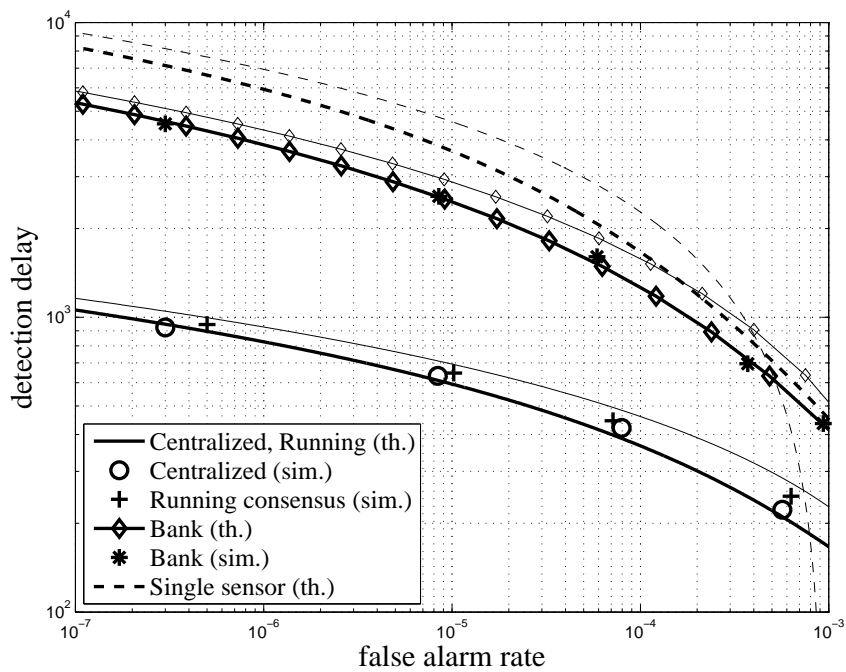}}
\caption{operating characteristics of the four systems considered in the work. The detection delay $D$ is depicted as a function of the false alarm rate $R$, for the same scenario of Figure~\ref{fig:sim2}. As for this latter, thinner lines represent the rougher approximations valid for large $\gamma$.}
\label{fig:sim1}
\end{figure}

In absence of such correction, the practical impact is that one is not able to set accurately the threshold, to guarantee a precise $R$. On the other hand, once a threshold has been selected, the corresponding pair of detection delay and false alarm rate stays quite close to the optimal curve. This analysis is provided in Figure~\ref{fig:sim1}, where we run the different tests over a range of detection thresholds, and report in the plot the values of the false alarm rate and of the detection delay, estimated via simulations. 

The curves pertaining to the theoretical formulas in Figure~\ref{fig:sim1} have been drawn in the following way.
The bold solid curve results by combining the first approximation of~(\ref{eq:centrthresh}) with that in~(\ref{eq:centrM}), while the thinner solid line is obtained when the corresponding expressions for large $\gamma$ are used. 
As to the bank detector, the bold curve with diamond markers follows by combining~(\ref{eq:integral}) with the first (more accurate) approximation in~(\ref{eq:bankthresh}); the rougher~(\ref{eq:operb}) is depicted as the thinner curve with diamond markers. 
Finally, as to the single sensor case, the bold dashed curve follows from the first expressions in~(\ref{eq:centrthresh1}) and in~(\ref{eq:centr1}), while the thinner dashed curve is depicts~(\ref{eq:opers}).

Comfortably, from Figure~\ref{fig:sim1} we see that all the simulation points stay very close to the respective theoretical prediction. Figure~\ref{fig:sim1} also shows that the small discrepancies between theory and simulation tend to vanish as $R$ decreases, as the arguments exploited in deriving the approximate formulas anticipated.
In addition, we note that the operating characteristic of the bank approaches that of a single sensor, 
when the false alarm rate increases: with large $R$, using $M$ processors organized in a bank architecture provides negligible improvements with respect to the simple single test.
On the other hand, for any value of false alarm rate, it is worth noting how the running consensus provides substantial improvements with respect to the bank, in terms of detection performances. We have already pointed out that the situation is reversed in terms of communication burdens: the bank might represent a viable choice whenever the usage of the running scheme is hampered by the energy costs.

\begin{figure}
\centerline{\includegraphics[width = 0.95\textwidth,angle=0]{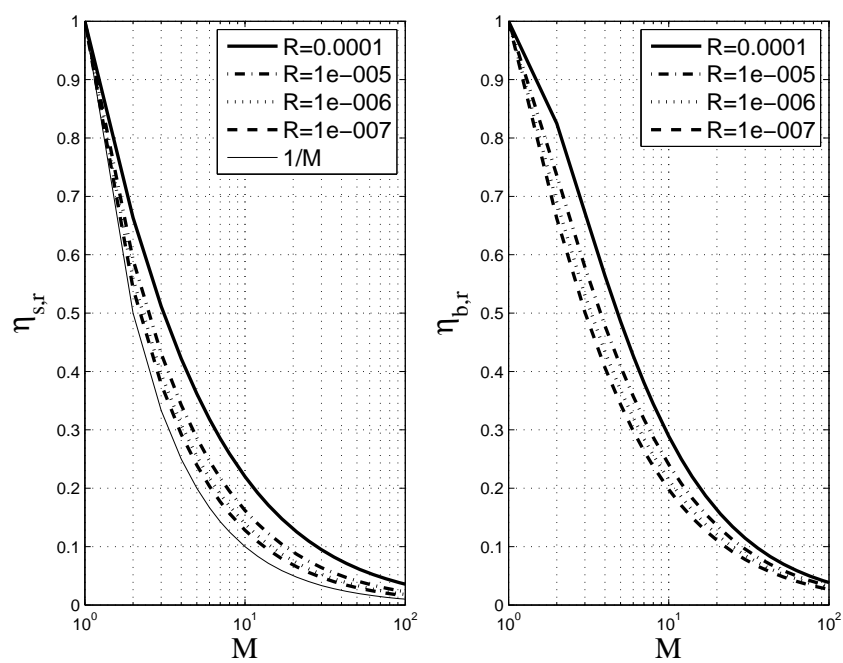}}
\caption{Relative efficiencies $\eta_{sr}$ and $\eta_{br}$ versus the number of sensors, for several values of the false alarm rate $R$.}
\label{fig:RE1}
\end{figure}

\begin{figure}
\centerline{\includegraphics[width = 0.95\textwidth,angle=0]{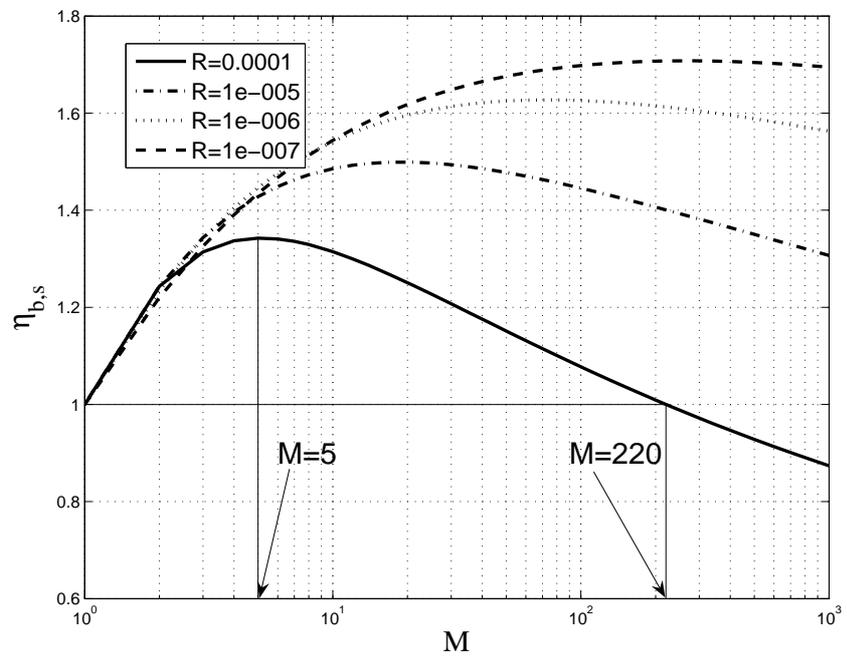}}
\caption{Relative efficiency $\eta_{bs}$ versus the number of sensors, for several values of the false alarm rate $R$.}
\label{fig:RE2}
\end{figure}

In Figures~\ref{fig:RE1} and~\ref{fig:RE2}, we display the theoretical relative efficiencies $\eta_{s,r}(R)$, $\eta_{b,r}(R)$ and $\eta_{b,s}(R)$, as function of the number of sensors $M$, for several values of the false alarm rate $R$. 
We see that $\eta_{s,r}(R)$ is always less than unity as we expected, and the curves approach that of $1/M$, also reported for comparison, for small false alarm rates. As a matter of fact, this behavior can be easily predicted by inspecting directly eq.~(\ref{eq:etasr}).

From $\eta_{b,r}(R)$ we observe that the running scheme always outperforms the bank of filters, as the arguments below~(\ref{eq:etabr}) suggested. It is also worth noting how $\eta_{b,r}(R)$  approaches zero for any choice of $R$, thus revealing how the fusion of the sensors' data during the detection test significantly improves the performances. 

The behavior of $\eta_{b,s}(R)$ in Figure~\ref{fig:RE2} reveals that the bank of Page's processor exhibits an optimum value of $M$ at which the detection performances are maximized, when compared with those of a single sensor. There also exists a maximum value of $M$ beyond which $\eta_{b,s}(R)$ falls below unity: interestingly, for moderately large $M$
there is no benefit at all in using a bank of Page's detectors with respect to a single unit, in terms of system performances. 
As an example, with a false alarm rate of $10^{-4}$, it turns out that a bank made of only $5$ filters provides the best performance improvements with respect to a single filter. 
The relative efficiency decreases for $M>5$ and at about $M=220$ falls below unity, meaning that a bank of more than $220$ filters performs worse than a single detection unit.

However, this limiting value of $M$ seems to increase significantly with decreasing false alarm rate, falling in a region where the number of sensors might be prohibitively large for practical purposes. We see that, at false alarm rates below $10^{-5}$ it is always convenient to use the bank, and the optimal $M$ that maximizes the performances takes values in the hundreds. 

Note, finally, that the curves in Figures~\ref{fig:RE1} and~\ref{fig:RE2} are obtained with the more accurate expressions derived in the previous sections. Actually, using the simpler
relationships~(\ref{eq:etasr}) and (\ref{eq:etabr}) in Figure~\ref{fig:RE1}, would change nothing since the curves would be practically indistinguishable
from those given in the figure. As to $\eta_{b,s}$ in Figures~\ref{fig:RE2}, eq.~(\ref{eq:etabs}) does give the precise behavior of the efficiency but, to address the quantitative analysis of the limiting values of $M$, the more accurate formulas are certainly preferable.

\chapter{Conclusions}
The number of scientific works concerning sensor network problems increased in the last few decades. Many applications with new challenges were born and a general definition of sensor network seems impossible\footnote{In~\cite{sadler-tutorial} Sadler affirms: ``What is a sensor network? We postulate that, given any definition of a sensor network, there exists a counter
example''.}, it seems that a common design goal does exist: To organize a multitude of simple, tiny,
cheap, (singly) unreliable, and low-capability sensors under a global architecture, in
such a way that --through cooperation-- the aggregate system would behave better than the
sum of its constituent.

A key aspect of such systems is the global reliability. Despite of the fact that single
sensors can be, and usually are, quite unreliable, the global system is required to be
very robust to  faults. The network must work properly even in the case that a large part
of the system is impaired or destroyed. Leading to one extreme this concept, we would
like that all the relevant information collected by the network (the sufficient
statistic, in the jargon of inference-oriented sensor networks that we consider in the
thesis) would be stored at any
single sensor.

To make a concrete example, in a hypothesis testing problem we would like that all the
sensors would store the same decision that would be made if all the observations
collected by the sensors were available, without any distortion, at a single site. This
would ensure that the ``global'' decision can be recovered by any single sensor of the
system.

In this work we consider networks with a completely flat architecture, in which the nodes
cooperate with each other without any hierarchical structure. As a consequence the
network can be seen as a single robust entity, where a node failure does not compromise
the
system functionality. The so-called consensus algorithms, that are key to our development
here, were indeed introduced in the scientific community exactly with these goals in mind.

The idea of a distributed algorithm that organizes some agents
to reach the consensus originally arises from the social-economic
community, where the study was focused on the human behavior.
Consensus strategies are applied in many engineering fields~\cite{Olfati-Saber}:
\begin{itemize}
	\item computer science;
	\item systems and control theory;
	\item distributed signal processing;
\end{itemize}
in particular there are subjects such as collective behavior of flocks and swarms, sensor fusion, random networks, synchronization
of coupled oscillators, algebraic connectivity of complex networks, asynchronous distributed algorithms, formation control for multi-robot systems, optimization-based cooperative control, dynamic graphs, complexity of coordinated tasks, and consensus-based belief propagation in Bayesian networks.

Within the context briefly described above, thesis presents a new paradigm for sensor networks that we call running consensus. It can be thought as a sort of time-varying consensus, in which the remote nodes sense the environment and, at the same time, implement agreement algorithms to
corroborate measurements made at different sites. The state of each node acts as a detection (or estimation) statistic to infer the
unknown state of the nature.

Typical inference problem, such as
estimation, detection and change point detection are considered. When compared with an
ideal, centrally computed, detection statistic, the local states are affected by an error
term
that, as we prove here, converges to zero. This motivates the
investigation of whether the running consensus attains the same
performances of such idealized entity, or not. To answer properly
a precise asymptotic framework is needed and we develop that in
due detail and rigor.
Mathematical properties are proved showing the asymptotic
equivalence of the running consensus with the idealized system,
in terms of detection/estimation capabilities. For what concerns
the change detection problem, an asymptotic equivalence is only
argued and the formal mathematical study left for future work.

When the running consensus is applied to parameter estimation,
the performances are computed in terms of mean square error, and
the asymptotic optimality is proved by the bounds on the running
consensus statistic. In the fixed-sample size (FSS) detection case, the theory based
on Pitman Asymptotic Relative Efficiency (ARE) and on the asymptotic normality of the
detection
statistics is used. On the other hand, in the sequential case more
advanced tools, concerning the asymptotic behavior of sequential
decision rules, are used. The convergence toward Wiener processes
is required. These theoretical results are the main findings of the
thesis, and provide additional practical significance to the running
consensus paradigm. In simple terms they emphasize how the decentralized
algorithms based on running consensus reach the same
performances of the centralized schemes, as long as the system
evolves for a sufficiently large amount of time.

A distinct feature of the running consensus scheme is the speed
of convergence of the algorithm. In fact, this is substantially different
from the exponential law that governs the classic consensus
algorithms~\cite{boyd-gossip-IT}. Furthermore, in our setup, the specific network
topology/connectivity (which rules the system eigenvalues) is less
crucial with respect to the classical consensus. The scaling law of running consensus is
$n^{-1}$, and only the value of the rate coefficient can be tuned by the eigenvalues.
This phenomenon seems in fact related to the rate of data acquisition, which is just
$n^{-1}$

Summarizing, this thesis proposes a new paradigm of consensus, called running consensus,
that is particularly suited to applications where the system operates in dangerous
environments. The basic idea is to combine the two stages of sensing and of data fusion
into a single step. This overcomes design troubles related to the necessity of extending
as much as possible the sensing stages, to collect more measurements.

The introduction of the running consensus with the above characteristics represents the
main contribution of this work, from a conceptual viewpoint. Technically, the main
contribution is in providing a rigorous assessment of the asymptotic convergences of the
running consensus, for typical inference applications: binary detection and nonrandom
parameter estimation.

\appendix
\chapter{}
\section{Proof of Proposition \ref{prop:2}}
\label{sec:lemma}
Recall that $\mu$ and $\sigma^2$ are the first two moments of $t(x)$, and 
\beqa
\xi^3&=&\E\left[\left\|\bt(\bx)- \mu\bone\right\|^3\right]\\
\be_n&=& M\ \sum_{i=1}^n 
\widetilde{\bPhi}_{n,i}\  \bt(\bx_i) \dfz \sum_{i=1}^{n} \by_{i}(n)
\label{eq:yerr}
\eeqa
where $\widetilde{\bW}_n = \bW_n - \bone\bone^T/M$, $\forall\,n\geq 1$ and 
$\widetilde{\bPhi}_{n,i} =\widetilde{\bW}_n\widetilde{\bW}_{n-1}\dots \widetilde{\bW}_i$. 
Let $\lambda_U$ be the the maximum eigenvalue of the matrix $\E[{\bf\widetilde W}^T {\bf\widetilde W}]$, and assume that $\lambda_U<1$.
From the Proposition~\ref{prop:1} the error term has zero mean, \emph{i.e.}
$\E[e_{n,s}]=0$ for $s=1,\dots,M$. 

In~(\ref{eq:yerr}) we defined
\beqa
\by_{i}(n) &=&\widetilde{\bPhi}_{n,i}\  M\ \bt(\bx_i)\nonumber\\
&=&M\,\widetilde{\bPhi}_{n,i}  [\bt(\bx_i) - \mu\bone]\dfz
M\,\widetilde{\bPhi}_{n,i}\  \widetilde{\bt}(\bx_i), \nonumber
\eeqa
where, for ease of notation, the dependence of $\widetilde{\bt}(\bx)$ upon $\mu$ has been skipped.
We now prove some properties of $\by_i(n)$, making use of the relationship~\cite{boyd-gossip-IT} 
\beq
\E\left[\left\|\widetilde{\bPhi}_{n,i}\bz\right\| ^2 \right] \leq 
\lambda_U^{(n-i+1)} \|\bz\|^2.
\label{eq:seminalBoyd}
\eeq
for any vector $\bz$.
Thanks to the independence between the random connection matrices and the observations $\bx_i$, the above yields
\beq
\E\left[\left\|\by_{i}(n)\right\| ^2 \right] \leq 
M^2\,\lambda_U^{(n-i+1)} \E \left[\left\|\widetilde{\bt}(\bx)\right\|^2\right].
\label{eq:boundlemma1}
\eeq
Since $\|\by_i(n)\|^2\geq y^2_{i,s}(n)$, where $y_{i,s}(n)$ is the $s^{th}$ entry of vector $\by_i(n)$, we further have 
\beq
\E\left[\,\left|y_{i,s}(n)\right|^2\,\right] \leq
M^3\,\lambda_U^{(n-i+1)} \sigma^2,
\label{eq:boundlemma3}
\eeq
where we explicited the definition of $\E[\|\widetilde{\bt}(\bx)\|^2]$.

Now, the maximum eigenvalue of the product $\bPhi_{n,i}^T \bPhi_{n,i}$ is equal to unity, $\bPhi_{n,i}$ being doubly stochastic by construction. This implies (see, {\em e.g.},~\cite{running-cons}) that the maximum eigenvalue of the product matrix $\widetilde{\bPhi}_{n,i}^T \widetilde{\bPhi}_{n,i}$ is less than or equal to unity. 
Hermitianity of this latter matrix yields, for all vectors $\bz$:
\beq
\bz^T\widetilde{\bPhi}_{n,i}^T\widetilde{\bPhi}_{n,i}\bz=
\left\|\widetilde{\bPhi}_{n,i}\  \bz\right\|^2\leq \left\|\bz\right\|^2.
\label{eq:z}
\eeq
From the above inequality, and using the definition of $\by_i(n)$, we can write
\beqa
\E\left[\left\|\by_{i}(n)\right\|^3 \right]&=&
M^3\,\E\left[\|\widetilde{\bPhi}_{n,i}\widetilde{\bt}(\bx_i)\|^3\right]\nonumber\\
&\leq&
\E\left[\|\widetilde{\bPhi}_{n,i}\widetilde{\bt}(\bx_i)\|^2
\,\|\widetilde{\bt}(\bx_i)\|\right]. \nonumber
\eeqa
Using the bound in eq.~(\ref{eq:seminalBoyd}), we have
\[
\E\left[|y_{i,s}(n)|^3 \right]
\leq
M^3\,\lambda_U^{n-i+1}\E\left[\|\widetilde{\bt}(\bx)\|^3\right]
=
M^3\,\lambda_U^{n-i+1}\xi^3,
\]
where we further referred to the vector entry $y_{i,s}$.

Before proving eqs.~(\ref{eq:Varerr}) and~(\ref{eq:thirderr}), we need the further properties
\beqa
&&\E\left[y_{j,s}(n)y_{k,s}(n)\right]=0,
\label{eq:Ezero1} \\
&&\E\left[\,\left|y_{i,s}(n)\right|\,\times y_{j,s}(n)\,y_{k,s}(n)\right]=0,
\label{eq:Ezero}
\eeqa
for all $j\neq k$.
We address eq.~(\ref{eq:Ezero}), since eq.~(\ref{eq:Ezero1}) can be proved similarly. 
Without loss of generality, suppose $k\neq i$.
Conditioning on the connection matrices, we have
\beqa
\lefteqn{
\E\left[
\,\left|y_{i,s}(n)\right|\,\times y_{j,s}(n)\,y_{k,s}(n)
\left|
\widetilde{\bW}_1,\dots,\widetilde{\bW}_n\right.
\right]}\nonumber\\
&=&
\E\left[
\,\left|y_{i,s}(n)\right|\times y_{j,s}
\left|
\widetilde{\bW}_1,\dots,\widetilde{\bW}_n\right.
\right]\nonumber\\
&\times&
M\,\widetilde{\bPhi}_{n,k}\E\left[\widetilde{\bt}(\bx_k)\right]=0, \nonumber
\eeqa
where the matrix $\widetilde{\bPhi}_{n,k}$ should be intended as the one corresponding to the conditioning matrices $\widetilde{\bW}_1,\dots,\widetilde{\bW}_n$.
Removing the conditioning clearly proves eq.~(\ref{eq:Ezero}).

We are now ready to prove eqs.~(\ref{eq:Varerr}) and~(\ref{eq:thirderr}).
Indeed, from eq.~(\ref{eq:yerr}), we know that $e_{n,s}=\sum_{i=1}^n y_{i,s}(n)$, yielding
\beqa
&&\E\left[e_{n,j}^2\right]=
\sum_{i=1}^{n}\E\left[y_{i,j}^2(n) \right]
\nonumber\\
&&\leq M^3\,\sum_{i=1}^{n} \lambda_U^{n-i+1}\, \sigma^2
\leq
M^3\,\frac{\lambda_U}{1-\lambda_U}\,\sigma^2, \nonumber
\eeqa
where we neglected the cross-products thanks to~(\ref{eq:Ezero1}), and used the bound~(\ref{eq:boundlemma3}). Therefore, eq.~(\ref{eq:Varerr}) follows.

Let us switch to the third-order bound~(\ref{eq:thirderr}). It is expedient to write
\[
|e_{n,s}|^3
=
\left(
\sum_{i=1}^n y_{i,s}(n)
\right)^2
\; 
\left|
\sum_{i=1}^n y_{i,s}(n)
\right|.
\]
Application of the triangle inequality to the last factor yields
\beqa
|e_{n,s}|^3
&\leq&
\left(
\sum_{i=1}^n y_{i,s}(n)
\right)^2
\sum_{i=1}^n |y_{i,s}(n)|
\nonumber\\
&=&\sum_{i,j,k}|y_{i,s}(n)|\,y_{j,s}(n)\,y_{k,s}(n).
\eeqa
Taking expectation yields
\beq
\E[|e_{n,s}|^3]
\leq
\sum_{i=1}^n \E[|y_{i,s}(n)|^3]
+
\sum_{i=1}^n\sum_{j\neq i} \E[|y_{i,s}(n)|\,y^2_{j,s}(n)],
\label{eq:en3}
\eeq
where the cross-terms corresponding to $j\neq k$ have been deleted by virtue of eq.~(\ref{eq:Ezero}).
Suppose now $i<j$. We have
\beqa
|y_{i,s}(n)|&\leq&\|\by_i(n)\|=M\,\|\widetilde{\bPhi}_{n,j}\,\widetilde{\bPhi}_{j-1,i}\,\widetilde{\bt}(\bx_i)\|\nonumber\\
&\leq& M\,\|\widetilde{\bPhi}_{j-1,i}\,\widetilde{\bt}(\bx_i)\|=\|\by_i(j-1)\| \nonumber
\eeqa
in view of eq.~(\ref{eq:z}).
This allows writing
\[
\left\{
\begin{array}{ll}
|y_{i,s}(n)|\,y^2_{j,s}(n)\leq
\|\by_i(j-1)\|\,y_{j,s}^2(n),& \textnormal{for $i<j$}
,\\
\\
|y_{i,s}(n)|\,y^2_{j,s}(n)\leq
|y_{i,s}(n)|\,\|\by_j(i-1)\|^2,&\textnormal{for $i>j$}.
\end{array}
\right.
\]
Taking expectations yields
\beq
\left\{
\begin{array}{ll}
\E[|y_{i,s}(n)|\,y^2_{j,s}(n)]\leq
\E[\|\by_i(j-1)\|]\,\E[y_{j,s}^2(n)], i<j
,\\
\\
\E[|y_{i,s}(n)|\,y^2_{j,s}(n)]\leq
\E[|y_{i,s}(n)|]\,\E[\|\by_j(i-1)\|^2], i>j,
\end{array}
\right.
\label{eq:systemofbounds}
\eeq
where we used the fact that $\by_i(j-1)$ and $\by_j(n)$ involve different observations as well as different connection matrices, whence they are statistically independent. The same holds for $\by_i(n)$ and $\by_j(i-1)$. 

The bounds in eqs.~(\ref{eq:boundlemma1}) and~(\ref{eq:boundlemma3}), by Jensen's inequality (being  $\sqrt{x}$ a convex-$\cap$ function), give
\beq
\E\left[\,\left|y_{i,s}(n)\right|\,\right] \leq
\E\left[\,\left\|\by_{i}(n)\right\|\,\right] \leq 
M^{\frac 3 2}\,\lambda_U^{\frac 1 2 (n-i+1)} \sigma.
\label{eq:boundlemma2}
\eeq
Putting now eqs.~(\ref{eq:boundlemma1}),~(\ref{eq:boundlemma3}) and~(\ref{eq:boundlemma2}) into
eqs.~(\ref{eq:systemofbounds}) gives
\[
\left\{
\begin{array}{ll}
\E[|y_{i,s}(n)|\,y^2_{j,s}(n)]\leq
M^{\frac 9 2}\,\sigma^3
\left[
\lambda_U^{\frac{j-i}{2}}\,\lambda_U^{n-j+1}
\right]
& i<j
,\\
\\
\E[|y_{i,s}(n)|\,y^2_{j,s}(n)]\leq
M^{\frac 9 2}\,\sigma^3
\left[
\lambda_U^{\frac{n-i+1}{2}} \,\lambda_U^{i-j}
\right]
& i>j.
\end{array}
\right.
\]
Using these inequalities into eq.~(\ref{eq:en3}) finally yields
\beqa
\lefteqn{\sum_{i=1}^n\sum_{j\neq i} \E[|y_{i,s}(n)|\,y^2_{j,s}(n)]\leq M^{\frac 9 2}\,\sigma^3}
\nonumber\\
&\times&
\left(\lambda_U\,\sum_{i=1}^n \lambda_U^{\frac{n-i}{2}}
\sum_{j=i+1}^n \lambda_U^{\frac{n-j}{2}}
+
\sum_{i=1}^n \lambda_U^{\frac{n-i}{2}}
\sum_{j=1}^{i-1} \lambda_U^{n-j}\right)
\nonumber\\
&\leq&
M^{\frac 9 2}\,\sigma^3
\left(
\frac{\lambda_U}{(1-\sqrt{\lambda_U})^2}
+
\frac{1}{(1-\lambda_U)(1-\sqrt{\lambda_U})}
\right),
\eeqa
where the summations have been bounded with the pertinent geometric series, although
they could be evaluated explicitly. The first term at RHS of~(\ref{eq:en3}) can be easily bounded similarly, and~(\ref{eq:thirderr}) follows.

\section{Proof of Theorem \ref{prop:3}}
The covariance matrix can be expressed as 
\beqa
\bC_n
&=&E\left[\left(\be_n+\frac{\bone \bone^T}{n} \bs_n\right)
\left(\be^T_n+\bs^T_n\frac{\bone \bone^T}{n}\right)
\right]\nonumber\\
&=&E[\be_n\be^T_n] + \sigma^2_{ctr}(t) \, \bone\bone^T, \label{eq:C}
\eeqa
where the last equality follows from straightforward algebra.
Now we need to compute the term $E[\be_n\be^T_n]$.
According to eq.~(\ref{eq:yerr_2}) and recalling that the matrices $\widetilde \bW_n$'s are iid
and independent of the sensors' observations, we have
\beqa
\lefteqn{E[\be_n\be^T_n]=\frac{1}{n^2}\sum_{h=1}^n E[\widetilde\bPhi_{n,h} 
E[\widetilde \bx_h\widetilde \bx^T_h]\widetilde\bPhi^T_{n,h}]} \nonumber \\
&=&\frac{\sigma^2}{n^2}\sum_{h=1}^n 
E[\widetilde\bPhi_{n,h}\widetilde\bPhi^T_{n,h}]
\dfz
\frac{\sigma^2}{n^2}
\sum_{h=1}^n 
\bar{\bPhi}_{n-h},
\label{eq:coverr}
\eeqa
having exploited $E[\widetilde \bx_h\widetilde \bx^T_h]=(\bI-\bone\bone^T/M ) \, \sigma^2$. 

To elaborate, we need the following results.
First, recalling the definition of $\widetilde\bPhi_{n,h}$, we have
\beqa
\lefteqn{\bar{\bPhi}_{n-h}= 
E[\widetilde \bW_n \dots \widetilde \bW_{h+1} \widetilde \bW_{h} \widetilde \bW^T_{h}} \nonumber \\ 
&& 
\qquad \widetilde \bW^T_{h+1}\dots \widetilde \bW^T_n] \nonumber \\
&\stackrel{(i)}{=} &
E \left [ \widetilde \bW_1 \dots \widetilde \bW_{n-h} \widetilde \bW_{n-h+1} \widetilde \bW^T_{n-h+1} \right .\nonumber \\ 
&& \left . \widetilde W^T(t-h)\dots \widetilde W^T(1) \right ]\nonumber \\
&\stackrel{(ii)}{=} &
E \left [ \widetilde \bW_1 \dots \widetilde \bW_{n-h} \; E[\widetilde \bW_{n-h+1} \widetilde \bW^T_{n-h+1}] \right .\nonumber \\ 
&& \left . \widetilde W^T(n-h)\dots \widetilde \bW^T_1 \right ]\nonumber \\
&\stackrel{(iii)}{=} &
E \left [ \widetilde \bW_1 \dots \widetilde \bW_{n-h} \; E[\widetilde \bW \widetilde \bW^T] 
\right .\nonumber \\ 
&& \left . \widetilde \bW^T_{n-h}\dots \widetilde \bW^T_1 \right ] \label{eq:Phi}
\eeqa
In the above: $(i)$ follows from the independence of the matrices $\bW_n$'s: for any selection of $n-h$ 
different time indices, the statistical expectation is the same; 
$(ii)$ is a straightforward application of the iteration property of conditional expectation,
that also exploits again the independence of the $\widetilde \bW_n$'s; in $(iii)$ we only simplify the notation
since the inner expectation is independent upon the time index.

Furthermore, let us consider an arbitrary (deterministic) vector, say $\bz_0$, orthogonal to~$\bone$, and define
$\bz_1\dfz \widetilde \bW^T_1 \bz_0$, $\bz_2 \dfz \widetilde \bW^T_2 \bz_1$, $\dots$,
$\bz_{n-h+1} \dfz\widetilde \bW^T_{n-h+1} \bz_{n-h}$. All these vectors still remain orthogonal to $\bone$, and we have the following properties: for $k=1,\dots,n-h+1$, 
\beq
E \left [||\bz_k||^2 \right ]= E \left [ \bz^T_{k-1} E[\widetilde \bW\widetilde \bW^T] \bz_{k-1}  \right ], \label{eq:norm}
\eeq
and, for $k=1,\dots,n-h+1$, 
\beq
\lambda_L \; E[||\bz_k||^2] \le
E \left [\bz^T_k \; E[\widetilde \bW\widetilde \bW^T]\; \bz_k \right ] \le
\lambda_U \; E[||\bz_k||^2] .
\label{eq:bounds_2}
\eeq
Equation~(\ref{eq:norm}) is obvious. 
As to eq.~(\ref{eq:bounds_2}), note first that the matrix $E[\widetilde \bW\widetilde \bW^T]$ 
has all its eigenvalues equal to those of $E[\bW\bW^T]$, 
but for the eigenvalue $1$, which is replaced by $0$. 
Thus, the maximum eigenvalue of $E[\widetilde \bW\widetilde \bW^T]$ is $\lambda_U$, whence the upper bound in eq.~(\ref{eq:bounds_2}). 
Then, since $E[\widetilde \bW\widetilde \bW^T]$ has a zero eigenvalue with eigenvector $\bone$ and recalling that $\bz_k \bot \bone$, we can lower bound the quadratic form with the successive eigenvalue of 
$$E[\widetilde \bW\widetilde \bW^T]$$ (which is not necessarily nonzero), namely with $\lambda_L$, see \emph{e.g.,}~\cite[Th.\ 4.2.2]{Johnson-Horn}.

Armed with the above results, we can write
\beqa
\lefteqn{\bz^T_0 \bar\bPhi_{n-h} \bz_0} \nonumber\\
&\stackrel{(i)}{=}&\bz^T_0 E\left[
\widetilde \bW_1 \dots \widetilde \bW_{n-h}
\;E[\widetilde \bW\widetilde \bW^T] \right .\nonumber \\
&& \qquad \left .\widetilde \bW^T_{n-h}\dots \widetilde \bW^T_1
\right]
\bz_0
\nonumber \\
&\stackrel{(ii)}{=}& E\left [ \bz^T_1 E\left[
\widetilde \bW_2 \dots \widetilde \bW_{n-h} \; E[\widetilde \bW\widetilde \bW^T] \; \right . \right . \nonumber \\
&& \qquad \left . \left .\widetilde \bW^T_{n-h}\dots \widetilde \bW^T_1
\right] 
\bz_2 \right ] \nonumber \\
&& \dots \nonumber\\
&=& E \left [ \bz^T_{n-h} \; E[\widetilde \bW\widetilde \bW^T]\; \bz_{n-h} \right ] \nonumber \\
&\stackrel{(iii)}{\le}& \lambda_U \; E \left [ ||\bz_{n-h}||^2 \right ]\nonumber \\
&\stackrel{(iv)}{=}& \lambda_U \; E \left [\bz^T_{n-h-1} E[\widetilde \bW\widetilde \bW^T] \bz_{n-h-1} \right ]\nonumber \\
&\le& \lambda_U^2 \;E \left [ ||\bz_{n-h-1}||^2 \right ] \nonumber \\ && \dots \nonumber \\
&=& \lambda_U^{n-h+1} \; ||\bz_0||^2 . \nonumber
\eeqa
In the above: $(i)$ is simply eq.~(\ref{eq:Phi}); 
$(ii)$ and the successive equalities result form the definition of the sequence $\bz_k$;
the upper bound in eq.~(\ref{eq:bounds_2}) implies the inequality $(iii)$;
$(iv)$ is a consequence of eq.~(\ref{eq:norm}); the remainder of the chain
is obtained by repeating the last two steps.

A similar result is obtained by using the lower bound in eq.~(\ref{eq:bounds_2}), instead of the upper bound. 
Thus, denoting $\bz_1$ simply by $\bz$, one obtains:
\beq
\lambda_L^{n-h+1}\; ||\bz||^2
\leq
\bz^T\; \bar\bPhi_{n-h+1} \; \bz
\leq
\lambda_U^{n-h+1}\; ||\bz||^2.
\label{eq:boundsfinal}
\eeq
As $\bz$ is an arbitrary vector orthogonal to~$\bone$, we can set $\bz=e_i-\bone /M$ in eq.~(\ref{eq:boundsfinal}), 
where $e_i$ denotes a vector of zeros with only the $i^{th}$ entry equal to~1. This yields
\[
\lambda_L^{n-h+1} (M-1)/M 
\leq
(\bar\bPhi_{n-h})_{ii}
\leq
\lambda_U^{n-h+1} (M-1)/M. 
\]

Now, as seen in eq.~(\ref{eq:coverr}) the diagonal entries $(\bC_{n})_{ii}$ of the covariance matrix in eq.~(\ref{eq:C})
involve just the terms $(\bar\bPhi_{n-h})_{ii}$, which can be bounded as above. 
In this way, after straightforward algebra, we finally get
\beq
\sigma^2_{ctr}(n) \, 
\left [1+ (M-1) \psi^L_n \right ] \le
(\bC_{n})_{ii} \le \sigma^2_{ctr}(n) \, 
\left [1+ (M-1) \psi^U_n \right ] \label{eq:cii}
\eeq
and the claim in eq.~(\ref{eq:claim1}) immediately follows.

Similarly, setting $\bz=e_i-e_j$ ($i \not = j$) in eq.~(\ref{eq:boundsfinal}), yields 
\[
\lambda_L^{n-h+1} \le
\frac{ (\bar\bPhi_{n-h})_{ii}+(\bar\bPhi_{n-h})_{jj} }{2} - (\bar\bPhi_{n-h})_{ij} 
\le \lambda_U^{n-h+1} , 
\]
from which we obtain
\[
M \; \sigma^2_{ctr}(n) \, \psi^L_n \le
\frac{(\bC_{n})_{ii} + (\bC_{n})_{jj}}{2} - (\bC_{n})_{ij}
\le
M\; \sigma^2_{ctr}(n) \, \psi^U_n .
\]
Recalling the definition of $\rho_n$, and further using eq.~(\ref{eq:cii}), 
the above immediately gives eq.~(\ref{eq:claim2}).
The proof is now complete.~\hfill~$\triangle$

\chapter{} 
\section{Proof of Theorem 2}
\label{app:th1}
The proof is simple and relies on establishing that $s_{n,j}$ is asymptotically normal:
\[
\begin{array}{lcl}
&&\displaystyle{\frac{s_{n,j}-n\,M\mu(\theta_0)}{\sqrt{n\,M}\,\sigma(\theta_0)}}
\stackrel{f_{\theta_0}}{\longrightarrow} {\cal N}(0,1),
\\
&&\displaystyle{\frac{s_{n,j}-n\,M\mu(\theta_n)}{\sqrt{n\,M}\,\sigma(\theta_n)}}
\stackrel{f_{\theta_n}}{\longrightarrow} {\cal N}(0,1),
\end{array}
\]
with exactly the same normalizing functions as for $s^{(c)}_n$, implying that the two statistics share the same efficacy.
We work under $\theta=\theta_0$, the proof being exactly the same for $\theta=\theta_n$.
From eq.~(\ref{eq:error_cons}) we can write
\beq
\frac{
s_{n,j}-n\,M\,\mu(\theta_0)}{\sqrt{n\,M}\,\sigma(\theta_0)}=
\frac{s^{(c)}_n-n\,M\,\mu(\theta_0)}{\sqrt{n\,M}\,\sigma(\theta_0)}
+\frac{e_{n,j}}{\sqrt{n\,M}\,\sigma(\theta_0)}.
\label{eq:buonalaprimafix_CH2}
\eeq
It is easy to show that the second moment of 
$e_{n,j}$ is uniformly bounded with respect to $n$, see eq.~(\ref{eq:Varerr}).
Then, direct application of Chebyshev's inequality ensures that $e_{n,j}/\sqrt{n}$ goes to zero in probability, as $n$ diverges.
By observing that, thanks to eqs.~(\ref{eq:centrasynorm_CH2}), the centralized statistic is asymptotically normal, and invoking Slutsky's theorem~\cite{Lehmann-large-sample}, the claim of the theorem immediately follows.~\hfill$\bigtriangledown$

\section{Proof of Theorem 3}
\label{app:S}

The technical conditions assumed in Theorem~2 essentially cover conditions $1$-$5$ in~\cite{lai-AMS} and are therefore 
sufficient to conclude that a sequential test based on~(\ref{eq:decseq_CH2}), for $T_n=s_n^{(c)}$, with thresholds~(\ref{eq:ar_br_CH2}), attains asymptotic performances given by eqs.~(\ref{eq:pfpd_CH2}) and~(\ref{eq:ASN_CH2}). 
In order to prove that the same holds true for $s_{n,j}$, it suffices to prove that the convergence to Wiener in~(\ref{eq:asyWiener_CH2}), as well as the additional condition~(\ref{eq:addcond_CH2}), hold true with $s^{(c)}_n$ replaced by $s_{n,j}$.
We have
\[
\frac{s_{[rt],j} - [rt]\,M\,\eta_r}{\sqrt{r\,M}\,\sigma(\theta_0)}=
\frac{s^{(c)}_{[rt]} - [rt]\,M\,\eta_r}{\sqrt{r\,M}\,\sigma(\theta_0)}
+
\frac{e_{[rt],j}}{\sqrt{r\,M}\,\sigma(\theta_0)}.
\]
Was the involved convergence that between random variables, it would suffice to prove that the last term converges to zero in probability. However, here we work with convergence of random process, so that the additional requirement of {\em tightness}~\cite{billingsley-book} is needed. 
This condition (which is reminiscent of an uniform convergence with respect to $t$) can be expressed in a simple way in our case (see lemma A1 in~\cite{hall-loynes-AMS}):
\[
\displaystyle{\max_{k\leq [rT]}\ \frac{e_{k,j}}{\sqrt{r\,M}\,\sigma(\theta_0)}} \longrightarrow 0 \quad\textnormal{for any $T>0$},
\]
where the convergence is now in probability, under the pertinent distribution.
Working for instance under $\theta_r$ (the case $\theta_0$ is managed similarly), this condition is verified, in that, by joint application of the union bound and Markov inequality: 
\beqa
\lefteqn{\textnormal{P}_{\theta_r}\left[\max_{k\leq [rT]}\ \frac{e_{k,j}}{\sqrt{r\,M}\,\sigma(\theta_0)}>\epsilon \right]}\nonumber\\
&\leq& \sum_{k=1}^{[rT]} \textnormal{P}_{\theta_r}\left[\left|\frac{e_{k,j}}{\sqrt{r\,M}\,\sigma(\theta_0)}\right|^3>\epsilon^3\right]\nonumber\\ 
&\leq& \frac{1}{\epsilon^3\sqrt{M}\sigma^3(\theta_0)}\sum_{k=1}^{[rT]} \frac{\E_{\theta_r}\left[ \left|e_{k,j}\right|^3\right]}{r^{\frac 3 2}}\nonumber \\
&\leq& 
\frac{C_1(M,\lambda_U)\xi^3(\theta_r)+C_2(M,\lambda_U)\sigma^3(\theta_r)}{\epsilon^3\sqrt{M}\sigma^3(\theta_0)}
\,\frac{[rT]}{r^{\frac 3 2}}, \nonumber
\eeqa
where the last inequality follows by eq.~(\ref{eq:thirderr}) proved in Appendix~\ref{sec:lemma}, such that convergence to zero is proved.

Let us switch now to the additional conditions~(\ref{eq:addcond_CH2}). We address only the first one, the other being treated similarly. We can write
\beqa
\lefteqn{\textnormal{P}_{\theta_0}\left[\frac{s_{[rt],j}-[r\,t]\,M\,\mu(\theta_0)}{\sqrt{r\,M}\,\sigma(\theta_0)}>\epsilon t\right]}
\nonumber\\
&\leq&
\textnormal{P}_{\theta_0}\left[\frac{s^{(c)}_{[rt]}-[r\,t]\,M\,\mu(\theta_0)}{\sqrt{r\,M}\,\sigma(\theta_0)}>\frac{\epsilon t}{2}\right]\nonumber\\
&+&
\textnormal{P}_{\theta_0}\left[\frac{e_{[rt],j}}{\sqrt{r\,M}\,\sigma(\theta_0)}>\frac{\epsilon t}{2}\right]\nonumber\\
&\leq&
g_\epsilon(t/2) + 
\frac{4}{[\epsilon\,t\,M\,\sigma(\theta_0)]^2}\frac{\E_{\theta_0}[e^2_{[rt],j}]}{r}
\nonumber\\
&\leq&
g_\epsilon(t/2) + \frac{4\,C_1(M,\lambda_U)}{(\epsilon\,t\,M)^2},\nonumber
\eeqa
where the last bound follows by jointly applying Chebyshev inequality and eq.~(\ref{eq:Varerr}).
The integrability of the RHS in the last equation trivially follows by that of $g_\epsilon(t)$ and of the function $1/t^2$.
~\hfill$\bigtriangledown$




\end{document}